\documentclass[a4paper,12pt]{article}
\pdfoutput=1
\usepackage{jheppub}
\usepackage{amsmath}
\usepackage{graphicx}
\usepackage{hyperref}
\usepackage{verbatim}

\usepackage{soul}

\usepackage{amsmath,amssymb,graphicx,color,bbold}

\newcommand{\fr}[1]{\mathfrak{#1}}

\def\qfr{\mathfrak{q}}
\def\wfr{\mathfrak{w}}
\def\mfr{\mathfrak{m}}
\def\m{{\mathfrak m}}

\def\qfr{\mathfrak{q}}
\def\wfr{\mathfrak{w}}

\def\coeff#1#2{{\textstyle {\frac {#1}{#2}}}}

\def\Nf{{N_{\rm f}}}
\title{Hydrodynamics with multiple charges and holography}

\author[a]{Liam Gladden,}
\author[b]{Victor  Ivo,}
\author[c]{Pavel K. Kovtun}
\author[a]{and Andrei O. Starinets}

\affiliation[a]{Rudolf Peierls Centre for Theoretical Physics, University of Oxford, \\ Parks Road,  
Oxford, OX1 3PU, UK}

\affiliation[b]{ Jadwin Hall, Princeton University, Princeton, NJ 08540, USA}

\affiliation[c]{Department of Physics \& Astronomy, University of Victoria, \\ PO Box 1700 STN CSC, Victoria,
BC, V8W 2Y2, Canada}

\emailAdd{vi6083@princeton.edu}

\emailAdd{pkovtun@uvic.ca}

\emailAdd{andrei.starinets@physics.ox.ac.uk}

\abstract{
We establish the connection between thermodynamic and dynamical instabilities in relativistic hydrodynamics with multiple flavours of conserved $U(1)$ charges. In theories with positive hydrodynamic entropy production, where the underlying perfect fluid has a positive speed of sound squared and satisfies the null energy condition, we show that hydrodynamic instabilities can arise only through negative diffusion coefficients associated with the $U(1)$ charges. The onset of such instabilities is governed by the eigenvalues of the thermodynamic Hessian matrix, while the flavour-space polarisations of the unstable diffusion modes are determined by the corresponding eigenvectors. We illustrate this connection using strongly coupled ${\cal N}=4$ supersymmetric Yang-Mills theory at finite densities of the three $U(1)$ R-charges. In the dual holographic description, the five-dimensional STU black brane exhibits unstable quasinormal modes precisely at the onset of thermodynamic instability. We derive analytic expressions for the R-charge diffusion coefficients in several representative cases, including the configuration with three equal chemical potentials.
}

\keywords{Gauge-string duality, relativistic fluid dynamics, quasinormal spectrum, finite density, strongly interacting quantum theories}

\begin{document}
\maketitle
\flushbottom

\section{Introduction}
\label{intro}
Relativistic hydrodynamics provides an effective macroscopic description of relativistic matter. It plays a crucial role in the phenomenology of heavy-ion collisions and the physics of the quark-gluon plasma~\cite{Romatschke:2017ejr, Heinz:2024jwu}, and also finds important applications in astrophysics~\cite{Baiotti:2016qnr, Annala:2019puf, Lovato:2022vgq}. The fundamental equations of relativistic hydrodynamics are the local conservation laws of energy and momentum, which are compactly expressed as the covariant conservation of the energy-momentum tensor~\cite{LL6}.

Depending on the physical system and the symmetries of the underlying microscopic theory, additional conservation laws may need to be incorporated into the hydrodynamic description. For instance, ongoing and upcoming experiments probing the quark-gluon plasma at finite baryon density—such as those at RHIC~\cite{rhic}, NICA~\cite{nica,MPD:2022qhn}, and FAIR~\cite{fair}—especially in the vicinity of the conjectured critical point of nuclear matter, require the inclusion of the baryon-number current,   which arises due to a global $U(1)$ baryon-number symmetry of the microscopic theory~\cite{Lovato:2022vgq}.  Similarly, in magnetohydrodynamics, conservation of magnetic flux must be accounted for~\cite{Goedbloed_Keppens_Poedts_2019}, and other symmetries may give rise to further conserved quantities.

Here we are interested in relativistic hydrodynamics of systems which have multiple $U(1)$ symmetries, for $\Nf$ species (flavours) of conserved quantities. An example from subnuclear matter is hydrodynamics with $\Nf=3$, corresponding to the conservation of baryon number, strangeness, and electric charge~\cite{Greif:2017byw, Plumberg:2024leb}. An example from astrophysics are two-fluid magneto-hydrodynamic theories with $\Nf=2$, corresponding to the conservation of electron and ion numbers~\cite{Most:2021uck}. The example which we will study later in the paper is the ${\cal N}=4$ supersymmetric Yang-Mills theory which has $\Nf=3$, corresponding to the conservation of three Abelian R-charges. This theory provides a particularly clean theoretical laboratory, allowing for analytical derivations of both thermodynamic and hydrodynamic parameters.

The goal of this paper is two-fold: to develop the theory of hydrodynamic modes in relativistic theories with a global $U(1)^\Nf$ symmetry, and to apply the results to strongly coupled ${\cal N}=4$ supersymmetric Yang-Mills (SYM) theory via its dual holographic description~\cite{Aharony:1999ti}. In doing so, we will elucidate the connection between thermodynamic and hydrodynamic stability of equilibrium states in field theory. Via the holographic duality, this translates to the connection between thermodynamic and hydrodynamic stability for charged anti-de Sitter (AdS) black branes. % We will expand on the results of~\cite{Gladden:2024ssb}, analyzing the hydrodynamic low-temperature instability in more detail.  

Physically, the connection between thermodynamic and dynamical instabilities arises naturally within the theory of fluctuations~\cite{callen-greene, Kubo:1991,zubarev-book, kvasnikov-book-3, McQuarrie:2000,pathria-book}. Dynamically, instabilities manifest as perturbations of a gravitational background that grow uncontrollably in time—or, in Fourier space with dependence $\sim e^{-i \omega t}$, as quasinormal modes whose imaginary part crosses into the upper half-plane of complex frequency. Thermodynamically, instabilities can be viewed as growing fluctuations in a system in thermal equilibrium. Recall that the probability of a fluctuation in a thermodynamic system in thermal equilibrium is given, in the microcanonical ensemble, by Einstein's formula~\cite{Einstein-1910}:
\begin{equation}
w_\Delta \propto e^{\Delta S}\,,
\label{einstein-1}
\end{equation}
where $\Delta S = S' - S$ is the difference between the entropy $S'$ of a near-equilibrium state resulting from the fluctuation and $S$, the entropy of the system in thermal equilibrium. The fluctuations are assumed to be characterised by sufficiently long time and large spatial scales to allow the formation of local thermodynamic equilibrium. Since these are precisely the conditions under which the hydrodynamic regime is valid, the dynamics of the fluctuations is governed by hydrodynamic variables. In holography, these hydrodynamic modes correspond to gapless quasinormal modes of the dual gravitational background. Thus, Eq.~\eqref{einstein-1} should allow one to predict which dynamical perturbations of the gravitational background become unstable, and at what parameter values this occurs.

For small fluctuations characterised by parameters $\xi_1, \ldots, \xi_n$, the difference $\Delta S$ can be expanded in a Taylor series around the equilibrium point $\xi_i = 0$, $i = 1, \ldots, n$, resulting in
\begin{equation}
w_\Delta \sim \exp{\left\{ - \frac{1}{2} \sum\limits_{ik} \lambda_{ik} \xi_i \xi_k \right\}}\,,
\label{fluctuation-01}
\end{equation}
where $\lambda_{ik} = -\left(\partial^2 S / \partial \xi_i \partial \xi_k \right)\big|_{\xi_i=0}$. For a stable thermodynamic equilibrium, the quadratic form $\sum\limits_{ik} \lambda_{ik} \xi_i \xi_k$ must be positive definite.  Introducing the thermodynamic variables $y_i \equiv (s, n_k)$, where $s$ is the entropy density and $n_k$ are the densities of conserved charges\footnote{We assume a spatially homogeneous system and normalise all extensive parameters by volume.}, Eq.~\eqref{fluctuation-01} becomes (see Appendix~\ref{einstein-formula-appendix})
\begin{equation}
w_\Delta \sim \exp{\left\{ - \frac{1}{2 T} \sum\limits_{ij} \frac{\partial^2 \varepsilon}{\partial y_i \partial y_j} \Delta y_i \Delta y_j \right\}}\,,
\label{fluctuation-02}
\end{equation}
where $\varepsilon$ is the internal energy density. The eigenvalues and eigenvectors of the Hessian
\begin{equation}
H^\varepsilon = \frac{\partial^2 \varepsilon}{\partial y_i \partial y_j}
\label{hessian-s}
\end{equation}
identify the unstable hydrodynamic modes and the corresponding dual quasinormal modes of the gravitational background. Our goal is to make this connection precise.

We will start in Section~\ref{hydro} by developing hydrodynamics of relativistic theories with $\Nf$ flavours of $U(1)$  currents. Perfect-fluid hydrodynamics requires the equation of state, say, for the pressure $p=p(T,\mu_a)$, where $T$ is temperature, and $\mu_a$ are the chemical potentials, $a=1\dots\Nf$. Thermodynamic stability conditions place constraints on the form of the equation of state which we discuss. Leading-order dissipative hydrodynamics further requires the input of transport coefficients: shear viscosity $\eta$, bulk viscosity $\zeta$, and the $\Nf\times\Nf$ conductivity matrix $\sigma_{ab}$, which are all in general functions of $T$ and $\mu_a$. Positivity of hydrodynamic entropy produciton places positivity constraints on hydrodynamic transport coefficients~\cite{LL6}. We then proceed to discuss small hydrodynamic perturbations of the thermal equilibrium state, and obtain dispersion relations of hydrodynamic modes of the form  $\omega = \omega(q)$, where $\omega$ is the frequency, and $q$ is the wave vector. Hydrodynamic stability of equilibrium can be expressed as ${\rm Im}\, \omega(q) < 0$ at small real $q$. The dispersion relations depend on the derivatives of the pressure $p$, as well as on $\eta$, $\zeta$, $\sigma_{ab}$. Thus, hydrodynamics provides an explicit relation between thermodynamic stability conditions (which constrain the derivatives of $p$) and the actual dynamic stability of equilibrium. 

As a simple example relating thermodynamic and hydrodynamic stability, consider diffusion of a $U(1)$ charge density $n$ in a charge-neutral state with $\mu=0$. The diffusion coefficient is $D = \sigma/\chi$, where $\sigma$ is the charge conductivity, and $\chi = (\partial n/\partial\mu)_{\mu=0}$ is the static charge susceptibility~\cite{Kovtun:2012rj}. In  momentum space, diffusion will manifest as a mode with frequency $\omega(q) = - i D q^2 + O(q^4)$. For positive hydrodynamic entropy production ($\sigma>0$), in a state which is thermodynamically stable ($\chi>0$), one has ${\rm Im}\, \omega(q) < 0$ at real $q$. On the other hand, a violation of thermodynamic stability ($\chi<0$) together with positive entropy production ($\sigma>0$) gives rise to ${\rm Im}\, \omega(q) > 0$, a dynamic instability caused by violation of thermodynamic stability. In Section~\ref{hydro}, we will generalise this simple  $\Nf=1$, $\mu=0$ example to relativistic hydrodynamics (including shear and sound modes) with $\Nf>1$, $\mu_a \neq 0$. We will show that in theories where perfect fluids have positive speed of sound squared, and obey the null energy condition, hydrodynamic instabilities can only appear through negative diffusion coefficients for the $U(1)$ charges.

We will continue in Section~\ref{stu-model-overview} with the holographic description of $SU(N_c)$, ${\cal N}=4$ SYM theory at large $N_c$ and at strong coupling~\cite{Maldacena:1997re, Gubser:1998bc, Witten:1998qj}. The SYM theory possesses an $SU(4)$ R-symmetry group, and $\Nf=3$ chemical potentials $\mu_a$ can be introduced as grand canonical variables conjugate to the global charges in the Cartan subalgebra $U(1)^3$ of $SU(4)$~\cite{Yamada:2006rx}. In the ten-dimensional string theory dual, the corresponding background involves rotating black three-branes \cite{Cai:1998ji, Cvetic:1999ne, Cvetic:1999rb, Harmark:1999xt}. Upon dimensional reduction on $S^5$, rotating black brane solutions in ten dimensions become charged black brane solutions in five dimensions supported by the gauge and neutral scalar fields of five-dimensional $\mathcal{N}=8$ supergravity \cite{Cvetic:1999xp, Cvetic:2000nc}.

The simplest solution of the reduced five-dimensional theory describing equilibrium states of ${\cal N}=4$ SYM theory at nonzero temperature and three chemical potentials, was discovered by Behrndt, Cvetic, and Sabra \cite{Behrndt:1998jd} and is known as the STU background. This solution includes the metric, three $U(1)$ gauge fields, and two independent real scalar fields. In the special case where all three chemical potentials are equal, the background reduces to the Reissner–Nordstr\"om black brane in five dimensions. 

The STU background exhibits low-temperature thermodynamic instabilities at finite non-zero values of $\mu_a/T$ \cite{Cvetic:1999ne, Cvetic:1999rb, Cai:1998ji, Gubser:2000ec, Gubser:2000mm, Son:2006em, Buchel:2010gd, Gentle:2012rg, Henriksson:2019zph, Anabalon:2024lgp}. The instability persists when all $\mu_a$ are taken equal, and the Reissner–Nordstr\"om black brane in the STU truncation is both thermodynamically and dynamically unstable at low temperature~\cite{Gladden:2024ssb}.

In this paper, we consider ${\cal N}=4$ SYM in the grand canonical ensemble in Minkowski space, rather than on a sphere. The latter implies the absence of the Hawking--Page transition and the equivalence of all ensembles in the thermodynamic limit.  Building on the results of Section~\ref{hydro}, we discuss the connection between thermodynamic and (hydro)dynamic instabilities of the STU background, extending our earlier results in ref.~\cite{Gladden:2024ssb}. Historically,  the equivalence between thermodynamic and dynamic instabilities for black branes was conjectured by Gubser and Mitra \cite{Gubser:2000ec,Gubser:2000mm} and subsequently extensively discussed (see, for example~\cite{Reall:2001ag, Buchel:2005nt, Emparan:2008eg}). Following the development of real-time correlation functions in holography~\cite{Son:2002sd, Kovtun:2005ev}, it became clear that instabilities of the backgrounds which describe thermal states in dual field theories manifest themselves as quasinormal modes crossing into the upper half-plane of complex frequency (at real wave vector).  

The equation of state corresponding to the STU background is known analytically~\cite{Son:2006em}, the bulk viscosity $\zeta$ vanishes by conformal symmetry of the SYM theory, and the shear viscosity is proportional to the entropy density, $\eta = s/4\pi$ \cite{Son:2006em, Mas:2006dy}, reflecting the holographic universality~\cite{Kovtun:2003wp,Kovtun:2004de,Buchel:2003tz,Starinets:2008fb}. What is not known is the analytic form of the R-charge conductivity matrix $\sigma_{ab}(T, \mu_1, \mu_2, \mu_3)$. In this paper, we will derive analytic expressions for the conductivity matrix in several illustrative cases, obtaining $\sigma_{ab}(T, \mu, 0, 0)$, $\sigma_{ab}(T, \mu, \mu, 0)$, and $\sigma_{ab}(T, \mu, \mu, \mu)$. With the conductivity matrix at hand, we can find the dispersion relations of hydrodynamic modes. This is discussed in Sections~\ref{single-kappa-section}, \ref{section-two-kappas}, and \ref{3-equal-kappas-section} for the three respective cases. As expected, we find that R-charge diffusion coefficients become negative at the onset of thermodynamic instability. Negative diffusion coefficients signify charge clumping for near-equilibrium perturbations in ${\cal N}=4$ SYM theory at low temperature. In particular, extremal $T=0$ AdS black branes in the STU truncation are both thermodynamically and dynamically unstable.

We discuss our results in Section~\ref{sec:discussion}.

%
%To extend this discussion to the case of several chemical potentials, we 
%need to develop  relativistic hydrodynamics with multiple charges. This is done in section \ref{hydro}. 
%

%The unstable modes we discuss here were also observed in \cite{Gentle:2012rg} at zero spatial momentum.

%%%%%%%%%%%%%%%%%%%%%%%%%%%%%%%%%%%%

\section{Hydrodynamics with several conserved charges}
\label{hydro}

\subsection{Thermodynamic stability}
We will study field theories with conserved $U(1)$ charges $Q_a$ of species (flavours) $a = 1, \dots ,\Nf$, so that $Q_a$ are the generators of a global symmetry $U(1)^\Nf$. We focus on thermodynamic relations for a system at rest, in volume $V$ in the thermodynamic limit $V\to\infty$, with charge densities $n_a \equiv \langle Q_a \rangle /V$ fixed.  The equation of state in the grand canonical ensemble is given by the pressure, $p = p(T,\mu_a)$, with 
\begin{align}
\label{eq:dp-gc}
   dp = s dT + n_a d\mu_a \,.
\end{align}
Here $T$ is temperature, $\mu_a$ are chemical potentials, $s = s(T,\mu_1, \dots,\mu_\Nf)$ is the entropy density, and $n_a = n_a(T,\mu_1, \dots, \mu_\Nf)$ are charge densities.  There is an implied summation over the repeated flavour indices. 
The equation of state in the micro-canonical ensemble is given by $s = s(\epsilon, n_a)$, where $\epsilon$ is the density of energy, and 
\begin{align}
\label{eq:ds-mc}
  ds  = \frac{1}{T} d\epsilon - \frac{\mu_a}{T} dn_a \,.
\end{align}
Here $T = T(\epsilon, n_1, \dots, n_\Nf)$ and $\mu_a = \mu_a(\epsilon, n_1, \dots, n_\Nf)$. 
Equations \eqref{eq:dp-gc} and \eqref{eq:ds-mc} are related through $\epsilon+p = Ts + \mu_a n_a$. 

Thermodynamic stability conditions~\cite{Callen} imply that the matrix of the second derivatives of entropy density $s(\epsilon, n^a)$ is negative definite in a stable equilibrium state, reflecting the maximum of entropy. Similarly, the matrices of the second derivatives of $\epsilon(s, n_a)$ and of  $p(T,\mu_a)$ are positive definite, reflecting the minimum of energy, and of the grand-canonical free energy. We define Hessian matrices $h^\epsilon \equiv \partial^2 \epsilon / \partial y_i \partial y_j$, $h^p \equiv \partial^2 p / \partial z_i \partial z_j$, where $y_i=(s, n_a)$, and $z_i = (T, \mu_a)$. Thermodynamic relations \eqref{eq:dp-gc} and \eqref{eq:ds-mc} imply $h^\epsilon = \partial z_i /\partial y_j$, $h^p = \partial y_i /\partial z_j$, and thus $h^p = (h^\epsilon)^{-1}$. In particular, in a thermodynamically stable system, $ T \partial s /\partial T = ({\partial\epsilon}/{\partial T}) -  {\mu_a} ({\partial n_a}/{\partial T}) > 0$, and the matrix  $\chi_{ab} \equiv (\partial n_a /\partial\mu_b)_T$ of charge susceptibilities must be positive definite. 
Passing to variables $T$ and $\gamma_a \equiv \mu_a/T$, we have another useful susceptibility matrix
\begin{align}
\label{eq:XXmatr}
  {\cal X} = \begin{pmatrix}
  T \!\left( \frac{\partial \epsilon}{\partial T} \right)_\gamma & \left( \frac{\partial\epsilon}{\partial \mu_a} \right)_{\!T} \\[8pt]
  T \!\left( \frac{\partial n_a}{\partial T} \right)_\gamma & \left( \frac{\partial n_a}{\partial \mu_b} \right)_{\! T}
  \end{pmatrix} 
  =
    \begin{pmatrix} T & \mu_a \\ 0_a & {\bf 1} \end{pmatrix}
    \begin{pmatrix} \frac{\partial^2 p}{\partial T^2} & \frac{\partial^2 p}{\partial T \partial \mu _a} \\[8pt]\frac{\partial^2 p}{\partial \mu_a \partial T} & \frac{\partial^2 p}{\partial\mu_a \partial\mu_b} \end{pmatrix}
    \begin{pmatrix} T & 0_a \\ \mu_a & {\bf 1} \end{pmatrix} \,.
\end{align}
The matrix ${\cal X}$ is positive-definite because $h^p$ is. In the grand canonical ensemble, the probability distribution is proportional to $\exp[-(H - \mu_a Q_a)/T]$, where $H$ is the Hamiltonian, thus
\begin{align}
   & T \!\left( \frac{\partial \epsilon}{\partial T} \right)_{\!\gamma} = \frac{1}{V T} \langle H^2 \rangle_{\rm conn} \,, \\[5pt]
   & \left( \frac{\partial\epsilon}{\partial \mu_a} \right)_{\!T} = T \!\left( \frac{\partial n_a}{\partial T} \right)_{\!\gamma} = \frac{1}{V T} \langle H Q_a \rangle_{\rm conn} \,, \\[5pt]
   &  \left( \frac{\partial n_a}{\partial \mu_b} \right)_{\! T} = \frac{1}{V T} \langle Q_a Q_b \rangle_{\rm conn} \,,
\end{align}
where the subscript signifies connected averages, $\langle H^2 \rangle_{\rm conn} = \langle H^2 \rangle - \langle H \rangle^2$ etc. This gives a more direct perspective on the positive-definiteness of ${\cal X}$.

\subsection{First-order hydrodynamics}

We next consider first-order relativistic hydrodynamics for theories with $\Nf$ flavours of $U(1)$ charges, in $d$ spatial dimensions. The hydrodynamic equations are the conservation laws for the energy-momentum tensor $T^{\mu\nu}$ and the $U(1)$ currents $J_a^\mu$, 
 \begin{align}
 \label{eq:TJ-conserv}
   \partial_\mu T^{\mu\nu} = 0, \ \ \ \ \partial_\mu J^\mu_a = 0 \,,
\end{align}
where $a = 1, \dots, \Nf$. 
The conservation laws must be supplemented by the constitutive relations. In the conventions of Landau and Lifshitz~\cite{LL6}, these are
\begin{align}
  & T^{\mu\nu} = \epsilon u^\mu u^\nu + p \Delta^{\mu\nu} - \eta S^{\mu\nu} - \zeta \Delta^{\mu\nu} \partial_\lambda u^\lambda + O(\partial^2) \,,\\
\label{eq:Ja-const-rel}
  & J^\mu_a = n_a u^\mu - \sigma_{ab} T \Delta^{\mu\nu} \partial_\nu \gamma_b + O(\partial^2) \,,
\end{align}
with implied summation over the repeated flavour indices. 
Here $u^\mu$ is the fluid velocity, $\Delta^{\mu\nu} \equiv g^{\mu\nu} + u^\mu u^\nu$ is the projector onto the space orthogonal to the fluid velocity, $g^{\mu\nu}$ is the inverse metric, and $S^{\mu\nu} \equiv \Delta^{\mu\alpha} \Delta^{\nu\beta} (\partial_\alpha u_\beta + \partial_\beta u_\alpha - \frac{2}{d} g_{\alpha\beta} \partial_\lambda u^\lambda)$ is the shear tensor.  
The thermodynamic functions $\epsilon$ (energy density), $p$ (pressure) and $n_a$ (charge densities) are functions of $T$ and $\gamma_a \equiv \mu_a/T$, whose explicit form is given by the equilibrium equation of state. The one-derivative transport coefficients $\eta$ (shear viscosity), $\zeta$ (bulk viscosity) and $\sigma_{ab}$ (charge conductivity matrix) are in general functions of $T$ and $\gamma_a$. 

The conductivity matrix $\sigma_{ab}$ in eq.~\eqref{eq:Ja-const-rel} is not arbitrary. If the microscopic system respects time-reversal symmetry (which we will assume), $\sigma_{ab}$ will be symmetric by the Onsager relations. Further, the divergence of the entropy current has an additive contribution $\sigma_{ab} \Delta^{\mu\nu} \partial_\mu \gamma_a \partial_\nu \gamma_b$. Hence, the matrix $\sigma_{ab}$ must be positive semi-definite, in order to ensure non-negative entropy production out of equilibrium~\cite{LL6}. Similarly, non-negative entropy production demands $\eta\geq 0$, $\zeta \geq 0$. Alternatively, linear response theory gives $\eta$, $\zeta$, and $\sigma_{ab}$ in terms of correlation functions of $T^{\mu\nu}$ and $J^\mu_a$. In relativistic hydrodynamics, the corresponding Kubo formulae are~\cite{Kovtun:2012rj}
\begin{align}
  & \eta = \frac{1}{2T} \lim_{\omega\to0} G_{T^{xy} T^{xy}} (\omega, {\bf q}=0) \,,\\[5pt]
  & \zeta = \frac{1}{2T d^2} \lim_{\omega\to0} G_{T^{ii} T^{kk}} (\omega, {\bf q}=0)\,, \\[5pt]
  & \sigma_{ab} = \frac{1}{2T d} \lim_{\omega\to0} G_{J_a^{i} J_b^{i}} (\omega, {\bf q}=0) \,,
\label{eq:Kubo-sigma}
\end{align}
where $G_{O_1 O_2}(\omega, {\bf q})$ denotes the Fourier transform of the symmetrized correlation function of $O_1$ and $O_2$ in equilibrium. 
The positivity conditions for $\eta$, $\zeta$, and $\sigma_{ab}$ then follow from the positivity conditions for the corresponding correlation functions of Hermitean operators $T^{ij}$ and $J_a^i$.

Let us now work out how derivatives of thermodynamic functions appear in the dispersion relations of hydrodynamic modes. To do so, consider small fluctuations about the equilibrium state of the fluid at rest with constant temperature $\bar T$ and chemical potentials $\bar \mu_a$. Namely, we take $u^\lambda = (1, {\bf 0}) + \delta u^\lambda$, $T = \bar T + \delta T$,  $\gamma_a = \bar \gamma_a + \delta \gamma_a$, assume that all fluctuations only depend on $t$ and $z$, and linearise.  The conservation equations \eqref{eq:TJ-conserv} for energy, momentum, and $U(1)$ currents then give linear equations for fluctuations $\delta T$, $\delta u^i$, and $\delta \gamma_a$. Transverse velocity fluctuations $\delta u^{i}$ with $i\neq z$ decouple from the rest, and satisfy 
\begin{align}
\label{eq:E-shear}
  \left( (\epsilon{+}p) \partial_t  - \eta \partial_z^2 \right) \delta u^i = 0 \,.
\end{align}
Longitudinal velocity fluctuations $\delta u^z$ couple to $\delta T$ and $\delta \gamma_a$, and satisfy
\begin{align}
\label{eq:E-linear-2}
      & \left( \frac{\partial \epsilon}{\partial T}\right)_{\!\!\gamma} \partial_t \delta T + T \! \left( \frac{\partial \epsilon}{\partial \mu_a}\right)_{\!\!T} \partial_t \delta \gamma_a + (\epsilon {+} p) \partial_z \delta u^z  = 0 \,,\\
\label{eq:P-linear-2}
    &  \left( \epsilon{+}p \right) \partial_t \delta u^z + \left( \frac{\partial p}{\partial T}\right)_{\!\!\gamma} \partial_z \delta T + T \! \left( \frac{\partial p}{\partial \mu_a}\right)_{\!\!T} \partial_z \delta \gamma_a - (\coeff{2d-2}{d} \eta + \zeta) \partial_z^2 \delta u^z = 0\,, \\
\label{eq:J-linear-2}
    & \left( \frac{\partial n_a}{\partial T}\right)_{\!\!\gamma}  \partial_t \delta T +  n_a \, \partial_z \delta u^z + T \left( \left(\frac{\partial n_a}{\partial\mu_b}\right)_{\!T}  \partial_t - \sigma_{ab}\, \partial_z^2  \right) \delta \gamma_b  = 0 \,,
\end{align}
where the coefficients are evaluated in equilibrium, and we omit the bar. Introducing the vector $V_A\equiv(\delta T/T^2, \delta u^z/T, \delta \gamma_a)$, these linearised equations can be written in the following matrix form: 
\begin{align}
\label{eq:E-linear-m}
  \left[ X_{AB} \partial_t + Y_{AB} \partial_z - Z_{AB} \partial_z^2 \right] V_B = 0 \,,
\end{align}
where the matrices $X$, $Y$, $Z$ are
\begin{align}
  X = \begin{pmatrix} C & 0 & \upsilon_a \\ 0 & \epsilon{+}p & 0_a \\ \upsilon_a & 0_a & \chi_{ab} \end{pmatrix} ,\ \ \ \ 
  Y = \begin{pmatrix} 0 & \epsilon{+}p & 0_a \\ \epsilon{+}p & 0 & n_a \\ 0_a & n_a & 0_{ab} \end{pmatrix} ,\ \ \ \ 
  Z = \begin{pmatrix} 0 & 0 & 0_a \\ 0 & \eta_s & 0_a \\ 0_a & 0_a & \sigma_{ab} \end{pmatrix} .
\end{align}
where $\eta_s \equiv \frac{2d-2}{d}\eta + \zeta$. The elements of the matrix $X$ are thermodynamic derivatives: 
\begin{align}
  C \equiv T \left( \frac{\partial\epsilon}{\partial T} \right)_{\!\!\gamma} \,, \ \ \ \ 
  \upsilon_a \equiv \left( \frac{\partial \epsilon}{\partial\mu_a}\right)_{\!T}  = T \left( \frac{\partial n_a}{\partial T} \right)_{\!\!\gamma} , \ \ \ \ 
  \chi_{ab} \equiv \left(\frac{\partial n_a}{\partial\mu_b}\right)_{\!T} .
\end{align}
The matrix $X_{AB}$ is symmetric; in a stable equilibrium state it is also positive definite, cf.~\eqref{eq:XXmatr}.
The matrix $Y_{AB}$ comes from the fluxes in perfect-fluid hydrodynamics. It has $\Nf$ vanishing eigenvalues, reflecting $\Nf$ diffusion modes which are invisible in perfect-fluid hydrodynamics. It also has two non-zero eigenvalues,
reflecting two sound modes which {\em are} visible in perfect-fluid hydrodynamics. The matrix $Z_{AB}$ comes from one-derivative terms in the fluxes; it is degenerate, and has one vanishing eigenvalue reflecting the Landau-frame convention for the constitutive relations. 

Next, consider plane-wave fluctuations, with $\delta T$, $\delta u^z$, and $\delta\gamma$ proportional to $e^{-i\omega t + i q z}$, where $q$ is the component of the three-momentum ${\bf q}$  along $z$,    ${\bf q} = (0,0,q)$. The linearised equations \eqref{eq:E-linear-m} then become:
\begin{align}
\label{eq:E-linear-m-3}
  \left( -\omega X + q Y - i q^2 Z \right) V = 0 \,.
\end{align}
Thus, the eigenfrequencies $\omega(q)$ can be found by solving the equation 
\begin{align}
  \det \left[ -\omega X + q Y - i q^2 Z \right] = 0\,,
\end{align}
with the above matrices $X$, $Y$, $Z$. In hydrodynamics, we work in the small-$q$ expansion, and our main interest is the small-$q$ behaviour of $\omega(q)$. We now discuss the hydrodynamic modes in more detail. In $d+1$ spacetime dimensions, there are $d+1+\Nf$ hydrodynamic modes.

\subsubsection*{Shear modes}
Shear modes are fluctuations of the transverse fluid velocity (or momentum density). Their dispersion relations follow from eq.~\eqref{eq:E-shear}, and are given by
\begin{align}
\label{eq:w-shear}
  \omega_{(i)}^{\rm shear}(q) = -i \frac{\eta}{\epsilon+p} q^2 + \dots \,,
\end{align}
where $i=2,\dots,d$, reflecting $d{-}1$ shear modes with identical frequencies and different polarizations. The decoupling of shear modes from the longitudinal fluctuations is a consequence of rotation symmetry. Both $\eta$ and $\epsilon{+}p$ in eq.~\eqref{eq:w-shear} are in principle functions of $T$ and $\mu_a$.

\subsubsection*{Diffusion modes}
The diffusion modes are contained in the coupled fluctuations of $\delta T$, $\delta u^z$, and $\delta \gamma_a$. There are $\Nf$ diffusion modes, whose eigenfrequencies are
\begin{align}
\label{eq:wD}
   \omega_{(a)}^{\rm diffusion}(q) = - i D_{(a)} q^2 + \dots \,,
\end{align}
where $D_{(a)}$ are diffusion coefficients. In a neutral state ($\mu_a = 0, n_a = 0$ in equilibrium), eq.~\eqref{eq:J-linear-2} implies that $\delta\gamma_a$ decouple from $\delta T$, $\delta u^z$, and the diffusion coefficients $D_{(a)}$ are given by the eigenvalues of the $\Nf \times \Nf$ diffusion matrix 
\begin{align}
  {\cal D} = \chi^{-1} \sigma \,.
\end{align}
In a state with $\mu_a$, $n_a$ non-vanishing in equilibrium, fluctuations $\delta\gamma_a$ couple to $\delta T$, $\delta u^z$. However, even then, the diffusion modes may be separated from the sound modes at small~$q$. 
Denote the $\Nf$ eigenvectors of $Y$ with zero eigenvalues as $W^{(a)}$,  
\begin{align}
 W^{(1)} = (n_1, 0, -(\epsilon{+}p), 0, \dots, 0) \,, \ \ \ \
 W^{(2)} = (n_2, 0, 0, -(\epsilon{+}p), 0, \dots, 0) \,, \ \ \ \  {\rm etc},
\end{align}
up to $W^{(\Nf)} = (n_\Nf, 0, \dots, 0, -(\epsilon{+}p))$. 
These vectors describe diffusive fluctuations to leading order in the small-$q$ expansion; in other words, the diffusion modes are the fluctuations in which
\begin{align}
  \frac{\epsilon + p - n_1 \mu_1 }{T} \delta T + n_1 \delta \mu_1  = 0 \,,\ \ \ \ 
  \frac{\epsilon + p - n_2 \mu_2 }{T} \delta T + n_2 \delta \mu_2  = 0 \,,\ \ \ \  {\rm etc},
\end{align}
or linear combinations thereof. For $\Nf=1$, the single diffusive mode has $\delta p = s \delta T + n \delta \mu =0$, while for $\Nf>1$, the diffusive modes have $\delta p \neq 0$ (in a state with all non-vanishing $n_a$). 
Let us pass to the basis of $W^{(a)}$, and define the following $\Nf \times \Nf$ matrices:
\begin{align}
  (M_X)^{ab} = W^{(a)} {\cdot} X {\cdot} W^{(b)}\,,\ \ \ \ 
  (M_Z)^{ab} = W^{(a)} {\cdot} Z {\cdot} W^{(b)} \,,
\end{align}
with $a,b = 1, \dots, \Nf$.
The $\Nf$ diffusion eigenfrequencies $\omega_{(a)}$ are again given by eq.~\eqref{eq:wD},
where $D_{(a)}$ are eigenvalues of the $\Nf \times \Nf$ diffusion matrix 
\begin{align}
\label{eq:D-matr}
  {\cal D} \equiv M_X^{-1} M_Z \,.
\end{align}
In a thermodynamically stable equilibrium state, $X$ is positive definite, and so is $M_X$. The matrix $M_Z = (\epsilon{+}p)^2 \sigma$ is positive semi-definite because the matrix $\sigma$ is. Thus, the matrix ${\cal D}$ is positive semi-definite,  so its eigenvalues $D_{(a)}$ are non-negative. Note that ${\cal D}$ does not depend on the viscosities. In a state with $n_a=0$, the diffusion matrix \eqref{eq:D-matr} reduces to ${\cal D} = \chi^{-1} \sigma$.

\subsubsection*{Sound modes}
The sound modes are also contained in the coupled fluctuations of $\delta T$, $\delta u^z$, and $\delta \gamma_a$. There are two sound modes, whose eigenfrequencies are 
\begin{align}
\label{eq:wS}
  \omega_\pm^{\rm sound}(q) = \pm c_s |q| - \frac{i}{2} \Gamma q^2 + \dots \,,
\end{align} 
where $c_s$ is the speed of sound, and $\Gamma$ is the sound damping coefficient, which depends on both the viscosities and charge conductivities.
The $\pm c_s$ in eq.~\eqref{eq:wS} are the two non-zero eigenvalues of $X^{-1} Y$. Using the two eigenvectors of $Y$ with non-zero eigenvalues, \begin{align}
  S^{(1,2)} = (\epsilon{+}p, \pm((\epsilon{+}p)^2 + n_a n_a)^{1/2}, n_1, \dots, n_\Nf),
\end{align}
one can write $c_s^2$ as
\begin{align}
\label{eq:cs2-1}
  c_s^2 = \frac{{\rm det} \left( S^{(\alpha)} X^{-1} S^{(\beta)} \right)}{ S^{(1)} S^{(1)} + S^{(2)} S^{(2)} } \,.
\end{align}
This gives $c_s^2$ in terms of thermodynamic derivatives of the grand canonical equation of state $p=p(T,\mu_1, \mu_2, \dots)$.  Alternatively, $c_s^2$ may be written using the matrix $M_X$,
\begin{align}
\label{eq:cs2-2}
  c_s^2 = \frac{1}{(\epsilon{+}p)^{2(\Nf-1)}} \frac{\det M_X}{\det X} \,.
\end{align}
A more direct way to obtain the speed of sound is to use the equation of state in the form $p=p(\epsilon, n_1, n_2, \dots)$, which gives
\begin{align}
  c_s^2 = \left( \frac{\partial p}{\partial\epsilon} \right)_{\!\!n_a} + \sum_a \frac{n_a}{\epsilon+p} \left( \frac{\partial p}{\partial n_a} \right)_{\!\!\epsilon} \,.
\end{align}
Note that the equation of state $p=p(\epsilon, n_a)$, while suitable for perfect-fluid hydrodynamics, can not be used to close the equations of dissipative hydrodynamics because the constitutive relations of the latter are formulated using $T$ and $\mu_a$, not $\epsilon$ and $n_a$. 

The sound damping coefficient is given by the standard perturbation theory as 
\begin{align}
\label{eq:Gamma-1}
  \Gamma = 2 \frac{V_+ Z V_+}{V_+ X V_+} =  2 \frac{V_- Z V_-}{V_- X V_-} \,,
\end{align}
where $V_\pm$ are eigenvectors of $X^{-1}Y$ with eigenvalues $\pm c_s$. The contributions to $\Gamma$ from viscosity and the charge conductivity decouple. To see this explicitly, denote the components of the eigenvectors by $V_\pm = (V_\pm^\epsilon, V_\pm^\pi, V_\pm^a)$, with $V_+^\epsilon = V_-^\epsilon$, $V_+^a = V_-^a$. The eigenvectors can be normalized such that $V_\pm^\pi = \pm c_s \det(X)$. Then we have
 \begin{align}
\label{eq:Gamma-2}
  \Gamma = \frac{\coeff{2d-2}{d} \eta+\zeta}{\epsilon{+}p}+ \frac{V_\pm^a \sigma_{ab} V_\pm^b }{ (\epsilon{+}p) \det(X)^2 \, c_s^2 } \,.
\end{align}
In a state with $\mu_a=0$, $n_a=0$, we have $V_\pm^a=0$, and $\Gamma$ reduces to $(\frac{2d-2}{d}\eta+\zeta)/(\epsilon{+}p)$.

\subsection{Hydrodynamic stability}
We take the conditions of hydrodynamic stability of equilibrium to be 
\begin{align}
  {\rm Im} \, \omega(q) \leq 0 \,,
\end{align}
at real $q$, for all near-equilibrium hydrodynamic modes. It is then clear that hydrodynamic stability is determined by the thermodynamic susceptibilities and the transport coefficients. If the transport coefficients are such that the hydrodynamic entropy production is non-negative ($\eta\ge0$, $\zeta\geq0$, $\sigma_{ab}$ positive semi-definite), then thermodynamic stability implies hydrodynamic stability. 

Shear modes \eqref{eq:w-shear} are hydrodynamically stable provided $\eta\geq0$ and $\epsilon{+}p>0$. Sound modes \eqref{eq:wS} are hydrodynamically stable provided $c_s^2 \geq 0$. It is clear from the grand canonical expressions \eqref{eq:cs2-1}, \eqref{eq:cs2-2} that a positive-definite $X$ (thermodynamic stability) implies $c_s^2 >0$. 
The sound damping coefficient must be non-negative for hydrodynamic stability of equilibrium, $\Gamma \geq 0$. For fluids with non-negative hydrodynamic entropy production, eq.~\eqref{eq:Gamma-2} implies that sound modes will be stable if $\epsilon{+}p>0$ and $c_s^2 >0$. 
Diffusion modes \eqref{eq:wD} are stable provided $D_{(a)} \geq 0$. For fluids with non-negative hydrodynamic entropy production, a positive-definite $X$ (thermodynamic stability) implies $D_{(a)} \geq 0$. 

In general, for fluids with non-negative hydrodynamic entropy production, a violation of thermodynamic stability conditions (negative eigenvalues of $X$) will lead to hydrodynamic instability of equilibrium caused by one (or several) of the hydrodynamic modes. Which hydrodynamic mode  is causing the instability is determined by the eigenvectors of $X$ corresponding to negative eigenvalues. Later in the paper, we will discuss an explicit example of a system where the susceptibility matrix $X$ has negative eigenvalues with $c_s^2 >0$, $\Gamma>0$, and $D_{(a)}<0$.

\subsection{Examples of hydrodynamic dispersion relations}
\label{sec:examples}
To make the formulae above somewhat more concrete, we now give a few examples. 
For definiteness, we will assume that the theory has unbroken charge conjugation symmetry for each flavour.

\subsubsection*{Equilibrium with all $\mu_a$ vanishing}
Our first example is the ``neutral'' equilibrium state with all $\mu_a=0$. Charge conjugation symmetry then implies that $n_a=0$, $\upsilon_a = 0$, and $\chi_{ab} = {\rm diag} (\chi_{(1)}, \chi_{(2)}, \dots )$. Similarly, charge conjugation in the Kubo formula \eqref{eq:Kubo-sigma} implies that the conductivity matrix is diagonal, $\sigma_{ab} = {\rm diag}( \sigma_{(1)}, \sigma_{(2)}, \dots)$. The speed of sound squared and the sound damping coefficient are
\begin{align}
  c_s^2 = \frac{\epsilon{+}p}{T (\partial \epsilon/\partial T) } \,,\ \ \ \ 
  \Gamma = \frac{\coeff{2d-2}{d} \eta+\zeta}{\epsilon{+}p} \,,
\end{align}
while the diffusion coefficients are
\begin{align}
  D_{(a)} = \frac{ \sigma_{(a)} }{\chi_{(a)}} \,, \ \ \ \ a=1, \dots, \Nf \,.
\end{align}
The stability of hydrodynamic modes is ensured by the positivity of hydrodynamic entropy production, and by the positivity of thermodynamic susceptibilities.

\subsubsection*{Equilibrium with one of $\mu_a$ non-vanishing}
Next, consider an equilibrium state with $\mu_1\neq0$, with the rest of $\mu_a$ vanishing. Now, $n_1\neq0$, $\upsilon_1 \neq 0$ with the rest of $n_a$, $\upsilon_a$ vanishing. The charge susceptibility matrix is still diagonal: $\chi_{ab} = 0$ for $b\neq a$ by charge conjugation for the remaining $\Nf-1$ flavours, hence $\chi_{ab} = {\rm diag} (\chi_{(1)}, \chi_{(2)}, \dots )$.  
Similarly, charge conjugation implies that $\sigma_{ab} = {\rm diag}( \sigma_{(1)}, \sigma_{(2)}, \dots)$. The speed of sound squared is 
\begin{align}
  c_s^2 = \frac{C n_1^2 -2(\epsilon{+}p) n_1 \upsilon_1 + (\epsilon{+}p)^2 \chi_{(1)}}{ (\epsilon{+}p) (C \chi_{(1)} - \upsilon_1^2 ) } \,,
\end{align}
where $C = T (\partial \epsilon/\partial T)_{\mu} + \mu_1 (\partial\epsilon/\partial\mu_1)_T$, $\upsilon_1 = (\partial \epsilon/\partial\mu_1)_T$. The damping coefficient is 
\begin{align}
  \Gamma = \frac{\coeff{2d-2}{d} \eta+\zeta}{\epsilon{+}p} + \frac{\sigma_{(1)} }{ (\epsilon{+}p) c_s^2  } \frac{(C n_1 - (\epsilon{+}p)\upsilon_1)^2 }{(C \chi_{(1)} - \upsilon_1^2)^2} \,,
\end{align}
and the diffusion coefficient for the first flavour is
\begin{align}
\label{eq:D-1}
  D_{(1)} = \frac{ (\epsilon{+}p) \sigma_{(1)} }{ c_s^2 (C \chi_{(1)} - \upsilon_1^2) } \,.
\end{align}
The diffusion coefficients for the remaining $\Nf-1$ flavours are 
\begin{align}
\label{eq:Da-2}
  D_{(a)} = \frac{ \sigma_{(a)} }{\chi_{(a)}} \,, \ \ \ \ a=2, \dots, \Nf \,,
\end{align}
In general, all $D_{(a)}$ in \eqref{eq:Da-2} depend on $\mu_1$. As expected, the sound mode is stable for $c_s^2 >0$, while the stability of diffusion modes requires positive susceptibilities in the charge sector~\cite{Kovtun:2012rj}.

\subsubsection*{Equilibrium with $N$ of $\mu_a$ non-vanishing}
A similar discussion applies for equilibrium states with $N<\Nf$ non-vanishing chemical potentials, $\mu_{a\leq N} \neq 0$, with the rest $\mu_{a>N} = 0$. Charge conjugation symmetry for $\Nf - N$ flavours implies that the only off-diagonal elements of $\chi_{ab}$ which do not vanish are in the $N{\times}N$ upper-left submatrix; the $(\Nf{-}N)\times(\Nf{-}N)$ lower-right submatrix is ${\rm diag}(\chi_{(N+1)}, \dots, \chi_{(\Nf)})$. The conductivity matrix $\sigma_{ab}$ has the same structure as $\chi_{ab}$. Further, $n_1, \dots, n_N$ and $\upsilon_1, \dots, \upsilon_N$ are in general non-zero, while the rest $n_{a>N} = 0$, $\upsilon_{a>N} = 0$. Then eq.~\eqref{eq:J-linear-2} implies that $\delta\gamma_1, \dots, \delta\gamma_N$ couple to $\delta T$ and $\delta u^z$, but the rest $\delta\gamma_{a>N}$ decouple. 
The speed of sound and the damping coefficient are given by the general expressions \eqref{eq:cs2-1}, \eqref{eq:Gamma-2}, written for the first $N$ flavours. There are $N$ diffusion modes whose diffusion coefficients are given by the eigenvalues of the diffusion matrix \eqref{eq:D-matr}, written for the first $N$ flavours. 
For the $\Nf{-}N$ decoupled diffusive fluctuations $\delta\gamma_{a>N}$, the diffusion coefficients are
\begin{align}
\label{eq:Da-3}
  D_{(a)} = \frac{ \sigma_{(a)} }{\chi_{(a)}} \,, \ \ \ \ a=N{+}1, \dots, \Nf \,.
\end{align}
Again, the diffusion coefficients $D_{(a)}$ in \eqref{eq:Da-3} in general depend on $\mu_1, \dots, \mu_N$.

\subsubsection*{Equilibrium with equal charges}
As our next example, consider a system with $\Nf=2$, and equal chemical potentials, $\mu_1 = \mu_2 = \mu$ in equilibrium.  We further assume that the densities are equal, i.e.\ $n_1(T, \mu_1 {=} \mu, \mu_2 {=} \mu) = n_2(T, \mu_1 {=} \mu, \mu_2 {=} \mu) = n$, so that $\upsilon_1 = \upsilon_2 = \upsilon$. Such equality of charge densities can arise as a consequence of an extended  symmetry in the flavour space, as we will discuss later.  The current conservation equations \eqref{eq:J-linear-2} are two equations; taking their difference, the terms with $\delta T$ and $\delta u^z$ drop out, so that
\begin{align}
\label{eq:diff-2}
  L_{11} \delta\gamma_1 + L_{12} \delta\gamma_2 - L_{21} \delta\gamma_1 - L_{22} \delta\gamma_2 = 0 \,,
\end{align}
where $L_{ab} = \chi_{ab} \partial_t - \sigma_{ab} \partial_i \partial^i$ is the diffusion operator. Recall that $\sigma_{ab}$ and $\chi_{ab}$ are symmetric matrices, so $L_{12} = L_{21}$. If the extended symmetry which relates the charges also implies $\chi_{11} = \chi_{22}$ and $\sigma_{11} = \sigma_{22}$, then  $L_{11} = L_{22}$, and eq.~\eqref{eq:diff-2} becomes 
\begin{align}
\label{eq:diff-2a}
  \left( L_{11} - L_{12} \right) (\delta \gamma_1 - \delta\gamma_2) = 0,
\end{align}
or
$
  \left[ (\chi_{11} {-} \chi_{12}) \partial_t - (\sigma_{11} {-} \sigma_{12}) \partial_i \partial^i \right]  ( \delta\gamma_1 {-} \delta\gamma_2) = 0.
$
In other words, the fluctuation of $\gamma_1 {-} \gamma_2$ decouples, and obeys the diffusion equation, with the diffusion coefficient 
\begin{align}
\label{eq:d12}
  D_{(1-2)} = \frac{\sigma_{11} - \sigma_{12}}{\chi_{11} - \chi_{12}} \,.
\end{align}
Recall that $\chi_{11} {-} \chi_{12}$ is an eigenvalue of $\chi_{ab}$, and $\sigma_{11} {-} \sigma_{12}$ is an eigenvalue of $\sigma_{ab}$. Thus in a system where off-equilibrium entropy production is non-negative, the numerator of \eqref{eq:d12} is non-negative. The denominator of \eqref{eq:d12} is positive for a thermodynamically stable system, but can be negative for a thermodynamically unstable system. The diffusion coefficient for the second diffusive mode (coupled to sound) is
\begin{align}
\label{eq:D1plus2}
  D_{(1+2)} = \frac{(\epsilon{+}p) (\sigma_{11} {+} \sigma_{12})}{ c_s^2 (C(\chi_{11} {+} \chi_{12}) - 2\upsilon^2)} \,.
\end{align}
The speed of sound squared is 
\begin{align}
  c_s^2 = \frac{2C n^2 - 4 (\epsilon{+}p) n \upsilon + (\epsilon{+}p)^2 (\chi_{11} {+} \chi_{12}) }{(\epsilon{+}p) 
  \left( C (\chi_{11} {+} \chi_{12}) - 2 \upsilon^2\right)} \,.
\end{align}
The sound damping coefficient is
\begin{align}
    \Gamma = \frac{\coeff{2d-2}{d} \eta+\zeta}{\epsilon{+}p} + \frac{2 (\sigma_{11} {+} \sigma_{12}) }{ (\epsilon{+}p) c_s^2  } \frac{(C n - (\epsilon{+}p)\upsilon)^2 }{(C (\chi_{11} {+} \chi_{12} ) - 2\upsilon^2)^2} \,.
\end{align}
As before, a thermodynamically unstable equation of state ($X$ with negative eigenvalues) gives rise to unstable hydrodynamic modes. 

More generally, one may consider a system with $\Nf$ global $U(1)$'s, and all equal chemical potentials. If there is a symmetry that ensures that all $n_a(T,\mu)$ are the same, all diagonal $L_{ab}$ are the same, and all off-diagonal $L_{ab}$ are the same, then there will be $\Nf{-}1$ diffusive modes, where fluctuations $\delta\gamma_n - \delta \gamma_{n+1}$ decouple from the fluctuations of $\delta T$ and $\delta u^z$. All these $\Nf{-}1$ modes will have the same diffusion coefficient, given by eq.~\eqref{eq:d12}.

\subsection{Symmetry constraints}
\label{sec:symmetry-constraints}

\subsubsection*{Conformal symmetry}
If the microscopic theory happens to be conformal, there are additional constraints on thermodynamics and hydrodynamics. In a conformal theory, scale invariance implies that the grand canonical equation of state is given by $ p(T,\mu_a) = T^{d+1} f(\mu_a/T)$, with a theory-dependent function $f(\gamma_1, \dots, \gamma_\Nf)$. The thermodynamic derivatives which appear in matrices $X$ and $Y$ are then related by
\begin{align}
\label{eq:CFT-thermo-constraints}
  C = d(d{+}1) p\,,\ \ \ \ 
  \epsilon+p = (d{+}1)p \,,\ \ \ \ 
  \upsilon_a = d \, n_a \,.
\end{align}
Correspondingly, the speed of sound in a conformal theory is $c_s = 1/\sqrt{d}$. Further, the charge conductivity contribution to sound damping in eq.~\eqref{eq:Gamma-1} drops out, as a consequence of \eqref{eq:CFT-thermo-constraints}. Finally, the bulk viscosity $\zeta$ vanishes. As a result, sound dispersion relations in conformal field theory are simply 
\begin{align}
\label{eq:sound-cft}
    \omega^{\rm sound}_\pm(q) = \pm \frac{|q|}{\sqrt{d}} - i \frac{d{-}1}{d(d{+}1)} \frac{\eta}{p} q^2 + O(q^3) \,.
\end{align}
Conformal symmetry does not impose any relations among the components of the charge susceptibility matrix $\chi_{ab}$. Such relations can arise as consequences of a flavour symmetry, as we discuss next.

\subsubsection*{Non-abelian flavour symmetry}
If the flavour symmetry $U(1)^\Nf$ forms an Abelian subgroup of a larger global symmetry of the theory, $U(1)^\Nf \subset SU(\Nf{+}1)$, there are additional constraints on thermodynamics and hydrodynamics.

The grand canonical partition function is 
\begin{align}
  {\cal Z} = {\rm Tr} \exp[-\beta H + \gamma_a Q^a] =  {\rm Tr} \exp[-\beta H + \gamma_a (T^a)_{AB} Q_{AB}] \,,
\end{align}
where $T^a$ are the generators of $U(1)^\Nf \subset SU(\Nf+1)$. Here, capital Latin indices run from 1 to $\Nf{+}1$, and lower case Latin indices run from 1 to $\Nf$. 
The partition function ${\cal Z}$ is $SU(\Nf{+}1)$ invariant, and so its dependence on the generators $T^a$ must come in the form of traces, ${\rm tr} [(T^1)^k (T^2)^l (T^3)^m \dots]$. Such traces of products of the generators are products of $\delta_{ab}$'s, $d_{abc}$'s and $f_{abc}$'s, with various contractions, where $f_{abc}$ are the structure constants, and $d_{abc}$ are anomaly coefficients.%
\footnote{
Explicitly, for $SU(n)$ one has
$
  T^a T^b = \frac{1}{2n} \delta^{ab} {\bf 1} + \frac{1}{2} ( d^{abc} + i f^{abc}) T^c\,,
$
where the generators are normalized such that ${\rm tr}(T^a T^b) = \frac12 \delta^{ab}$, and $a,b,c$ run from $1$ to $n^2 {-} 1$.
}
For the Abelian subalgebra, $f_{abc} = 0$.
In the partition function, each $T^a$ will come accompanied by the corresponding $\gamma_a = \mu_a/T$, so ${\cal Z}$ will be a function of $\delta_{ab}$'s and $d_{abc}$'s contracted with $\mu_a$'s. 
Correspondingly, the dependence of the partition function on the chemical potentials will come from invariants such as 
\begin{eqnarray}\label{eq:invariants}
m_2 &=&  \mu_a \mu_a\,,   \\
m_3 &=& d_{abc} \mu_a \mu_b \mu_c\,, \\
m_4 &=& d_{abc} d_{afg} \mu_b \mu_c \mu_f \mu_g\,,
\end{eqnarray}
%$m_2 \equiv \mu_a \mu_a$, $m_3 \equiv d_{abc} \mu_a \mu_b \mu_c$, $m_4 \equiv d_{abc} d_{afg} \mu_b \mu_c \mu_f \mu_g$, 
    etc.\footnote{The number of such invariants is equal to  $\Nf$ (see Appendix \ref{cayley-hamilton} for details). } Hence, the pressure is $p = F(T, m_2, m_3, m_4, \dots)$.
This general form of $p$ allows one to express the charge susceptibility matrix $\chi_{ab} = \partial^2 p/\partial\mu_a \partial\mu_b$ in terms of the derivatives of~$F$ and contractions of $d_{abc}$. 

As an example,  we take $\Nf=3$, and choose the basis for the Cartan generators $T^a$ so that the coefficients $d_{abc}$ are only non-zero if $abc$ is a permutation of $123$. The pressure is $p = F(T, m_2, m_3^2, m_4)$, where $m_3$ appears squared due to charge conjugation invariance. With only one non-vanishing chemical potential $\mu_1$, and $\mu_2 = \mu_3 = 0$, the susceptibility matrix takes the following form:
\begin{align}
  \chi_{ab} = \begin{pmatrix}
  \chi_{11} & 0 & 0 \\
  0 & \chi_{22} & 0 \\
  0 & 0 & \chi_{22}
  \end{pmatrix} \,.
\end{align}
With $\mu_1 = \mu_2$, and $\mu_3 = 0$, the susceptibility matrix takes the following form: 
\begin{align}
\label{eq:chiab-2mu}
  \chi_{ab} = \begin{pmatrix}
  \chi_{11} & \chi_{12} & 0 \\
  \chi_{12} & \chi_{11} & 0 \\
  0 & 0 & \chi_{33}
  \end{pmatrix} \,.
\end{align}
With  $\mu_1 = \mu_2 = \mu_3$, the susceptibility matrix takes the following form: 
\begin{align}
\label{eq:chi-Nf3-equal-mu}
  \chi_{ab} = \begin{pmatrix}
  \chi_{11} & \chi_{12} & \chi_{12} \\
  \chi_{12} & \chi_{11} & \chi_{12} \\
  \chi_{12} & \chi_{12} & \chi_{11}
  \end{pmatrix} \,.
\end{align}
Correlation functions of conserved flavour currents will have the same structure, and so will the conductivity matrix, through the Kubo formula~\eqref{eq:Kubo-sigma}. Given our choice of Cartan generators, in a state with equal chemical potentials, the $SU(\Nf{+}1)$ symmetry ensures the simplifications discussed in the last example of sec.~\ref{sec:examples}, leading to the simple expressions \eqref{eq:d12} for the diffusion coefficients.

\subsection{All-orders hydrodynamics}
\label{sec:all-orders-hydro}
Let us return to the argument in the last example of sec.~\ref{sec:examples} about the decoupling of $\Nf-1$ diffusive modes in a state with $\bar n_1 = \dots = \bar n_\Nf$.
In that argument, we have used linearised relativistic hydrodynamics, to first order in derivatives. However, the decoupling of $\Nf-1$ diffusive modes in a symmetric state does not require hydrodynamics to be relativistic, and can  happen to all orders in derivatives, provided there is an extended flavour symmetry such as $SU(\Nf+1)$ discussed above. In order to see that,  we write down the linearised hydrodynamic equations in Fourier space,
\begin{align}
\label{eq:CCF}
  -i \omega \delta n_a + i q_l \, \delta J^l_a = 0 \,,
\end{align}
where we have taken $J^0_a = \bar n_a + \delta n_a$, expanding about the equilibrium state with constant charge densities $\bar n_a$.
 The constitutive relations for the spatial flux $\delta J^i_a$ must be written in terms of the hydrodynamic variables; we choose the latter as $\delta n_a$ (fluctuations of charge densities), $\delta\epsilon$ (fluctuation of the energy density), and $\delta \pi^i$ (fluctuation of momentum density). Rotation invariance then dictates%
\footnote{
   Depending on dimension, one can also include tems such as $\varepsilon^{lmn} q_m \delta \pi_n$, or $\varepsilon^{lm} q_m \delta \epsilon$. Such terms drop out of the current conservation equation, and can be ignored in linearised hydrodynamics.
}
\begin{align}
  \delta J^l_a = A_a \delta\pi^l + i q^l B_a \delta\epsilon + i q^l C_a i q_j \delta\pi^j - i q^l D_{ab} \delta n_b  \,,
\end{align}
where $A_a$, $B_a$, $C_a$, $D_{ab}$ are scalars under rotation, and are all functions of $\omega$, ${\bf q}^2$, and $\bar n_a$.  The terms $A_a$, $B_a$ and $C_a$ give rise to the coupling between $\delta n_a$ and $\delta\epsilon$, $\delta \pi^i$.

We shall  take the charges in question to correspond to the Cartan generators of $SU(4)$, with $\Nf=3$, as discussed in sec.~\ref{sec:symmetry-constraints}. The index $a$ is then a Lie algebra index, and the coefficients $A_a$, $B_a$, $C_a$, $D_{ab}$ are made out of contractions of $\bar n_a$ and the anomaly coeffcients $d_{abc}$. As $d_{abc}$ are completely symmetric, in a state with equal charges, $\bar n_1 = \bar n_2 = \bar n_3$, all $A_a$'s will be equal to each other, all $B_a$'s will be equal to each other, and all $C_a$'s will be equal to each other. We can then take pair-wise differences of current conservation equations with different flavours; 
the contributions proportional to $\delta\epsilon$ and $\delta\pi^i$ drop out, and we end up with decoupled equations for $\delta n_a$, just like we had earlier in eq.~\eqref{eq:diff-2}. For three flavours, we have
\begin{subequations}
\begin{align} 
  & -i\omega (\delta n_1 {-} \delta n_2) + {\bf q}^2 (D_{11} \delta n_1 + D_{12} \delta n_2 + D_{13} \delta n_3 - D_{21} \delta n_1 - D_{22} \delta n_2 - D_{23} \delta n_3) = 0 , \\
  & -i\omega (\delta n_2 {-} \delta n_3) + {\bf q}^2 (D_{21} \delta n_1 + D_{22} \delta n_2 + D_{23} \delta n_3 - D_{31} \delta n_1 - D_{32} \delta n_2 - D_{33} \delta n_3) = 0 , \\
  & -i\omega (\delta n_1 {-} \delta n_3) + {\bf q}^2 (D_{11} \delta n_1 + D_{12} \delta n_2 + D_{13} \delta n_3 - D_{31} \delta n_1 - D_{32} \delta n_2 - D_{33} \delta n_3) = 0 .
\end{align}
\end{subequations}
The diffusion matrix $D_{ab}$ will in general be given by a combination of terms such as $\bar n_a \bar n_b$, $d_{abc} \bar n_c$, $d_{acd} d_{bfg} \bar n_c \bar n_d \bar n_f \bar n_g$, etc.  We choose the basis for the Cartan generators so that $d_{abc}$ are only non-zero if $abc$ is a permutation of $123$. Then, in a state with all $\bar n_a$ equal, the diffusion matrix $D_{ab}$ will have all diagonal elements equal to each other, and all off-diagonal elements equal to each other, and thus 
\begin{subequations}
\begin{align} 
  & -i\omega (\delta n_1 {-} \delta n_2) + {\bf q}^2 (D_{11} - D_{12}) (\delta n_1 {-} \delta n_2) = 0 , \\
  & -i\omega (\delta n_2 {-} \delta n_3) + {\bf q}^2 (D_{11} - D_{12}) (\delta n_2 {-} \delta n_3) = 0 , \\
  & -i\omega (\delta n_1 {-} \delta n_3) + {\bf q}^2 (D_{11} - D_{12}) (\delta n_1 {-} \delta n_3) = 0 .
\end{align}
\end{subequations}
Of the above three equations, only two are independent. Thus we have two idential sets of modes, whose dispersion relations are determined by
\begin{align}
  -i \omega + D_{11}(\omega,{\bf q}^2) {\bf q}^2 - D_{12}(\omega,{\bf q}^2) {\bf q}^2 = 0 \,.
\end{align}
At small $\omega$ and ${\bf q}^2$, each set will give a diffusion mode $\omega = -i (D_{11} - D_{12}){\bf q}^2$.

%%%%%%%%%%%%%%%%%%%%%%%%%%%%%%%%%

\section{Gravity dual to \texorpdfstring{${\cal N}=4$}{N=4} SYM at finite density}
\label{stu-model-overview}
The ten-dimensional gravity dual to  ${\cal N}=4$ $SU(N_c)$ SYM at $T\neq 0$, $\mu_a\neq 0$ at
 infinite $N_c$ and infinite 't Hooft coupling is given by the near-horizon limit 
 of the set of $N_c$ rotating black three-branes \cite{Gubser:1998jb}, \cite{Cai:1998ji}, 
\cite{Chamblin:1999tk}, \cite{Cvetic:1999ne}. The dimensional reduction on $S^5$ gives a 
background which is a Reissner-Nordstr\"om-type solution of ${\cal N}=2$ supergravity in 
five dimensions.\footnote{A generic construction of  $D=5$ ${\cal N}=2$ supergravity can be 
found in ref.~\cite{Bergshoeff:2004kh}, see also \cite{Gunaydin:1999zx,Ceresole:2000jd,Gunaydin:1983bi}. } 
The relevant part of the five-dimensional Lagrangian density is%
\footnote{
   We use  normalisation of the gauge fields which is different from the one used e.g. in refs.~\cite{Behrndt:1998jd,Cvetic:1999ne}  by a factor of $\sqrt{2}/L$. Accordingly, the coefficients appearing in front of the terms involving gauge fields are different from the ones in refs.~\cite{Behrndt:1998jd,Cvetic:1999ne}. 
}
 \cite{Behrndt:1998jd}
\begin{eqnarray}
    \mathcal{L} &=& \sqrt{-g}\bigg(R+\frac{2}{L^{2}}\mathcal{V}-\frac{L^2}{4}G_{ab}F_{\mu \nu}^{a} F^{\mu \nu\, b}
    - G_{ab} \partial_{\mu} X^{a} \partial^{\mu} X^{b}\bigg)\nonumber \\
    &+& 
    \frac{L^3 \sqrt{2}}{96}\varepsilon^{\mu \nu \rho \sigma \lambda} C_{abc} F_{\mu \nu}^{a}F_{\rho \sigma}^{b}A_{\lambda}^{c}\,.
\label{lagrangian-stu}
\end{eqnarray}
The action contains the metric,  three Abelian gauge fields $A_\mu^a$ and three real scalar fields $X_a$. The symmetric coefficients $C_{abc}$ are proportional to the anomaly coefficients of ${\cal N} = 4$ SYM theory. For the STU solution\footnote{The name ``STU'' stems from the notations 
$X_{1}=S$,  $X_{2}=T$, $X_{3}=U$  and the prepotential condition ${\cal V}=STU=1$  \cite{Behrndt:1998jd}.}
   \cite{Behrndt:1998jd} we discuss below, the scalars are constrained by $X_{1}X_{2}X_{3}=1$,%
\footnote{
\label{c-y} 
  More generally, the three scalars are constrained by $C_{abc} X^a X^b X^c/6=1$ \cite{Behrndt:1998jd}. For the STU solution, the non-vanishing components of $C_{abc}$ (symmetric in all indices) are $C_{123}=1$ and its cyclic permutations.
}
the metric on the scalar manifold is given by
$$
  G_{ab} = {1\over 2} \mbox{diag} \left[ (X^1)^{-2}, \, (X^2)^{-2}, \,(X^3)^{-2}
\right]\,,
$$
and the scalar potential is
$$
 {\cal V} = 2 \sum\limits_{a=1}^3 {1\over X^a}\,.
$$
When all $X^a = 1$, the potential term in \eqref{lagrangian-stu} becomes $-2\Lambda$, where $\Lambda = -6/L^2$ is the five-dimensional cosmological constrant expressed in terms of the AdS radius $L$. 
Following \cite{Behrndt:1998jd}, \cite{Cvetic:1999ne},  we introduce new scalar fields $H_{a}$ via\footnote{Note a different notation w.r.t. ref.~\cite{Son:2006em}.  The product of all $H_a$ was denoted by ${\cal H}$ there,  while we use the notation $H$ for the same product in this paper and in ref.~\cite{Gladden:2024ssb}.}
\begin{equation}
X_{a} = H^{\frac{1}{3}}/H_{a}\,,
\label{fields-H}
\end{equation}
where $H \equiv H_1 H_2 H_3$. In terms of the fields $H_a$, the Lagrangian is 
\begin{eqnarray}
   \mathcal{L} 
   &=& 
   \sqrt{-g} \bigg( R+ \frac{4}{L^{2}} H^{-\frac{1}{3}} \sum_{a=1}^{3}H_{a} - \frac{L^2}{8}  H^{-\frac{2}{3}} \sum_{a=1}^3 H_{a}^{2}F_{\mu \nu}^{a} F^{\mu \nu \, a} - \frac{1}{3}\sum_{a=1}^{3} \frac{\partial_{\mu}H_{a}\partial^{\mu}H_{a}}{H_{a}^{2}} \nonumber \\ 
   &+& 
   \frac{1}{3} \sum_{a<b}^{3} \frac{\partial_{\mu}H_{a}\partial^{\mu}H_{b}}{H_{a}H_{b}}\bigg) + \frac{L^3 \sqrt{2}}{96}\varepsilon^{\mu \nu \rho \sigma \lambda} C_{abc} F_{\mu \nu}^{a} F_{\rho \sigma}^{b} A_{\lambda}^{c}\,.
   \label{lagrangian-stu-H}
 \end{eqnarray}
The equations of motion corresponding to  the Lagrangian \eqref{lagrangian-stu-H} read
\begin{eqnarray}
  &\,& R_{\mu \nu} = T_{\mu \nu}^{{\rm (m)}} - \coeff13 g_{\mu \nu}  T_{\lambda}^{{\rm (m)}\, \lambda} \,,  
\label{grav-eq}\\
  &\,& \partial_{\nu} \biggl( \sqrt{-g} H^{-\frac{2}{3}} H_{a}^{2} F^{\mu \nu a} \biggr) =
 \frac{L\sqrt{2}}{16} \varepsilon^{\mu \nu \rho \sigma \lambda} C_{abc} F_{\rho \sigma}^{b}F_{\nu \lambda}^{c}\,, \label{gauge-eq} \\
  &\,& \frac{1}{\sqrt{-g}} \partial_{\mu}\bigg[\sqrt{-g} \bigg( -\frac{2}{3}\frac{\partial^{\mu}H_{a}}{H_{a}}+
\frac{1}{3} \sum_{b \neq a} \frac{\partial^{\mu}H_{b}}{H_{b}}\bigg)\bigg] =-\frac{4 }{3 L^{2}}H^{-\frac{1}{3}} \sum_{b \neq a} H_{b} \nonumber \\ 
  &\,& + \frac{8}{3 L^{2}} H^{-\frac{1}{3}} H_{a} - \frac{L^2}{6} H^{-\frac{2}{3}} H_{a}^{2} F_{\mu \nu}^{a} F^{\mu \nu a}
+\frac{L^2}{12} H^{-\frac{2}{3}} \sum_{b \neq a} H_{b}^{2} F_{\mu \nu}^{b} F^{\mu \nu \, b}\,, 
\label{scal-eq} 
\end{eqnarray}
There is no summation over ``$a$'' in the above equations. The right-hand-side of Eq.~\eqref{grav-eq} contains the energy-momentum tensor of the scalars and the $U(1)$ fields, 
\begin{eqnarray}
    T_{\mu \nu}^{{\rm (m)}} &=& \frac{L^2 G_{ab}}{2} F_{\mu \rho}^{a} F_\nu^{\; \, \rho \, b} - \frac{L^2 g_{\mu \nu}}{8} G_{ab} F^a_{\; \rho \sigma} F^{b\; \rho \sigma} + G_{ab} \partial_{\mu} X^{a} \partial_{\nu}X^{b}\nonumber \\
    &-&\frac{g_{\mu \nu}}{2} G_{ab}\partial_{\rho}X^{a} \partial^{\rho}X^{b} + \frac{g_{\mu \nu}\mathcal{V}}{L^{2}}\,.
\end{eqnarray}
%
%\begin{eqnarray}
%&\,& R_{\mu \nu} =T_{\mu \nu}^{(m)}-\frac{1}{3}g_{\mu \nu}T^{(m)}\,,  \label{grav-eq}\\
%\begin{gathered}
% &\,&    \partial_{\nu}\biggl(\sqrt{-g}H^{-\frac{2}{3}}H_{i}^{2}F^{\mu \nu i}\biggr)
% =\frac{1}{8}\epsilon^{\mu \nu \rho \sigma \lambda}C_{i j k}F_{\rho \sigma}^{j}F_{\nu \lambda}^{k}\,, \label{gauge-eq} \\
%&\,& \frac{1}{\sqrt{-g}} \partial_{\mu}\bigg[\sqrt{-g}\bigg(-\frac{2}{3}\frac{\partial^{\mu}H_{i}}{H_{i}}
%+\frac{1}{3}\sum_{j \neq i}\frac{\partial^{\mu}H_{j}}{H_{j}}\bigg)\bigg] =-\frac{4 }{3 L^{2}}H^{-\frac{1}{3}} \sum_{j \neq i}H_{j}\\ 
%&\,& + \frac{8}{3 L^{2}}H^{-\frac{1}{3}} H_{i}-\frac{1}{3} H^{-\frac{2}{3}} H_{i}^{2}F_{\mu \nu}^{i}F^{\mu \nu i}
%+\frac{1}{6} H^{-\frac{2}{3}} \sum_{j \neq i} H_{j}^{2}F_{\mu \nu}^{j}F^{\mu \nu j}\,, \label{scal-eq} 
%\end{gathered} 
%\end{eqnarray}
%where the right-hand-side of Eq.~\eqref{grav-eq} contains the energy-momentum tensor of the 
%scalars and the $U(1)$ fields, 
%
%\begin{eqnarray}
 %   T_{\mu \nu}^{(m)}&=&G_{i j}F_{\mu \rho}^{i}F_\nu^{\; \, \rho j}
%    -\frac{g_{\mu \nu}}{4}
 %   G_{i j}F^i_{\; \rho \sigma}F^{j\; \rho \sigma}
 %   +G_{i j} \partial_{\mu} X^{i} \partial_{\nu}X^{j}
  %  -\frac{g_{\mu \nu}}{2} G_{i j}\partial_{\rho}X^{i}\partial^{\rho}X^{j}\nonumber \\ 
  %  &+& g_{\mu \nu}\frac{\mathcal{V}}{L^{2}}\,,
%\end{eqnarray}
%and $T^{(m)}= T_{\mu}^{(m)\, \mu}$. 
 
\subsection{The STU background}
\label{sec:STU-background}
The equations \eqref{grav-eq}, \eqref{gauge-eq}, \eqref{scal-eq} admit a solution known as the STU black brane background \cite{Behrndt:1998jd}  which is dual to ${\cal N}=4$ $SU(N_c)$ SYM theory (at infinite $N_c$ and infinite 't Hooft coupling) at finite temperature and with  three non-vanishing chemical potentials $\mu_1$,  $\mu_2$,  $\mu_3$,  conjugate to the global conserved R-charges in the Cartan subalgebra of $SU(4)_R$. The metric of the STU background is \cite{Behrndt:1998jd,Cvetic:1999ne}\footnote{We use the notations of ref.~\cite{Son:2006em},  which are slightly different from the ones 
in refs.~\cite{Behrndt:1998jd},\cite{Cvetic:1999ne}.  In particular, our radial coordinate is compact, $u=r_+^2/r^2$ and the parameters $\kappa_a$ are defined as $\kappa_a = q_a/r_+^2$ w.r.t.  ref.~\cite{Cvetic:1999ne}.}
\begin{equation}
ds^2_5 = - H^{-2/3}{(\pi T_0 L)^2 \over u}\,f(u) \, dt^2 
+    H^{1/3}{(\pi T_0 L)^2 \over u}\, \left( dx^2 + dy^2 + dz^2\right)
+  H^{1/3}{L^2 \over 4 f u^2} du^2\,.
\label{metric_stu_3}
\end{equation}
It depends on four integration constants which we denote $T_0$ and $\kappa_a$, with $a=1,2,3$.  When all $\kappa_a = 0$, the metric reduces to that of the AdS--Schwarzschild black brane in five dimensions, with Hawking temperature $T_0$. The parameters $\kappa_a$ are taken to be positive to ensure the regularity of the background and the positivity of the ADM mass\footnote{As defined for an asymptotically AdS space by Horowitz and Myers~\cite{Horowitz:1998ha}. }  \cite{Behrndt:1998jd,Cvetic:1999ne}.  The background for the scalar fields is
\begin{equation}
H_a = 1 + \kappa_a u\,,
\label{stu-scalars}
\end{equation}
hence $H(u)= (1+\kappa_1 u)(1+\kappa_2 u)(1+\kappa_3 u)$. The function $f(u)$ in the metric \eqref{metric_stu_3} is a cubic polynomial $f(u) =  H(u) - u^2  H(1) = (1 - u) \left(1 + (1 + \kappa_1 + \kappa_2 + \kappa_3) u - \kappa_1 \kappa_2 \kappa_3 u^2\right)$. The asymptotic boundary is at $u=0$, and the horizon is at $u=1$. 
The  background
gauge fields are given by
\begin{align}
\label{eq:A-bg}
  A_\mu^a(u) = \delta_\mu^t \left( \frac{1}{1+\kappa_a} - \frac{u}{H_a(u)} \right) \pi T_0 \sqrt{2\kappa_a} \sqrt{(1+\kappa_1)(1+\kappa_2)(1+\kappa_3)}\,,
\end{align}
where the integration constant was chosen to set $A_\mu^a(u=1)=0$. %The chemical potentials are then defined as $\mu_a = A_\mu^a (u=1)$. 

We identify the temperature $T$ of ${\cal N}{=}4$ SYM theory with the Hawking temperature of the black brane metric \eqref{metric_stu_3}, and the chemical potentials $\mu_a$ for the three $U(1)$ R-charges with the boundary values of the gauge fields \eqref{eq:A-bg}. This gives $T$ and $\mu_a$ as functions of the four integration constants $T_0$ and $\kappa_a$,
\begin{align}
\label{eq:TH-kappa}
  & T = \frac{2+\kappa_1 + \kappa_2 + \kappa_3 - \kappa_1 \kappa_2 \kappa_3}{2\sqrt{(1{+}\kappa_1) (1{+}\kappa_2) (1{+}\kappa_3)}} T_0 \,,\\[5pt]
\label{eq:mu-kappa}
  & \mu_a = \pi T_0 \frac{\sqrt{2\kappa_a}}{1{+}\kappa_a} \sqrt{1{+}\kappa_1} \sqrt{1{+}\kappa_2} \sqrt{1{+}\kappa_3} \,.
\end{align}
Next, we identify the entropy density $s$ of ${\cal N}{=}4$ SYM theory with the Bekenstein-Hawking entropy of the black brane metric \eqref{metric_stu_3} per unit three-volume. The equilibrium charge densities $n_a$ for the three $U(1)$ R-charges are identified from the asymptotic ``electric'' fields via the 
Gauss-Ostrogradsky theorem.\footnote{First discovered by Joseph Louis Lagrange in 1762 \cite{Lagrange1762, Lagrange2007}.} This gives $s$ and $n_a$ as functions of the four integration constants $T_0$ and $\kappa_a$,
\begin{align}
\label{eq:s-kappa}
  & s = \coeff12 \pi^2 N_c^2 T_0^3 \, \sqrt{1{+}\kappa_1} \sqrt{1{+}\kappa_2} \sqrt{1{+}\kappa_3}\,,\\[5pt]
\label{eq:na-kappa}
  & n_a = \coeff18 \pi N_c^2 T_0^3 \sqrt{2\kappa_a} \sqrt{1{+}\kappa_1} \sqrt{1{+}\kappa_2} \sqrt{1{+}\kappa_3} \,.
\end{align}
Given the expressions \eqref{eq:s-kappa} and \eqref{eq:na-kappa}  for the entropy density and the densities of charges, the pressure can be found by integrating the Gibbs-Duhem relation\footnote{Alternatively, the pressure can be computed holographically from the renormalized on-shell action, either as the one-point function of the spatial diagonal components of the energy--momentum tensor as in ref.~\cite{Policastro:2002tn},  or via Kubo formulae applied to the corresponding two-point functions \cite{Baier:2007ix,chiral-life}.}
\[
dP = s\, dT + \sum_a n_a\, d\mu_a\,,
\]
with the result
\begin{equation}
\label{eq:pressure-kappa}
   p = \frac{ \pi^2 N^2_c T_0^4}{8}\, \prod\limits_{a=1}^3  (1+\kappa_a)\,.
\end{equation}
The energy density is then computed from the Euler relation,
\[
\epsilon + p = T s + \sum_a \mu_a n_a\,,
\]
giving 
\begin{equation}
\label{eq:e-kappa}
   \epsilon = 3 p = \frac{3 \pi^2 N^2_c T_0^4}{8}\, \prod\limits_{a=1}^3  (1+\kappa_a)\,.
\end{equation}
The result  \eqref{eq:e-kappa}  implies that the speed of sound squared is $c_s^2=1/3$ as expected in a conformal theory. 
The integration constants $T_0$ and $\kappa_a$ can be eliminated from \eqref{eq:s-kappa}, \eqref{eq:na-kappa}, and \eqref{eq:e-kappa}, leading to the equation of state in the form (see e.g. ref.~\cite{Son:2006em} and ref.~\cite{Harmark:1999xt}):
 \begin{align}
\label{eq:eos-STU}
  \epsilon(s, n_1, n_2, n_3) = \frac{3}{2 (2\pi N_c)^{2/3}} \, s^{4/3} \prod\limits_{a=1}^3 \left(1+\frac{8\pi^2 n_a^2}{s^2} \right)^{1/3}\,.
\end{align}
%\hl{[[Where did this equation of state appear first?]]}

\subsection{Thermodynamic stability}
 In the grand canonical ensemble, a stable thermodynamic equilibrium can be determined by the following conditions on the potential $\Omega \equiv \omega V=-pV$:
 \begin{equation}
\left( \delta \Omega \right)_{T, \mu_a \mbox{\tiny fixed}}= 0\,, \qquad \qquad \left(\delta^2 \Omega \right)_{T, \mu_a \mbox{\tiny fixed}} >0\,,
\end{equation}
where variations are taken with respect to the parameters (such as $\varepsilon$ and $n_a$) which at fixed $T$ and $\mu_a$ 
(specifying the equilibrium state) can assume non-equilibrium values. 
%Here, $\Omega$ is the grand canonical thermodynamic 
%potential, with $\omega \equiv \Omega/V =-P = \varepsilon - T_H s -\sum \mu_a n_a$, 
%where $s$ and $\varepsilon$ are the (volume) entropy and energy densities, respectively,  and $n_i$ 
%are the (volume) densities of the three $R$-charges.  
Explicitly,
  \begin{eqnarray}
 &\,& \left( \delta \omega \right)_{T, \mu_a \mbox{\tiny fixed}} =   \frac{\partial \varepsilon (s,n)}{\partial s} \delta s +
   \frac{\partial \varepsilon (s,n)}{\partial n_a} \delta n_a - T \delta s - \mu_a \delta n_a =0\,, \\
  &\,& \left( \delta^2 \omega \right)_{T, \mu_a \mbox{\tiny fixed}} =   \frac{\partial^2 \varepsilon }{\partial s^2} (\delta s)^2 +
   2 \frac{\partial^2 \varepsilon }{\partial s \partial n_a} \delta s \delta n_a +  \frac{\partial^2 \varepsilon }{\partial n_a \partial n_b} \delta n_a \delta n_b > 0\,.\label{pos-def}
\end{eqnarray}
 Introducing the notation $y_i = (s, n_a)$,  the condition \eqref{pos-def} means that the quadratic form $h_{ij}\delta y_i \delta y_j$ must be positive definite.  In turn, this implies that the Hessian
\begin{align}
\label{eq:hessian-2}
  h^\epsilon_{ij} \equiv \frac{\partial^2 \epsilon}{\partial y_i \partial y_j}\,
\end{align}
 must be positive definite,  i.e.  its eigenvalues $\lambda_i$ should be all positive.  Technically,  when the equation of state $\epsilon(s, n_a)$ is a complicated function, the analysis of thermodynamic stability is aided by using   Sylvester's criterion according to which  a symmetric matrix is positive definite if and only if all of its leading principal minors are positive (details of analysis for various thermodynamic potentials can be found e.g. in  ref.~\cite{Tester-Modell}).

\begin{figure}[t!]
\centering
\includegraphics[width=0.55\textwidth]{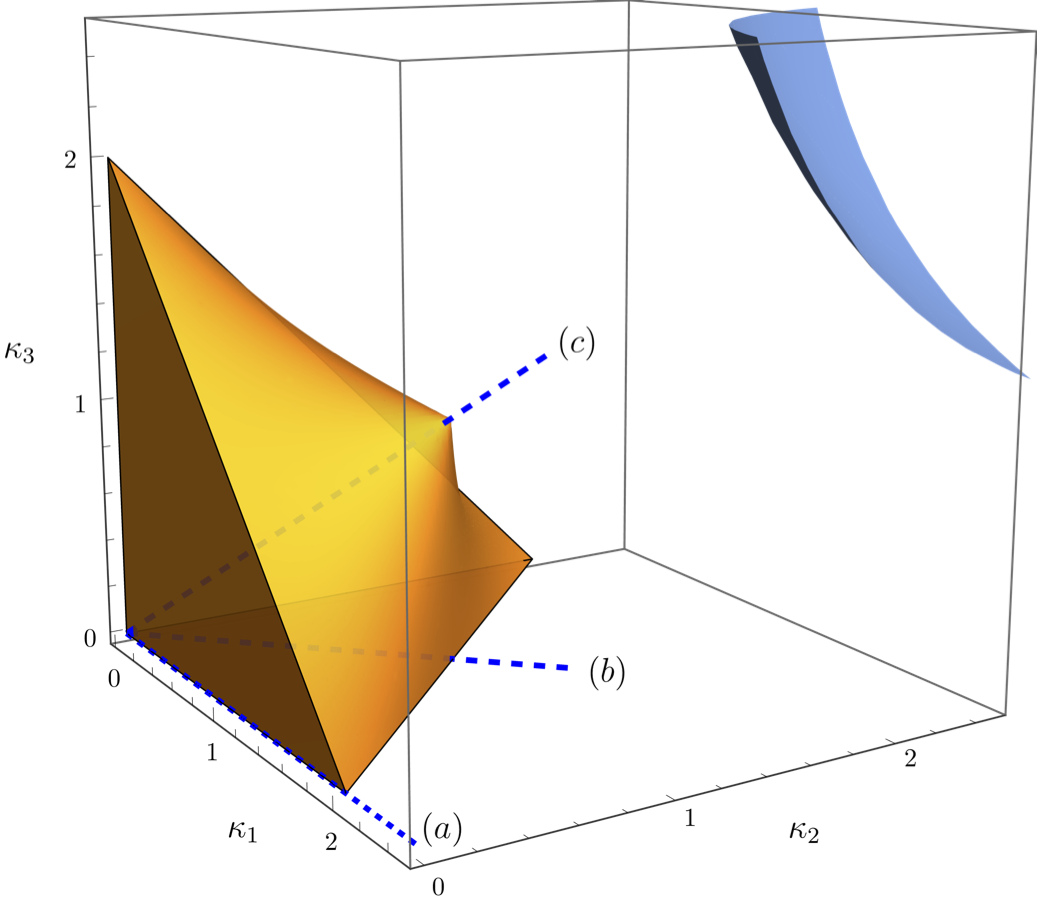}
\caption{{\small  The stability region \eqref{td-stability-region-boundary} - \eqref{td-stability-ineq}  (orange solid region) in the three-dimensional space of the black brane parameters $\kappa_1$, $\kappa_2$, $\kappa_3$. The dashed line marked {\it (a)} is the line $\kappa_2 =\kappa_3 = 0$. The dashed line marked {\it (b)} indicates $\kappa_3=0$, $\kappa_1 = \kappa_2$. The dashed line marked {\it (c)} indicates $\kappa_1=\kappa_2=\kappa_3$, and corresponds to the STU background with all $X_a = 1$, whose geometry is the Reissner-Nordstr\"{o}m AdS$_5$ black brane.  The blue surface in the upper-right corner is the surface $T=0$, determined by eq.~\eqref{eq:TH-kappa}. The zero-temperature surface lies outside the region of thermodynamic stability.}
}
\label{fig-kappa-space}
\end{figure}
%Mathematica file: Hessian-eigenvalues-picture.nb

%The determinant of $h_{ab}$, corresponding to the equation of state \eqref{eq:eos-STU} vanishes when
%\begin{align}
%\label{eq:zero-det}
%  2 - \kappa_1 - \kappa_2 - \kappa_3 + \kappa_1 \kappa_2 \kappa_3 = 0\,.
%\end{align}
%This equation determines the boundary region of thermodynamic stability, where at least one of the eigenvalues of $h_{ab}$ vanishes. 
The four eigenvalues of the conveniently normalised Hessian,  
\begin{equation}
\bar{h}^\epsilon_{ij}\equiv {\tiny \frac{N_c^2 T_0^2}{8}}\; h^\epsilon_{ij}\,,
\label{the-hessian}
\end{equation}
 are rather cumbersome functions of $\kappa_a$.  In contrast,  the four leading principal minors of $\bar{h}^\epsilon_{ij}$ are simple,  and  the conditions of positive-definiteness of $\bar{h}^\epsilon_{ij}$ may be written as
\begin{align}
  & 2 - \kappa_1 - \kappa_2 - \kappa_3 + \kappa_1 \kappa_2 \kappa_3 > 0 \,,   \label{ineq1}  \\
  & 3 - \kappa_1 - \kappa_2 - \kappa_3 - \kappa_1 \kappa_2 - \kappa_1 \kappa_3 - \kappa_2 \kappa_3 + 3 \kappa_1 \kappa_2 \kappa_3 >0 \,,   \label{ineq2} \\
  & 3 - \kappa_2 - \kappa_3 - \kappa_2 \kappa_3 > 0 \,,    \label{ineq3}   \\
  & 3-\kappa_3 > 0  \label{ineq4}  \,.
\end{align}
Inspection of  the inequalities \eqref{ineq1} - \eqref{ineq4}  determines the boundary of  the thermodynamic stability region \cite{Gladden:2024ssb}: 
\begin{align}
 & 2 - \kappa_1 - \kappa_2 - \kappa_3 + \kappa_1 \kappa_2 \kappa_3 = 0 \,,   \label{td-stability-region-boundary}\\
 & \kappa_1 + \kappa_2 + \kappa_3 < 3 \label{td-stability-ineq} \,,
 %\label{stability-region-kappa}
\end{align}
where the inequality \eqref{td-stability-ineq} indicates which part of the region bounded by the surface \eqref{td-stability-region-boundary} is stable.\footnote{The condition \eqref{td-stability-ineq} was overlooked in the analysis of ref.~\cite{Son:2006em}.}   The stability region is shown in Fig.~\ref{fig-kappa-space} together with the surface corresponding to the zero-temperature limit of the system. Note that with the equation of state \eqref{eq:eos-STU}, one cannot start in the thermodynamically stable high-temperature region, and cool the system down to zero temperature without encountering an instability.

\subsection{Fluctuations of the STU background}
We consider perturbations of the background \eqref{metric_stu_3}, \eqref{stu-scalars}, \eqref{eq:A-bg}, 
\begin{align}
  g_{\mu\nu} &= g^{(0)}_{\mu\nu} + h_{\mu\nu}\,,  \label{fluct-metric}\\
  A_\mu^a &= A_\mu^{a\; (0)} + \delta A_\mu^a\,, \label{fluct-gauge}\\
  H_a &= H_a^{(0)} + \delta H_a\,, \label{fluct-scalar}
\end{align}
and linearise the equations of motion \eqref{grav-eq}, \eqref{gauge-eq}, \eqref{scal-eq}. 
We choose the gauge $h_{\mu u} = 0$, $\delta A_u^a = 0$,
and write the perturbations as
\begin{align}
  \delta A_z^a (u,t,z) &= \pi T_0 \sqrt{2}\left(\prod_{a=1}^3 (1+\kappa_a)^{1/2}\right) e^{-i\omega t + iqz} a_z^a(u)\,,\\
  \delta A_t^a (u,t,z) &= \pi T_0 \sqrt{2} \left(\prod_{a=1}^3 (1+\kappa_a)^{1/2}\right) e^{-i\omega t + iqz}  a_t^a(u)\,, \\
\label{eq:sa-def}
  \delta H_a (u,t,z) & = e^{-i\omega t + iqz} s_a(u)\,,
\end{align}
where the $z$-axis was chosen to align with the direction of the spatial momentum  ${\bf q}=(0,0,q)$, due to the rotational invariance of the background.

Introducing the fields $H_a$ via Eq.~\eqref{fields-H} leads to a redundancy, which proves useful in deriving the equations of motion for the fluctuations. This is explained in detail in Appendix~\ref{scalar-and-redundancy}.

The rotational invariance of the STU background  implies that  the fluctuations can be classified into three channels --- scalar, shear, and sound (we follow the standard classification of refs.~\cite{Policastro:2002se,Policastro:2002tn,Kovtun:2005ev}). The scalar and shear fluctuations are discussed in ref.~\cite{chiral-life}. In this paper, we focus on fluctuations in the sound channel.

The general equations governing sound-channel fluctuations for arbitrary values of $\kappa_1$, $\kappa_2$, and $\kappa_3$ are rather cumbersome. For three special cases, namely $(\kappa_1, \kappa_2, \kappa_3) = (\kappa, 0, 0)$, $(\kappa, \kappa, 0)$, and $(\kappa, \kappa, \kappa)$, we present the corresponding equations of motion in Appendices~\ref{eoms-1-kappa}, \ref{fluct-eom-2-kappa}, and \ref{fluct-eom-3-kappa}, respectively.

\section{\texorpdfstring{${\cal N}=4$}{N=4} SYM with a single non-zero chemical potential}
\label{single-kappa-section}
We shall now study thermodynamic and dynamic stability in more detail. We start with the STU background of only one non-vanishing $\kappa_a$, taking 
$(\kappa_1,\kappa_2,\kappa_3) = (\kappa,0,0)$, corresponding to line {\it (a)} in Fig.~\ref{fig-kappa-space}. According to eq.~\eqref{eq:mu-kappa}, such backgrounds correspond to one non-vanishing chemical potential in ${\cal N}=4$ SYM theory, $(\mu_1,\mu_2,\mu_3)=(\mu,0,0)$.

\subsection{The STU background and thermodynamics}
\begin{figure}[t]
\centering
\includegraphics[width=0.45\textwidth]{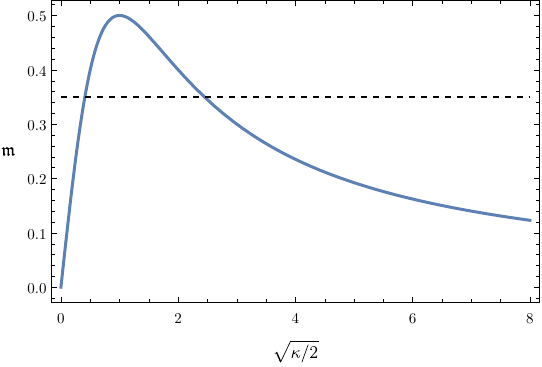}
\hspace{0.05\textwidth}
\includegraphics[width=0.45\textwidth]{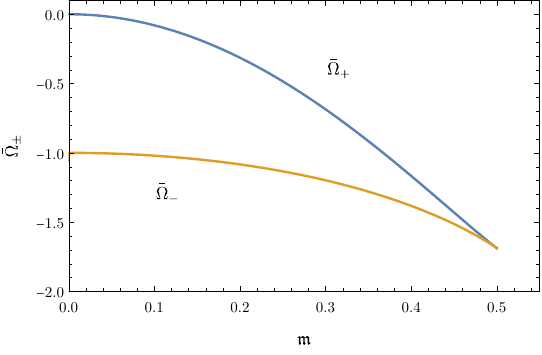}
\caption{\small The (normalised) chemical potential $\mfr = \mu/2\pi T_H$ as a function of $\sqrt{\kappa/2}$ (left panel). Two values of $\kappa=\kappa_\pm$ correspond to each fixed value of the chemical potential $\mfr$ as indicated by the  dashed line.  The normalised Gibbs thermodynamic potentials $\bar{\Omega}_\pm$ (Eq.~\eqref{Gibbs-single-kappa}) as functions of the (normalised) chemical potential $\mfr$. The potential $\bar{\Omega}_-$ (lower curve) corresponds to a stable physical equilibrium state. The two branches $\bar{\Omega}_\pm$ merge at the boundary of thermodynamic stability at $\mfr=1/2$ ($\kappa=2$).
}
\label{mu-kappa-plot}
\end{figure}

 The  STU background in the case of a single chemical potential is given by 
\begin{eqnarray}
ds^2_5 &=& - \mathcal{H}^{-2/3}{(\pi T_0 L)^2 \over u}\,f \, dt^2 
+   \mathcal{H}^{1/3}{(\pi T_0 L)^2 \over u}\, \left( dx^2 + dy^2 + dz^2\right) \nonumber \\
&+& \mathcal{H}^{1/3}{L^2 \over 4 f u^2} du^2\,,
\label{metric_u_3+} \\
 A_\mu(u) &\equiv& A_\mu^{(1)} = \delta_\mu^t \frac{1-u}{1+\kappa u}\, \pi T_0 \sqrt{\frac{2\kappa}{1+\kappa}}\,,  \qquad A_\mu^{(2)} = 0\,, \qquad A_\mu^{(3)} =0\,. \label{eq:A-bg-1} \\
 \mathcal{H} (u) &\equiv&  H_1(u) = 1+\kappa u\,,   \qquad H_2=H_3=1\,.
\end{eqnarray}
Here,  $f(u) = 1-u^2 + \kappa u (1-u)$,  and the integration constant has been chosen so that $A_t$ vanishes at the horizon. 
The temperature and the chemical potential can be expressed in terms of the parameters $T_0$ and $\kappa$ as
\begin{equation}
T = 
{2 + \kappa\over 
2\sqrt{1+\kappa}}\, T_0\,, \qquad  \mu = {\pi T_0 \sqrt{2 \kappa}\over \sqrt{1+\kappa}}\,, \qquad \mfr \equiv \frac{\mu}{2\pi T } =  {\sqrt{2 \kappa}\over \kappa +2}\,.
\label{mu-1-kappa}
\end{equation}
The requirement of thermodynamic stability \eqref{td-stability-region-boundary}, \eqref{td-stability-ineq} implies that $0\leq \kappa <2$ (and therefore  $\mfr \in [0,1/2)$).
The equilibrium energy density is
\begin{equation}
  \epsilon = \frac{6 \pi^2 N_c^2 (1+\kappa)^3}{(2+\kappa)^4}\, T^4 \,,
\label{eps-kappa-T}
\end{equation}
and the pressure is $p = \epsilon/3$. The equilibrium entropy density is
\begin{equation}
s = \frac{4 \pi^2 N_c^2 (1+\kappa)^2}{(2+\kappa)^3}\, T^3\,.
\label{s-kappa-T}
\end{equation}
The equilibrium charge densities are
\begin{equation}
  n_1 = \frac{\pi N_c^2 \sqrt{2\kappa} \, (1+\kappa)^2}{(2+\kappa)^3}\, T^3\,,\ \ \ \ 
  n_2 = n_3 = 0. 
\label{n-kappa-T}
\end{equation}
The equation \eqref{mu-1-kappa} implies that for each value of the (normalised) chemical potential $\mfr$ we have two values of the parameter $\kappa=\kappa_\pm = (1\pm\sqrt{1-4 \mfr^2})/2\mfr$ and hence two background solutions to the equations of motion - see  Fig.~\ref{mu-kappa-plot} (left panel).  The Gibbs potentials corresponding to the two solutions are
\begin{equation}
\bar{\Omega}_\pm \equiv \frac{\Omega_\pm}{Vp_0} = - \frac{(1+\kappa_\pm)^3}{(1+\kappa_\pm/2)^4}   = - \frac{16 \mfr^2 (1-\mfr^2 \pm\sqrt{1-4\mfr^2})}{1\pm\sqrt{1-4 \mfr^2}}\,,
\label{Gibbs-single-kappa}
\end{equation}
where $\Omega = - p V$,  $p$ is the pressure  \eqref{eq:pressure-kappa} and $V$ is the three-volume,   $p_0=\pi^2 N_c^2 T^4/8$ is the pressure at zero chemical potential.  They are shown in the right panel of  Fig.~\ref{mu-kappa-plot}.  The potential $\bar{\Omega}_- <\bar{\Omega}_+$ corresponds to a stable equilibrium and connects smoothly to the physical value $\bar{\Omega}_-=-1$ at zero chemical potential.  
The branch $\bar{\Omega}_+$ is 
unphysical\footnote{Some authors treat the branch $\bar{\Omega}_+$ as physical yet unstable,  and the point $\kappa=2$ as the point 
of a phase transition \cite{Maeda:2008hn,Buchel:2009mf,Buchel:2010gd,Natsuume:2010bs,DeWolfe:2011ts}.   We do 
not share this interpretation since the phase of the theory for $\kappa>2$ is unknown \cite{Son:2006em},  and it may well happen that 
$\kappa=2$ ($\mfr=1/2$) is the endpoint of the phase diagram for this theory.  At the same time, computing critical exponents at $\kappa\to 2^-$ is justified.}. The two branches 
merge at the boundary of thermodynamic stability at $\mfr=1/2$ ($\kappa=2$) and do not continue beyond that point. More details can be found in Appendix~\ref{app:tbb}.
%
%\begin{figure}[t]
%\centering
%\includegraphics[width=0.45\textwidth]{figs/single-kappa-mu-correspondence.pdf}
%\hspace{0.05\textwidth}
%\includegraphics[width=0.45\textwidth]{figs/single-kappa-two-omega-potentials.pdf}
%\caption{\small The (normalised) chemical potential $\mfr = \mu/2\pi T_H$ as a function of $\sqrt{\kappa/2}$ (left panel).
%}
%\label{mu-kappa-plot}
%\end{figure}
%
%
%Mathematica file: TD-stability-phase-transitions-gubser-buchel.nb

\subsection{Thermodynamic stability}
For a single non-vanishing chemical potential,  the Hessian \eqref{the-hessian} is given by
\begin{equation}
\bar{h}^\epsilon_{ij} =\begin{pmatrix}
\frac{2 + 5 \kappa - \kappa^2}{24 \pi^2 (1 + \kappa)^2} & \frac{( \kappa - 1) \sqrt{\kappa}}{3 \sqrt{2}\, \pi (1 + \kappa)^2} & 0 & 0 \\
\frac{(\kappa - 1) \sqrt{\kappa}}{3 \sqrt{2} \pi (1 + \kappa)^2} & \frac{3 - \kappa}{3 (1 + \kappa)^2} & 0 & 0 \\
0 & 0 & 1 & 0 \\
0 & 0 & 0 & 1
\end{pmatrix}\,.
\label{Hessian-1}
\end{equation}
The eigenvalues of $\bar{h}^\epsilon_{ij}$ are 
\begin{align}
     \lambda_{1} &= \lambda_2 = 1\,,\\[5pt]
     \lambda_{3,4} &= 
                      \frac{2 - 8 \pi^2 (-3 + \kappa) + 5 \kappa - \kappa^2}{48 \pi^2 (1 + \kappa)^2} \\ 
                   &\mp \frac{ \sqrt{96 \pi^2 (-2 + \kappa) (1 + \kappa)^2 + \left(-2 + 8 \pi^2 (-3 + \kappa) + (-5 + \kappa) \kappa \right)^2}}{48 \pi^2 (1 + \kappa)^2} \,,
 \label{eigen-3-4-kappa-1} \,
\end{align}
and the determinant is
\begin{equation}
\mbox{det}\, \bar{h}^\epsilon_{ij}=\frac{2-\kappa}{24\pi^2 (1+\kappa)^2}\,.
\end{equation}
The eigenvalue $\lambda_3$ is positive for $0\leq \kappa<2$, and negative for $\kappa>2$. The eigenvalue~$\lambda_4$ is positive for all $\kappa\geq 0$. Thus the region of thermodynamic stability for $(\kappa_1, \kappa_2, \kappa_3) = (\kappa, 0, 0)$ is $0\leq \kappa<2$, as can be seen from Fig.~\ref{fig-kappa-space}.

The corresponding eigenvectors are 
\begin{eqnarray}
V_1 &=& \left( 0, 0, 0, 1\right)\,,\\
V_2 &=& \left( 0, 0, 1, 0\right)\,, \\
V_3 &=& \left( p_{-}(\kappa), 1, 0, 0\right)\,,  \label{instability-eigenvector}\\
V_4 &=&  \left( p_+(\kappa), 1, 0, 0\right)\,,
\end{eqnarray}
where $p_{\pm}(\kappa)$ are functions of $\kappa$ whose explicit form we will not need.
%\begin{eqnarray}
%p_{\pm}&=&\frac{2 + 8 \pi^2 (-3 + \kappa) - (-5 + \kappa) \kappa}{8 \sqrt{2} \pi (-1 + \kappa) \sqrt{\kappa}} \nonumber \\
%&\pm& \frac{\sqrt{64 \pi^4 (-3 + \kappa)^2 + \left(-2 + (-5 + \kappa) \kappa\right)^2 
%+ 16 \pi^2 \left(-6 + \kappa (-5 + \kappa (-8 + 7 \kappa))\right)}}{8 \sqrt{2} \pi (-1 + \kappa) \sqrt{\kappa}}\,. \nonumber
%\end{eqnarray}
%
The eigenvector $V_3$ in  \eqref{instability-eigenvector} corresponding to the unstable mode,  mixes fluctuations of the entropy density $s$ and the charge density $n_1$. Accordingly, we expect that in the bulk description of the unstable mode, fluctuations of the metric couple to the fluctuations of the gauge field. We will now discuss the hydrodynamic modes.

\subsection{The hydrodynamic modes}
\subsubsection{Shear modes}
Dispersion relations for shear modes are given by the standard expression~\eqref{eq:w-shear}. 
The shear viscosity $\eta$ for the STU background was computed in ref.~\cite{Mas:2006dy, Son:2006em}. As expected, it obeys the universal relation $\eta/s = 1/4\pi$ \cite{Kovtun:2003wp,Kovtun:2004de,Buchel:2003tz,Starinets:2008fb}.  Given the thermodynamic functions \eqref{eps-kappa-T}, \eqref{s-kappa-T}, the shear mode dispersion relation is 
\begin{equation}
  \omega^{\rm shear}(q) = -i \frac{q^2}{8\pi T } \, \frac{2 + \kappa}{1 + \kappa} + O(q^4)\,.
\label{D-hydro-shear}
\end{equation}
Using Eq.~\eqref{mu-1-kappa}, this can be expressed in terms of $\mu/T$. For spatial momentum in the $z$-direction, shear mode singularities at $\omega = \omega^{\rm shear}(q)$ appear in the retarded two-point functions of shear-channel operators such as $G^{\rm ret.}_{T^{tx} T^{tx}}(\omega, q)$, $G^{\rm ret.}_{T^{tx} T^{zx}}(\omega, q)$, $G^{\rm ret.}_{J^{x}_1 J^{x}_1}(\omega, q)$, $G^{\rm ret.}_{J^{x}_1 T^{zx}}(\omega, q)$, etc.

\subsubsection{Sound modes}
Sound mode dispersion relations in a conformal theory are given by the standard expression \eqref{eq:sound-cft}. With the shear viscosity $\eta = s/4\pi$, one has
\begin{align}
   \omega^{\rm sound}_\pm(q) = \pm \frac{q}{\sqrt{3}} - i \frac{q^2}{12 \pi T }  \frac{2+\kappa}{ 1+\kappa } + O(q^3) \,.
\end{align}
Using Eq.~\eqref{mu-1-kappa}, this can be expressed in terms of $\mu/T$. For spatial momentum in the $z$-direction, sound mode singularities at $\omega = \omega^{\rm sound}(q)$ appear in the retarded two-point functions of sound-channel operators such as $G^{\rm ret.}_{T^{tt} T^{tt}}(\omega, q)$, $G^{\rm ret.}_{T^{tt} T^{tz}}(\omega, q)$, $G^{\rm ret.}_{T^{zz} T^{zz}}(\omega, q)$, $G^{\rm ret.}_{J^{z}_1 J^{z}_1}(\omega, q)$, $G^{\rm ret.}_{J^{z}_1 T^{tz}}(\omega, q)$, etc.

\subsubsection{Diffusion modes}
There are three diffusion modes \eqref{eq:wD}, whose diffusion coefficients are given by eq.~\eqref{eq:D-1} and \eqref{eq:Da-2}. Thermodynamic suscetibilities in eqs.~\eqref{eq:D-1},  \eqref{eq:Da-2} can be evaluated from the equation of state \eqref{eq:eos-STU}, and the relations between $s$, $n_a$ and $T_0$, $\kappa_a$. For the diffusion mode which mixes with sound, the diffusion coefficient is 
\begin{align}
\label{eq:D1-1}
  D_{(1)} = \frac{4}{N_c^2 T^2} \frac{2-\kappa}{1 +\kappa} \sigma_{(1)}\,.
\end{align}
The diffusion coefficients for the two diffusion modes which decouple from sound are 
\begin{align}
\label{eq:D2-1}
  D_{(2)} = \frac{2}{N_c^2 T^2} \frac{(2+\kappa)^2}{1+\kappa} \sigma_{(2)}\,,\ \ \ \ 
  D_{(3)} = D_{(2)} \,.
\end{align}
Assuming positive hydrodynamic entropy production ($\sigma_{(1)} > 0$, $\sigma_{(2)} >0$), the first diffusion mode becomes unstable at $\kappa>2$, due to the thermodynamic instability. 

The conductivities $\sigma_{(1)}$ and $\sigma_{(2)}$ are indeed positive and can be computed using standard holographic methods. In particular, $\sigma_{(1)}$ can be extracted from the results of ref.~\cite{Son:2006em}, which evaluated the two-point retarded correlators of $J^i_1$ at zero spatial momentum. Applying the Kubo formula to Eq.~(4.34) of Ref.~\cite{Son:2006em}, we find\footnote{The conductivity is also computed in Appendix~\ref{conductivity-matrix-app}.}
\begin{equation}
  \sigma_{(1)} =  - \frac{1}{\omega}\, \mbox{Im}\, G^{\rm ret.}_{J^z_1 J^z_1}(\omega, {\bf q} = 0) =\frac{N^2_c T (2+\kappa)}{32 \pi}\,.
\label{eq:sigma-1-1}
\end{equation}
Substituting into \eqref{eq:D1-1} gives the diffusion coefficient for the first diffusive mode,
\begin{equation}
  D_{(1)} = \frac{1}{2\pi T} \,\frac{4-\kappa^2}{4(1+\kappa)} \,.
\label{eq:D-11}
\end{equation}
At $\kappa= 0$, this becomes $D_{(1)}=1/2\pi T$, the diffusion coefficient for uncharged black branes~\cite{Policastro:2002se}. The diffusion coefficient \eqref{eq:D-11} becomes negative at $\kappa>2$, signalling a hydrodynamic instability of the STU background with $(\kappa_1, \kappa_2, \kappa_3) = (\kappa, 0, 0)$ at $\kappa>2$. Using eq.~\eqref{mu-1-kappa}, the diffusion coefficient $D_{(1)}$ in \eqref{eq:D-11} can be expressed in terms of $\mu/T$. We see that the first R-charge diffusion mode at $(\mu_1, \mu_2, \mu_3) = (\mu, 0, 0)$ is unstable, with a negative diffusion coefficient, for $\mu > \pi T$. 

%\begin{equation}
%D_R = \frac{2-\kappa}{6\pi T_H}+ \frac{(2-\kappa)^2}{72 \pi T_H} +\cdots
%\label{D-R-2}
%\end{equation}
%which agrees with numerical results in ref.~\cite{Buchel:2010gd}.
% $D_{(1)} = \frac{2-\kappa}{6\pi T}+ \frac{(2-\kappa)^2}{72 \pi T} +\dots$
%
%\begin{figure}
%\centering
%\includegraphics[width=0.46\textwidth]{figs/hessian-single-kappa-lamda-3.pdf}
%%{figs/diffusion-mode-pkk.pdf}
%\hspace{0.05\textwidth}
%\includegraphics[width=0.46\textwidth]{figs/hessian-single-kappa-lamda-4.pdf}
%\caption{
%\label{fig:eigenvalues-kappa=1}{\small The eigenvalues $\lambda_{3}$ and $\lambda_4$ of the Hessian \eqref{Hessian-1} in the case of a single non-vanishing chemical potential. The eigenvalue $\lambda_3$ changes sign at $\kappa=2$, whereas $\lambda_4$ is always positive.
%}}
%\end{figure}
%%

The conductivities $\sigma_{(2)}$ and $\sigma_{(3)} =\sigma_{(2)} $ can be evaluated analytically as well (see Appendix  \ref{conductivity-matrix-app}), and one finds
\begin{equation}
  \sigma_{(2)} =  - \frac{1}{\omega}\, \mbox{Im}\, G^{\rm ret.}_{J^z_2 J^z_2}(\omega, {\bf q} = 0) =\frac{N^2_c T}{8\pi} \frac{1}{2+\kappa}\,.
\label{eq:sigma-2-1}
\end{equation}
Substituting into \eqref{eq:D2-1}, we find the diffusion coefficients 
\begin{equation}
   D_{(2)} = \frac{1}{2\pi T} \,\frac{2+\kappa}{2(1+\kappa)} \,,\ \ \ \ 
   D_{(3)} = D_{(2)}\,.
\label{eq:D-21}
\end{equation}
These are positive for all $\kappa>0$, and the corresponding diffusion modes are stable. 
Using eq.~\eqref{mu-1-kappa}, the diffusion coefficients in \eqref{eq:D-21} can be expressed in terms of $\mu/T$.

\subsection{The unstable quasinormal mode}

So far, we have evaluated the diffusion coefficients $D_{(a)}$ in \eqref{eq:Da-kappa2} by combining the results for thermodynamic susceptibilities (evaluated from the equation of state) with the results for charge conductivities (evaluated using the Kubo formulas). Hydrodynamics predicts that dispersion relations such as \eqref{eq:w-shear}, \eqref{eq:wD}, \eqref{eq:wS} appear as singularities in real-time response functions of the corresponding operators, such as the energy-momentum tensor and R-currents. In the holographic description, such singularities in response functions manifest themselves as quasinormal modes of the corresponding bulk fields~\cite{Kovtun:2005ev}. 

Of particular note is the first diffusive mode with $\omega(q) = -i D_{(1)} q^2 + \dots$. Hydrodynamic expectation dictates that the STU background with $(\kappa_1, \kappa_2, \kappa_3) = (\kappa, 0, 0)$ has an unstable quasinormal mode when $\kappa>2$, with ${\rm Im}(\omega) > 0$ at small real $q$. This quasinormal mode arises from fluctuations of $A_\mu^1$ coupled to the fluctuations of the metric and of the scalar in the bulk. This unstable quasinormal mode was analyzed numerically by Buchel in \cite{Buchel:2010gd} in the vicinity of $\kappa=2$.%
\footnote{
  See also the brief summary in~\cite{Buchel:2010wk}. An earlier attempt \cite{Maeda:2008hn} to evaluate the diffusion coefficient $D_{(1)}$ in ${\cal N}=4$ SYM in a state with $(\mu_1, \mu_2, \mu_3) = (\mu, 0, 0)$ was unsuccessful because the authors assumed that $D_{(1)}$ was given by $\sigma_{(1)}/\chi_{(1)}$, which is not the correct expression for $D_{(1)}$ at $\mu\neq 0$.
}
Our analytic prediction \eqref{eq:D-11} near $\kappa=2$ gives $D_{(1)} = (2-\kappa)/(6\pi T) + (2-\kappa)^2/(72\pi T) + \dots$, in agreement with numerical results of~\cite{Buchel:2010gd}. The numerical analysis of the quasinormal spectrum at arbitrary $\kappa$ is technically challenging, and we leave it for future work. 

The fluctuations of $A_\mu^2$ and $A_\mu^3$ decouple from the fluctuations of the metric and of the scalar (see Appendix \ref{eoms-1-kappa}). The corresponding quasinormal spectrum contains the (stable) diffusion mode with diffusion coefficient $D_{(2)}$ in eq.~\eqref{eq:D-21}. The analysis of that quasinormal spectrum is straighforward, and we will not present it here.

%%%%%%%%%%%%  TWO CHEMICAL POTENTIALS  %%%%%%%%%%%%%%%%%%%%%%%%

\section{\texorpdfstring{${\cal N}=4$}{N=4} SYM with two equal non-zero chemical potentials}
\label{section-two-kappas}
We now focus on the STU background with two equal non-vanishing $\kappa_a$, taking $(\kappa_1,\kappa_2,\kappa_3) = (\kappa, \kappa, 0)$, corresponding to line {\it (b)} in Fig.~\ref{fig-kappa-space}. According to eq.~\eqref{eq:mu-kappa}, such backgrounds correspond to two equal chemical potentials%
\footnote{
Another way to obtain a background with $(\mu_1,\mu_2,\mu_3)=(\mu, \mu, 0)$ is to take $(\kappa_1,\kappa_2,\kappa_3) = (\kappa,  1/\kappa,0)$. However, such a configuration lies outside of the stability domain in Fig.~\ref{fig-kappa-space}. See Appendix~\ref{app:tbb} for more details. 
} 
in ${\cal N}=4$ SYM theory, $(\mu_1,\mu_2,\mu_3)=(\mu, \mu, 0)$.

\subsection{The STU background and thermodynamics}
The STU background \eqref{metric_stu_3}, \eqref{stu-scalars}, \eqref{eq:A-bg}  with  $(\kappa_1,\kappa_2,\kappa_3) = (\kappa,\kappa,0)$ is given by
\begin{align}
\label{stu-2-kappas-1}
   & ds_5^2 = -\frac{(\pi T_0 L)^2}{u} f(u)\mathcal{H}^{-4/3}\, dt^2 + \frac{(\pi T_0 L)^2}{u} \mathcal{H}^{2/3}\left(dx^2 {+} dy^2 {+} dz^2\right) 
  + \frac{L^2}{4f(u)u^2} \mathcal{H}^{2/3}\, du^2 , \\
\label{stu-2-kappas-2}
   & A_\mu^1 = A_\mu^2 = \pi T_0 \sqrt{2 \kappa}\, \frac{ 1- u}{1+\kappa u}\,, \qquad A_\mu^3 = 0\,, \\
   & \mathcal{H}\equiv H_1 = H_2 = 1+ \kappa u\,,  \qquad H_3 = 1,
\label{stu-2-kappas-3}
\end{align}
where $f(u)  = (1 - u) \left[1 + (1 + 2\kappa ) u \right]$. The Hawking temperature, the chemical potential and their ratio are given by
\begin{equation}
  T = T_0\,, \qquad  
  \mu = \pi T_0 \sqrt{2 \kappa}\,, \qquad 
  \mfr = \frac{\mu}{2\pi T } = \sqrt{\frac{\kappa}{2}}\,.
\label{mu-2-kappa}
\end{equation}
Note that the dependence of $\mfr$ on $\kappa$ is monotonic (unlike in the case of a single non-vanishing chemical potential), and thus each value of $\mu$ corresponds to a single STU background.  

The equilibrium energy density is
\begin{equation}
  \epsilon = \frac{3 \pi^2 N_c^2}{8} (1+\kappa)^2 \, T^4 \,,
\label{eps-kappa-T2}
\end{equation}
and the pressure is $p = \epsilon/3$. The equilibrium entropy density is
\begin{equation}
  s = \frac{\pi^2 N_c^2}{2} (1+\kappa)\, T^3\,.
\label{s-kappa-T2}
\end{equation}
The equilibrium charge densities are
\begin{equation}
  n_1 = n_2 = \frac{\pi N_c^2}{4\sqrt{2} } \sqrt{\kappa} \, (1+\kappa)\, T^3\,,\ \ \ \ 
  n_3 = 0. 
\label{n-kappa-T2}
\end{equation}
Thermodynamic stability conditions \eqref{td-stability-region-boundary}, \eqref{td-stability-ineq} imply that    $0\leq \kappa <1$,  and so  $\mfr \in [0,\sqrt{2}/2)$.
 
The normalised Gibbs potential is 
\begin{equation}
\bar{\Omega} \equiv \frac{\Omega}{Vp_0} = - \left( 1+\kappa\right)^2 = - \left( 1 + 2 \mfr^2\right)^2 \,,
\label{Gibbs-two-kappas}
\end{equation}
where $\Omega = - p V$,  $p$ is the pressure  \eqref{eq:pressure-kappa} and $V$ is the three-volume,   $p_0=\pi^2 N_c^2 T^4/8$ is the pressure at zero chemical potential.

\subsection{Thermodynamic stability}
\label{thermo-instab-2-kappa}
For two equal chemical  potentials, $(\kappa_1,\kappa_2,\kappa_3) = (\kappa,\kappa,0)$, the Hessian \eqref{the-hessian} is
\begin{equation}
  \bar{h}^\epsilon_{ij} =
  \begin{pmatrix}
  \frac{1 + 3 \kappa}{12 \pi^2 (1 + \kappa)^2} & \frac{ - \sqrt{\kappa}}{3 \sqrt{2} \pi (1 + \kappa)^2} & \frac{-\sqrt{\kappa}}{3 \sqrt{2} \pi (1 + \kappa)^2} & 0 \\
  \frac{-\sqrt{\kappa}}{3 \sqrt{2} \pi (1 + \kappa)^2} & \frac{3 - \kappa}{3 (1 + \kappa)^2} & \frac{2 \kappa}{3 (1 + \kappa)^2} & 0 \\
  \frac{-\sqrt{\kappa}}{3 \sqrt{2} \pi (1 + \kappa)^2} & \frac{2 \kappa}{3 (1 + \kappa)^2} & \frac{ 3 - \kappa}{3 (1 + \kappa)^2} & 0 \\
  0 & 0 & 0 & 1
\end{pmatrix}\,.
\end{equation}
The first two eigenvalues of this matrix are simple,
\begin{eqnarray}
  \lambda_{1} = \frac{1-\kappa}{(1+\kappa)^2}\,,\ \ \ \ \lambda_{2} = 1\,,  
 \label{eigen-kappa-2} 
\end{eqnarray}
whereas $\lambda_{3,4}$ are unilluminating, with $\lambda_{3,4}>0$.
%
%
%The eigenvalues of the Hessian  are 
%\begin{eqnarray}
%&\,&\lambda_{1} = \frac{1-\kappa}{(1+\kappa)^2}\,,\\
%&\,& \lambda_{2} = 1\,,  \\
%&\,& \lambda_{3,4} = \frac{1 + 12 \pi^2 + 3 \kappa + 4 \pi^2 \kappa}{24 \pi^2 (1 + \kappa)^2} \\ & 
%\mp & \frac{\sqrt{1 - 24 \pi^2 + 144 \pi^4 + 6 \kappa - 16 \pi^2 \kappa + 96 \pi^4 \kappa 
%+ 9 \kappa^2 - 24 \pi^2 \kappa^2 + 16 \pi^4 \kappa^2}}{24 \pi^2 (1 + \kappa)^2}\,,
% \label{eigen-kappa-2} \,
%\end{eqnarray}
Both the determinant
\begin{equation}
  \mbox{det}\, \bar{h}^\epsilon_{ij}=\frac{1-\kappa}{12\pi^2 (1+\kappa)^4}\,,
\end{equation}
and the eigenvalue $\lambda_1$ change sign at $\kappa=1$, implying a thermodynamic instability, whereas $\lambda_2,\lambda_3,\lambda_4$ stay positive for all $\kappa>0$. The corresponding eigenvectors are
\begin{eqnarray}
V_1 &=& \left( 0, -1, 1, 0\right)\,, \label{2-kappa-eigenvector-1}\\
V_2 &=& \left( 0, 0, 0, 1\right)\,, \\
V_3 &=& \left( q_{-}(\kappa), 1, 1, 0\right)\,,\\
V_4 &=&  \left( q_+(\kappa), 1, 1, 0\right)\,,
\end{eqnarray}
where $q_\pm(\kappa)$ are functions of $\kappa$ whose explicit form we will not need.
%\begin{eqnarray}
%q_\pm & = & 2 \sqrt{2} \pi \sqrt{\kappa} - \frac{1 - 12 \pi^2 + 3 \kappa +
% 12 \pi^2 \kappa }{4 \sqrt{2} \pi \sqrt{\kappa} }\nonumber  \\
%  &   \pm&  \frac{\sqrt{1 - 24 \pi^2 + 144 \pi^4 + 6 \kappa - 16 \pi^2 \kappa + 96 \pi^4 \kappa 
%  + 9 \kappa^2 - 24 \pi^2 \kappa^2 + 16 \pi^4 \kappa^2}}{4 \sqrt{2} \pi \sqrt{\kappa}}\,. \nonumber 
%\end{eqnarray}
The eigenvector $V_1$ corresponding to the unstable  mode suggests that the instability is driven by the density fluctuation $\delta (n_2-n_1)$. Accordingly, we expect that the bulk description of the unstable mode is given by the fluctuations of $A^2_\mu - A^1_\mu$. This will indeed be the case. We now discuss the hydrodynamic modes.

\subsection{The hydrodynamic modes}

\subsubsection{Shear modes}
Dispersion relations for shear modes are given by the standard expression~\eqref{eq:w-shear}. 
The shear viscosity $\eta$ for the STU background was computed in ref.~\cite{Mas:2006dy, Son:2006em}. As expected, it obeys the universal relation $\eta/s = 1/4\pi$  \cite{Kovtun:2003wp,Buchel:2003tz,Kovtun:2004de}.
Given the thermodynamic functions \eqref{eps-kappa-T2}, \eqref{s-kappa-T2}, the shear mode dispersion relation is 
\begin{equation}
  \omega^{\rm shear}(q) = -i \frac{q^2}{4\pi T } \, \frac{1}{1 + \kappa} + O(q^4)\,.
\label{D-hydro-shear2}
\end{equation}
Using Eq.~\eqref{mu-2-kappa}, this can be expressed in terms of $\mu/T$. For spatial momentum in the $z$-direction, shear mode singularities at $\omega = \omega^{\rm shear}(q)$ appear in retarded two-point functions of shear-channel operators such as $G^{\rm ret.}_{T^{tx} T^{tx}}(\omega, q)$, $G^{\rm ret.}_{T^{tx} T^{zx}}(\omega, q)$, $G^{\rm ret.}_{J^{x}_1 J^{x}_1}(\omega, q)$, $G^{\rm ret.}_{J^{x}_1 T^{zx}}(\omega, q)$, etc.

\subsubsection{Sound modes}
Sound mode dispersion relations in a conformal theory are given by the standard expression \eqref{eq:sound-cft}. With the shear viscosity $\eta = s/4\pi$, one has
\begin{align}
   \omega^{\rm sound}_\pm(q) = \pm \frac{q}{\sqrt{3}} - i \frac{q^2}{6 \pi T }  \frac{1}{ 1+\kappa } + O(q^3) \,.
\end{align}
Using Eq.~\eqref{mu-2-kappa}, this can be expressed in terms of $\mu/T$. For spatial momentum in the $z$-direction, sound mode singularities at $\omega = \omega^{\rm sound}(q)$ appear in retarded two-point functions of sound-channel operators such as $G^{\rm ret.}_{T^{tt} T^{tt}}(\omega, q)$, $G^{\rm ret.}_{T^{tt} T^{tz}}(\omega, q)$, $G^{\rm ret.}_{T^{zz} T^{zz}}(\omega, q)$, $G^{\rm ret.}_{J^{z}_1 J^{z}_1}(\omega, q)$, $G^{\rm ret.}_{J^{z}_1 T^{tz}}(\omega, q)$, etc.

\subsubsection{Diffusion modes}

There are three diffusion modes \eqref{eq:wD}, whose diffusion coefficients can be obtained using the analysis of sec.~\ref{hydro}. The first diffusion mode describes fluctuations $\delta n_1 + \delta n_2$ which couple to sound; the diffusion coefficient is given by eq.~\eqref{eq:D1plus2}. The second diffusion mode decouples from sound, and describes fluctuations $\delta n_1 -  \delta n_2$; this is the mode which will become unstable when $\kappa > 1$, with the diffusion coefficient given by eq.~\eqref{eq:d12}. The third diffusion mode describes fluctuations $\delta n_3$, with the diffusion coefficient given by eq.~\eqref{eq:Da-3}.

In the STU background with $(\kappa_1, \kappa_2, \kappa_3) = (\kappa, \kappa, 0)$, the charge susceptibility matrix $\chi_{ab}\equiv \partial^2 p/\partial\mu_a \partial\mu_b$ takes the general form~\eqref{eq:chiab-2mu}, and similarly the conductivity matrix takes the form
\begin{align}
\label{eq:sigma-2kappa}
  \sigma_{ab} = \begin{pmatrix}
  \sigma_{11} & \sigma_{12} & 0 \\
  \sigma_{12} & \sigma_{11} & 0 \\
  0 & 0 & \sigma_{33}
  \end{pmatrix} .
\end{align}
In the matrix $X_{AB}$ which appears in the linearised hydrodynamic equations \eqref{eq:E-linear-m}, the mixed derivatives $\upsilon_a \equiv \partial\epsilon/\partial\mu_a$ are such that $\upsilon_1 = \upsilon_2 = \upsilon$, and $\upsilon_3 = 0$. Substituting the thermodynamic derivatives into \eqref{eq:D1plus2}, \eqref{eq:d12}, \eqref{eq:Da-3}, we find the diffusion coefficients
\begin{align}
  D_{(1)} = \frac{8}{N_c^2 T^2} (\sigma_{11} {+} \sigma_{12})\,,\ \ \ \ 
  D_{(2)} = \frac{8}{N_c^2 T^2} \frac{1{-}\kappa}{(1{+}\kappa)^2} (\sigma_{11} {-} \sigma_{12})\,,\ \ \ \ 
  D_{(3)} = \frac{8}{N_c^2 T^2} \sigma_{33} \,.
\end{align}
The conductivity matrix $\sigma_{ab}$ in eq.~\eqref{eq:sigma-2kappa} can be evaluated using the standard holographic methods (see Appendix \ref{conductivity-matrix-app}).  Substituting $\sigma_{ab}$ in terms of $\kappa$, one finds
\begin{align}
\label{eq:Da-kappa2}
  D_{(1)} = \frac{1}{2 \pi T} \frac{1}{1+\kappa} \,,\ \ \ \ 
  D_{(2)} = \frac{1}{2\pi T} \frac{1-\kappa}{1+\kappa}\,,\ \ \ \ 
  D_{(3)} = \frac{1}{2 \pi T} \frac{1}{1+\kappa} \,.
\end{align}
As expected, the second diffusion coefficient becomes negative at $\kappa > 1$, signalling a dynamic instability of the state with $(\mu_1, \mu_2, \mu_3) = (\mu, \mu, 0)$ at $\mu > \sqrt{2}\, \pi T$.

%The charge susceptibility matrix $\chi_{ab}\equiv \partial^2 p/\partial\mu_a \partial\mu_b$ evaluates to
%\begin{align}
%  \chi_{ab} = \frac{N^2_c T^2}{8}\, \begin{pmatrix}
% \frac{1+2\kappa-\kappa^2}{1-\kappa} & \frac{2\kappa^2}{\kappa - 1} &0 \\[5pt]
%   \frac{2\kappa^2}{\kappa -1} &  \frac{1+2\kappa-\kappa^2}{1-\kappa}  & 0 \\[5pt]
%  0 & 0 & 1
%  \end{pmatrix} .
%\label{sus-matrix-2-kappas}
%\end{align}
%Its eigenvalues are $\frac{1}{8} N^2_c T^2 (1+3\kappa)$,  $\frac{1}{8} N^2_c T^2 (1+\kappa)^2/(1-\kappa)$, and $\frac{1}{8} N^2_c T^2$. The second eigenvalue of $\chi_{ab}$ becomes negative for $\kappa>1$, violating thermodynamic stability conditions. The corresponding eigenvvectors are $(1,1,0)$, $(-1,1,0)$, and $(0,0,1)$. 

\subsection{The unstable quasinormal mode}
\label{dynamic-instab-2-kappa}
So far, we have evaluated the diffusion coefficients $D_{(a)}$ in \eqref{eq:Da-kappa2} by combining the results for thermodynamic susceptibilities (evaluated from the equation of state) with the results for charge conductivities (evaluated using the Kubo formulas). Hydrodynamics predicts that dispersion relations such as \eqref{eq:w-shear}, \eqref{eq:wD}, \eqref{eq:wS} appear as singularities in real-time response functions of the corresponding operators, such as the energy-momentum tensor and R-currents. In the holographic description, such singularities in response functions manifest themselves as quasinormal modes of the corresponding bulk fields~\cite{Kovtun:2005ev}. 

We will focus on the second diffusive mode with $\omega(q) = -i D_{(2)} q^2 + \dots$. Hydrodynamic expectation dictates that the STU background with $(\kappa_1, \kappa_2, \kappa_3) = (\kappa, \kappa, 0)$ has an unstable hydrodynamic quasinormal mode when $\kappa>1$, with ${\rm Im}(\omega) > 0$ at small real $q$. This quasinormal mode arises from fluctuations of $A_\mu^2 - A_\mu^1$ in the bulk. 

We consider the fluctuations \eqref{fluct-metric}, \eqref{fluct-gauge}, \eqref{fluct-scalar},  applied to the background \eqref{stu-2-kappas-1}, \eqref{stu-2-kappas-2}, \eqref{stu-2-kappas-3}  and linearise the equations of motion \eqref{grav-eq}, \eqref{gauge-eq}, \eqref{scal-eq}.  The resulting coupled linear equations are written in Appendix      \ref{fluct-eom-2-kappa}.

Following ref.~\cite{Kovtun:2005ev},  we also introduce the ``electric'' field  
\begin{equation}
\label{eq:Ez-def}
  E^a_z = \mathfrak{w}_0\,  a_z^a + \mathfrak{q}_0\,  a_t^a\,,
\end{equation}
where $\mathfrak{w}_0 {\equiv} {\omega}/{2\pi T_0}$, $\mathfrak{q}_0 {\equiv} {q}/{2\pi T_0}$,
and by \eqref{mu-2-kappa}, $\wfr_0=\wfr\equiv \omega/2\pi T$, $\qfr_0=\qfr \equiv q/2\pi T$.
Naively, the equations for $E^a_z$ and $s_a$ couple to the equations obeyed by the metric perturbations $h_{\mu\nu}$, the so-called ``sound channel'' of ref.~\cite{Kovtun:2005ev}. However, hydrodynamic analysis suggests that there are decoupled ``diffusion channels'' in the sound channel. 

Indeed, one expects that $E_z^3$ and $s_3$ decouple from other perturbations, and indeed they do. The corresponding quasinormal spectrum contains the (stable) diffusion mode with diffusion coefficient $D_{(3)}$ in eq.~\eqref{eq:Da-kappa2}. The analysis of the quasinormal spectrum is straighforward, and we will not present it here. 

Metric perturbations couple to the linear combination $E_z^1 + E_z^2$. Naively, the linearised Einstein equations also couple to $\mathfrak{S} \equiv s_1 + s_2 - 2(1+\kappa u)s_3$ and $s_3$, but in fact by taking linear combinations of the equations, it can be shown that $\mathfrak{S} = 0$ (see Appendix \ref{scalar-and-redundancy}).   The corresponding quasinormal spectrum contains sound waves and the (stable) diffusion mode with diffusion coefficient $D_{(1)}$ in eq.~\eqref{eq:Da-kappa2}. The analysis of the quasinormal spectrum is technically cumbersome, and we will not present it here.

Finally, the combinations 
\begin{eqnarray}
  &\,& \mathfrak{E} \equiv E_z^1 - E_z^2\,, \\
  &\,& \mathfrak{s} \equiv s_1 - s_2\,,
\end{eqnarray}
decouple from all other perturbations. The corresponding quasinormal spectrum contains the diffusion mode with diffusion coefficient $D_{(2)}$ in eq.~\eqref{eq:Da-kappa2}. The decoupled equations for $\mathfrak{E}$ and $\mathfrak{s}$ are
\begin{align}
\mathfrak{s}'' &+ \left(\frac{f'}{f} - \frac{1+3\kappa u}{u\mathcal{H}}\right)\cdot \mathfrak{s}' - \frac{2\kappa^{1/2}(1+\kappa)^2 u\mathfrak{q}_0\,\mathcal{H}}{\mathfrak{D}(u)}\cdot \mathfrak{E}' \nonumber\\
&+ \left(\frac{1}{u^2\mathcal{H}f} -\frac{\kappa(1+2\kappa)u}{\mathcal{H}f} + \frac{3\kappa u}{u^2 f} + \frac{\mathfrak{D}(u)}{uf^2} - \frac{4\kappa(1+\kappa)^2u\,\mathfrak{q}_0^2}{\mathcal{H}^2\mathfrak{D}(u)}\right)\cdot\mathfrak{s} = 0\label{eq:2chgScalarEqn} \,,
\end{align}
\begin{align}
\mathfrak{E}'' &- \left(\frac{2\kappa\,\mathfrak{q}_0^2\,f}{\mathcal{H}\,\mathfrak{D}(u)}-\frac{\mathcal{H}^2\,\mathfrak{w}_0^2\,f'}{\mathfrak{D}(u)\cdot f}\right)\cdot\mathfrak{E}' + \frac{2\,\sqrt{\kappa}\,\mathfrak{q}_0}{\mathcal{H}^3}\cdot \mathfrak{s}' + \frac{\mathfrak{D}(u)}{u\,f^2}\mathfrak{E} \nonumber \\
&-2\,\sqrt{\kappa}\,\mathfrak{q}_0\,\left(\frac{\kappa}{\mathcal{H}^4}- \frac{\mathfrak{w}_0^2\,f'}{\mathcal{H}f\,\mathfrak{D}(u)} + \frac{2\kappa\,\mathfrak{w}_0^2}{\mathcal{H}^2\mathfrak{D}(u)}\right)\cdot\mathfrak{s} =0, \label{eq:2chgGaugeEqn}
\end{align}
where $f(u) = (1+\kappa u)^2 - u^2(1+\kappa)^2$, and we have introduced  $\mathfrak{D}(u) \equiv \mathfrak{w}_0^2(1+\kappa u)^2 - \mathfrak{q}_0^2f$.
%\begin{align*}
%\mathfrak{D}(u) &\equiv \mathfrak{w}_0^2(1+\kappa u)^2 - \mathfrak{q}_0^2f \,,\\
%\mathcal{H}(u) &\equiv 1+ \kappa u. 
%\end{align*}
%
%
\begin{figure}
\centering
\includegraphics[width=0.46\textwidth]{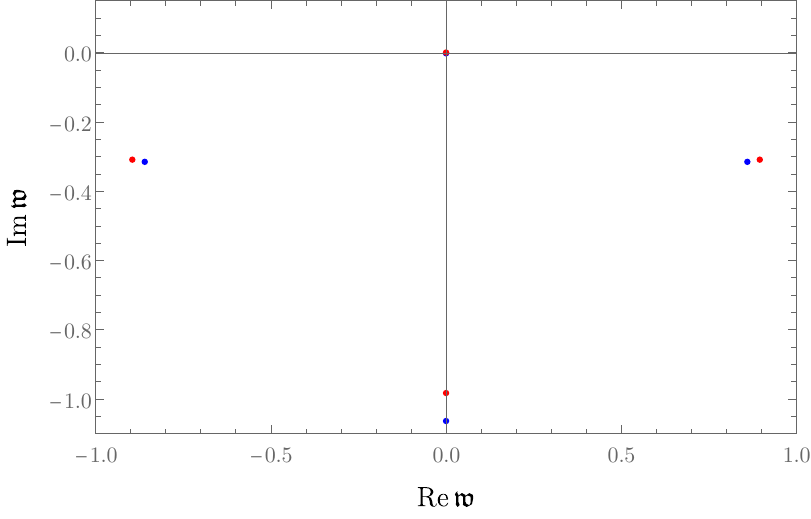}
%{figs/diffusion-mode-pkk.pdf}
\hspace{0.05\textwidth}
\includegraphics[width=0.46\textwidth]{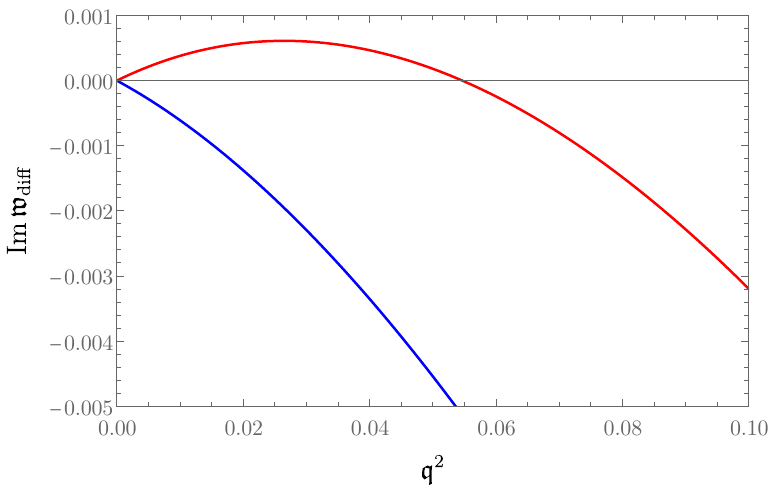}
\caption{
\label{fig:2chg-qnm-instability} The case of $(\kappa_1,\kappa_2,\kappa_3)=(\kappa,\kappa,0)$.  Left panel: The complex frequency plane showing the four highest quasinormal modes $\wfr_n=\wfr_n(\qfr^2)$ at $\qfr^2=0.025$, arising from the system (\ref{eq:2chgScalarEqn})---(\ref{eq:2chgGaugeEqn}). Blue dots correspond to modes at $\kappa = 0.9 < \kappa_c=1$, while red dots show modes at $\kappa = 1.1 > \kappa_c$. The purely imaginary diffusion modes crosses into the upper half-plane at $\kappa = \kappa_c = 1$, signifying the onset of an instability. 
Right panel: The dispersion relation of the diffusion mode with the diffusion coefficient $D_{(2)}$.   For $\kappa = 0.9 < \kappa_c$, shown in blue, $\text{Im}(\mathfrak{w}_\text{diff}) < 0$ for all $\mathfrak{q}^2$. However, for $\kappa = 1.1$, there is a critical region at small real momentum for which $\text{Im}(\mathfrak{w}_\text{diff}) > 0$.
}
\end{figure}
Quasinormal spectrum of the system \eqref{eq:2chgScalarEqn}, \eqref{eq:2chgGaugeEqn} can be found numerically.
%\footnote{
%  Our initial attempts to find an analytic solution for the diffusion mode were unsuccessful.
%}
%Since quasinormal solutions are proportional to $\exp[-i\omega t + iqz]$, $\text{Im}\, \omega>0$ corresponds to an exponentially growing, and therefore unstable, perturbation. 

The spectrum in the complex frequency plane for a fixed value of $\qfr^2=0.025$ and two values of $\kappa$ ($\kappa=0.9$ and $\kappa=1.1$)  is shown in the left panel of Fig.~\ref{fig:2chg-qnm-instability}.  The spectrum exhibits the hydrodynamic diffusion mode on the imaginary axis,  crossing into the upper half-plane for $\kappa>\kappa_c=1$, in agreement with the stability analysis prediction of section   \ref{thermo-instab-2-kappa}.

\begin{figure}
\centering
\includegraphics[width=0.46\textwidth]{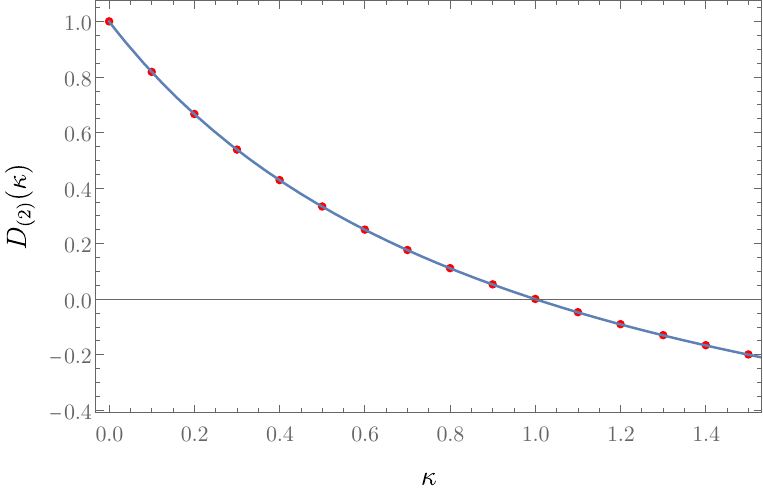}
\hspace{0.05\textwidth}
\includegraphics[width=0.46\textwidth]{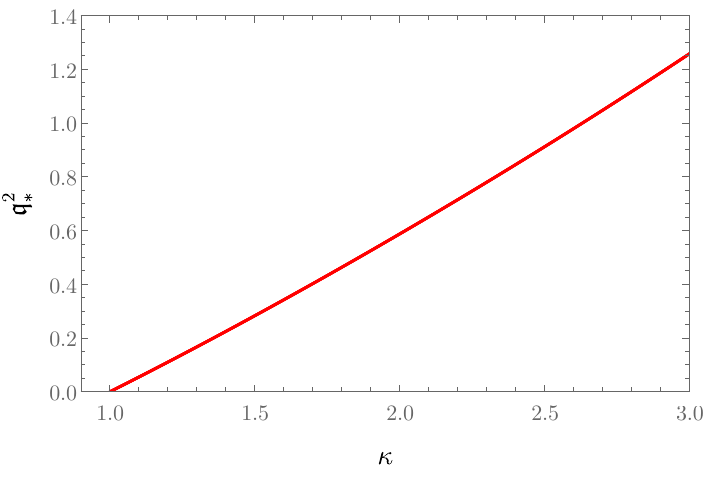}
\caption{ 
\label{fig:2chg-qnm-instability-2x}
 The case of $(\kappa_1,\kappa_2,\kappa_3)=(\kappa,\kappa,0)$.  Left panel: The analytic prediction for the diffusion coefficient $D_{(2)}$ (normalised by $1/2\pi T$) in Eq.~\eqref{diffusion-2-kappa} as a function of $\kappa$ is shown in blue. Numerical data are plotted in red, showing agreement. 
Right panel: The endpoint $\qfr_*^2$ of the instability domain as a function of $\kappa$.
}
\end{figure}

The dispersion relation of the unstable mode is shown in the right panel of Fig.~\ref{fig:2chg-qnm-instability}.  For sufficiently small $\mathfrak{q}^2$, the dispersion relation is linear in $\qfr^2$, which can be compared with our analytic result from eq.~\eqref{eq:Da-kappa2},
\begin{equation}
  \wfr^{\rm diffusion}_{(2)}(\qfr) = -i \, \frac{1-\kappa}{1+\kappa} \, \qfr^2 +\cdots\,.
\label{diffusion-2-kappa}
\end{equation}
This comparison is illustrated in the left panel of Fig.~\ref{fig:2chg-qnm-instability-2x}.

For  sufficiently large $\mathfrak{q}^2$,  the hydrodynamic mode crosses back into the lower half-plane, and becomes stable again for $\qfr^2 > \qfr_*^2$,  as shown in the right panel of Fig.~\ref{fig:2chg-qnm-instability}.  The dependence of $\qfr_*^2$ on $\kappa$ is shown in the right panel of Fig.~\ref{fig:2chg-qnm-instability-2x}. The value of $\qfr^2_*$ is an increasing function of $\kappa$.

%
%In addition to the hydrodynamic mode, there are other modes present on the imaginary axis of frequency
% (see Fig.~\ref{fig:2chg-qnm-instability}, left panel). However, unlike in the three-charge case 
% \cite{Gladden:2024ssb}, if these cross into the upper half-plane they do so at very large $\kappa$; no such behaviour
%  has been observed up to $\kappa = 10^4$. 
%\hl{[I don't understand the last sentence. Does it aply only to QNM on the imaginary axis, or to all QNM? And for which values of $\qfr$?]]}
%
%
%
%%%%%%%%%%%%%%%%%%%%%%%%%%%%%%%%%%%%%%%%%%%%%%%%%
\section{\texorpdfstring{${\cal N}=4$}{N=4} SYM with three equal non-zero chemical potentials}
\label{3-equal-kappas-section}
Finally, we consider STU backgrounds with three equal non-vanishing $\kappa_a$, taking $(\kappa_1,\kappa_2,\kappa_3) = (\kappa, \kappa, \kappa)$, corresponding to line {\it (c)} in Fig.~\ref{fig-kappa-space}. According to eq.~\eqref{eq:mu-kappa}, such backgrounds correspond to three equal chemical potentials%
\footnote{
As pointed out in ref.~\cite{Buchel:2025tjq},  another way to obtain  the state with $(\mu_1,\mu_2,\mu_3)=(\mu, \mu, \mu)$ is to take $(\kappa_1,\kappa_2,\kappa_3) = (\kappa, \kappa,  1/\kappa)$. However, such a configuration lies outside of the stability domain in Fig.~\ref{fig-kappa-space}. See Appendix~\ref{app:tbb} for more details. 
}
in ${\cal N}=4$ SYM theory, $(\mu_1,\mu_2,\mu_3)=(\mu, \mu, \mu)$. We have previously considered this case in ref.~\cite{Gladden:2024ssb}; here, we will provide additional details.

\subsection{The STU background and thermodynamics}
The STU background \eqref{metric_stu_3}, \eqref{stu-scalars}, \eqref{eq:A-bg} for $(\kappa_1,\kappa_2,\kappa_3)=(\kappa,\kappa,\kappa)$  reduces to 
\begin{eqnarray}
    ds_{5}^{2} &=& -\frac{(\pi T_{0}L)^{2}}{u {\cal H}^{2}} f dt^{2}+\frac{(\pi T_{0}L)^{2} {\cal H}}{u}\big(dx^{2}+dy^{2}+dz^{2}\big)
+\frac{{\cal H} L^{2}}{4 f u^{2}} du^{2}\,,
\label{metric_u_3_diag_x} \\
A_t (u) &\equiv& A^1_t (u) = A^2_t (u) = A^3_t (u)  = \frac{1-u}{{\cal H}(u)} \pi T_0 \sqrt{2 \kappa (1+\kappa)}\,, 
\label{gauge-bg_x}\\
{\cal H}(u) &\equiv&  H_1 (u) = H_2 (u) =H_3(u) = 1+\kappa u \,,
\label{scal_gauge_u_3_diag_x}
\end{eqnarray}
where $f(u) = (1-u)\left(1+(1+3 \kappa) u - \kappa^3 u^2\right)$. 
%The background is characterised by  two parameters,  $Q$ and $T_0$, where $Q$ is the charge and $T_0$ is the Hawking temperature of the metric in the limit $Q\to 0$ (i.e.  the AdS-Schwarzschild black brane metric),
%\begin{equation}
% ds^{2} = -\frac{(\pi T_{0}L)^{2}}{u} f_0 dt^{2}+\frac{(\pi T_{0}L)^{2} }{u}\big(dx^{2}+dy^{2}+dz^{2}\big)
%+\frac{ L^{2}}{4 f_0 u^{2}} du^{2}\,,
%\end{equation}
%where $f_0(u)=1-u^2$.  The parameter $\kappa$ is related to $Q$ and $T_0$ via $\kappa= Q/(\pi T_0 L^2)^2$. 
The Hawking temperature is given by
\begin{equation}
  T = \frac{(2  -\kappa ) \sqrt{1+\kappa} }{2} \, T_0\,.
\label{T-H-diag_x}
\end{equation}
The chemical potentials $\mu_a = \mu$ are related to the parameter $\kappa$ by
\begin{equation}
  \mu_a  = \pi T_0 \sqrt{2\kappa (1+\kappa)}\,,\ \ \ \ 
  \mfr \equiv \frac{\mu}{2\pi T } =  {\sqrt{ \kappa/2}\over 1- \kappa /2 }\,.
\label{mu-diag}
\end{equation}
The dependence \eqref{mu-diag} is monotonic for $\kappa \in [0,2]$; that is, in contrast to the case of a single non-vanishing chemical potential, and similarly to the case of two equal chemical potentials, each physical chemical potential corresponds to a unique gravity background.

The equilibrium energy density is 
\begin{equation}
  \epsilon = 6 \pi^2 N_c^2 \, \frac{(1+\kappa)}{(2-\kappa)^4} \, T^4 \,,
\label{eps-kappa-T3}
\end{equation}
and the pressure is $p = \epsilon/3$. The equilibrium entropy density is 
\begin{equation}
  s =  \frac{4\pi^2 N_c^2}{(2-\kappa)^3} \, T^3\,.
\label{s-kappa-T3}
\end{equation}
The equilibrium charge densities are 
\begin{equation}
  n_1 = n_2 = n_3 = \pi N_c^2  \frac{\sqrt{2\kappa}}{(2-\kappa)^3} \, T^3\,.
\label{n-kappa-T3}
\end{equation}
%The dependence $\mfr=\mfr (\kappa)$ is monotonic for $0\leq\kappa<2$.
% in the domain of stability $\kappa\in [0,1)$ (or $\mfr \in [0,\sqrt{2}))$:
%\begin{equation}
%  \mfr = \frac{\mu}{2\pi T_H } =  {\sqrt{ \kappa/2}\over 1- \kappa /2 }\,.
%\label{mu-T-diag_x}
%\end{equation}
The normalised Gibbs potential is 
\begin{equation}
\bar{\Omega} \equiv \frac{\Omega}{Vp_0} = - \frac{1+\kappa}{(1-\kappa/2)^4} 
= -\frac{16 \mfr^7 \left(\sqrt{1+4 \mfr^2}+\mfr-1\right)}{(\sqrt{1+4 \mfr^2}-1)^4}\,,
\label{Gibbs-three-kappas}
\end{equation}
where $\Omega = - p V$,  $p$ is the pressure  \eqref{eq:pressure-kappa} and $V$ is the three-volume,    $p_0=\pi^2 N_c^2 T^4/8$ is the pressure at zero chemical potential.  

The metric (\ref{metric_u_3_diag_x}) can be written in the standard $\text{AdS}_5$-Reissner-Nordstr\"om form 
\begin{equation}
    ds_{5}^{2}=-\frac{(\pi T_{0} L)^{2}(1+\kappa)}{\Tilde{u}} \Tilde{f} dt^{2}+\frac{(\pi T_{0} L)^{2}(1+\kappa)}{\Tilde{u}}(dx^{2}+dy^{2}+dz^{2})+\frac{L^{2}}{4 \Tilde{f} \Tilde{u}^{2}} d\Tilde{u}^{2}\,,
\end{equation}
where $\Tilde{f}=1-\Tilde{u}^{2}(1+\kappa)+\kappa \Tilde{u}^{3}$, 
by making the change of variables  ${\Tilde{u}} = u H(1)/H(u)$. The gauge field becomes
\begin{equation}
A_t  = {\sqrt{2 Q (1+\kappa) } {\Tilde{u}}\over L^2} = \mu {\Tilde{u}}\,,
\label{scal_gauge_u_3_diag_tilde}
\end{equation}
and the scalar field is\footnote{
In the original $S^5$ reduction ansatz, the STU fields $X_a$ are expressed as exponentials of the two scalars (see, e.g., ref.~\cite{Cvetic:1999xp}). In the case of three equal chemical potentials, we have $X_1 = X_2 = X_3 = 1$, and the background scalar fields are trivial.
}
\begin{equation}
{\cal H}(\Tilde{u}) = \frac{1+\kappa}{1+\kappa(1-\Tilde{u})}\,.
\end{equation}
Formulas \eqref{T-H-diag_x} - \eqref{mu-diag} remain valid, as expected.  The parameter $\kappa$ is related to $Q$ and $T_0$ via $\kappa= Q/(\pi T_0 L^2)^2$.

 \subsection{Thermodynamic stability}
 \label{td-stability}
For three equal chemical potentials, the thermodynamic stability condition is $\kappa<1$, and we expect $\kappa_c=1$ to correspond to the onset of the dynamic instability. The energy Hessian $\bar{h}^\epsilon_{ij}$ in eq.~\eqref{the-hessian} is given by
\begin{equation}
  \bar{h}^\epsilon_{ij} =
  \begin{pmatrix}
  \frac{2 + 5 \kappa}{24 \pi^2 (1 + \kappa)} & -\frac{\sqrt{\kappa}}{3\sqrt{2}\pi (1 + \kappa)} & -\frac{\sqrt{\kappa}}{3\sqrt{2}\pi (1 + \kappa)} & -\frac{\sqrt{\kappa}}{3\sqrt{2}\pi (1 + \kappa)} \\
-\frac{\sqrt{\kappa}}{3\sqrt{2}\pi (1 + \kappa)} & \frac{3 - \kappa}{3 (1 + \kappa)^2} & \frac{2 \kappa}{3 (1 + \kappa)^2} & \frac{2 \kappa}{3 (1 + \kappa)^2} \\
-\frac{\sqrt{\kappa}}{3\sqrt{2}\pi (1 + \kappa)} & \frac{2 \kappa}{3 (1 + \kappa)^2} & \frac{3 - \kappa}{3 (1 + \kappa)^2} & \frac{2 \kappa}{3 (1 + \kappa)^2} \\
-\frac{\sqrt{\kappa}}{3\sqrt{2}\pi (1 + \kappa)} & \frac{2 \kappa}{3 (1 + \kappa)^2} & \frac{2 \kappa}{3 (1 + \kappa)^2} & \frac{3 - \kappa}{3 (1 + \kappa)^2}
\end{pmatrix} .
\end{equation}
The eigenvalues of this matrix are
\begin{eqnarray}
\lambda_{1} &=& \lambda_2 = \frac{1-\kappa}{(1+\kappa)^2} \,,  \label{eigen-1-2}\\
\lambda_{3,4} &=&  \frac{2+ 24 \pi^2 + 5 \kappa \mp \sqrt{576 \pi^4 +48 \pi^2 (3\kappa-2) +(5\kappa+2)^2}}{48 \pi^2 (1+\kappa)}\label{eigen-3-4} \,.
\end{eqnarray}
The eigenvalues $\lambda_{1}$ and $\lambda_2$ become negative for $\kappa>1$ whereas the eigenvalues $\lambda_{3,4}$ remain positive for all $0\leq\kappa<2$. Thus, the system is thermodynamically unstable for $\kappa>1$,  although the  determinant of the Hessian is positive for $\kappa>1$.

The eigenvectors corresponding to the eigenvalues $\lambda_{1,2,3,4}$ are given by\footnote{The role of the eigenvalues and eigenvectors of the Hessian corresponding to the unstable modes was first emphasised in ref.~\cite{Gubser:2000ec}. They were identified for the STU model in ref.~\cite{Gentle:2012rg}.} 
\begin{eqnarray}
V_1 &=& \left( 0, -1, 0, 1\right)\,,\\
V_2 &=& \left( 0, -1, 1, 0\right)\,, \\
V_3 &=& \left( r_{-}(\kappa), 1, 1, 1\right)\,,\\
V_4 &=&  \left( r_+(\kappa), 1, 1, 1\right)\,,
\end{eqnarray}
where $r_\pm$ are functions of $\kappa$ whose explicit form we will not need.
The eigenvectors $V_{1,2}$ corresponding to the unstable  modes suggest that the instability is driven by the density fluctuation $\delta (n_2-n_1)$, $\delta (n_3-n_1)$, $\delta (n_3-n_2)$ (only two of which are independent). Accordingly, the bulk description of the unstable mode will be given by the fluctuations of $A^2_\mu - A^1_\mu$, $A^3_\mu - A^1_\mu$, $A^3_\mu - A^2_\mu$.

Alternatively, the eigenvectors of $\bar{h}^\epsilon_{ij}$ corresponding to the degenerate eigenvalues $\lambda_{1,2}$ may be chosen as two linear combinations of the three vectors 
\begin{eqnarray}
 W_1 &=& \left( 0, 2, -1, -1\right)\,,  \label{eigenvectors-hessian-1} \\
 W_2 &=& \left( 0, -1, 2, -1\right)\,,  \label{eigenvectors-hessian-2} \\
 W_3 &=& \left( 0, -1, -1, 2\right)\,.
\label{eigenvectors-hessian-3}
\end{eqnarray}
The eigenvectors $W_{1,2,3}$ are not independent, $W_1+W_2+W_3=0$. Thus we may also view the unstable modes as the fluctuations $2\delta n_1 - \delta n_2 - \delta n_3$, $2\delta n_2 - \delta n_1 - \delta n_3$, $2\delta n_3 - \delta n_1 - \delta n_2$. Accordingly, the bulk description of the two unstable modes will be given by the fluctuations of $2A^1_\mu - A^2_\mu - A^3_\mu$, $2A^2_\mu - A^1_\mu - A^3_\mu$, $2A^3_\mu - A^1_\mu - A^2_\mu$.
We now discuss the hydrodynamic modes.

\subsection{The hydrodynamic modes}

\subsubsection{Shear modes}
Dispersion relations for shear modes are given by the standard expression~\eqref{eq:w-shear}. 
The shear viscosity $\eta$ for the STU background was computed in ref.~\cite{Mas:2006dy, Son:2006em}. As expected, it obeys the universal relation $\eta/s = 1/4\pi$  \cite{Kovtun:2003wp,Buchel:2003tz,Kovtun:2004de}.  Given the thermodynamic functions \eqref{eps-kappa-T3}, \eqref{s-kappa-T3}, the shear mode dispersion relation is 
\begin{equation}
  \omega^{\rm shear}(q) = -i \frac{q^2}{8\pi T } \, \frac{2 - \kappa}{1 + \kappa} + O(q^4)\,.
\label{D-hydro-shear3}
\end{equation}
Using Eq.~\eqref{mu-diag}, this can be expressed in terms of $\mu/T$. For spatial momentum in the $z$-direction, shear mode singularities at $\omega = \omega^{\rm shear}(q)$ appear in retarded two-point functions of shear-channel operators such as $G^{\rm ret.}_{T^{tx} T^{tx}}(\omega, q)$, $G^{\rm ret.}_{T^{tx} T^{zx}}(\omega, q)$, $G^{\rm ret.}_{J^{x}_1 J^{x}_1}(\omega, q)$, $G^{\rm ret.}_{J^{x}_1 T^{zx}}(\omega, q)$, etc.

\subsubsection{Sound modes}
Sound mode dispersion relations in a conformal theory are given by the standard expression \eqref{eq:sound-cft}. With the shear viscosity $\eta = s/4\pi$, one has
\begin{align}
   \omega^{\rm sound}_\pm(q) = \pm \frac{q}{\sqrt{3}} - i \frac{q^2}{12 \pi T }  \frac{2-\kappa}{ 1+\kappa } + O(q^3) \,.
\end{align}
Using eq.~\eqref{mu-diag}, this can be expressed in terms of $\mu/T$. For spatial momentum in the $z$-direction, sound mode singularities at $\omega = \omega^{\rm sound}(q)$ appear in retarded two-point functions of sound-channel operators such as $G^{\rm ret.}_{T^{tt} T^{tt}}(\omega, q)$, $G^{\rm ret.}_{T^{tt} T^{tz}}(\omega, q)$, $G^{\rm ret.}_{T^{zz} T^{zz}}(\omega, q)$, $G^{\rm ret.}_{J^{z}_1 J^{z}_1}(\omega, q)$, $G^{\rm ret.}_{J^{z}_1 T^{tz}}(\omega, q)$, etc.

\subsubsection{Diffusion modes}
There are three diffusion modes \eqref{eq:wD}, whose diffusion coefficients can be obtained using the analysis of sec.~\ref{hydro}. The first diffusion mode describes fluctuations $\delta n_1 + \delta n_2 + \delta n_3$ which couple to sound. The second and the third diffusion modes decouple from sound, and describe (two of the) fluctuations $\delta n_1 -  \delta n_2$, $\delta n_1 -  \delta n_3$, $\delta n_2 -  \delta n_3$; these are the modes which will become unstable when $\kappa > 1$, with the diffusion coefficient given by eq.~\eqref{eq:d12}. 

In the STU background with $(\kappa_1, \kappa_2, \kappa_3) = (\kappa, \kappa, \kappa)$, the charge susceptibility matrix $\chi_{ab}\equiv \partial^2 p/\partial\mu_a \partial\mu_b$ takes the general form~\eqref{eq:chi-Nf3-equal-mu}, and similarly the conductivity matrix takes the form
\begin{align}
\label{eq:sigma-3kappa}
  \sigma_{ab} = \begin{pmatrix}
  \sigma_{11} & \sigma_{12} & \sigma_{12} \\
  \sigma_{12} & \sigma_{11} & \sigma_{12} \\
  \sigma_{12} & \sigma_{12} & \sigma_{11}
  \end{pmatrix} .
\end{align}
In the matrix $X_{AB}$ which appears in the linearised hydrodynamic equations \eqref{eq:E-linear-m}, the mixed derivatives $\upsilon_a \equiv \partial\epsilon/\partial\mu_a$ are such that $\upsilon_1 = \upsilon_2 = \upsilon_3$. 
The three diffusion coefficients $D_{(a)}$ may be obtained as the eigenvalues of the diffusion matrix \eqref{eq:D-matr}. 
Substituting the thermodynamic derivatives, we find 
\begin{align}
  & D_{(1)} = \frac{4}{N_c^2 T^2} (1+\kappa) (2+\kappa) (\sigma_{11} + 2\sigma_{12})\,,\\
  & D_{(2)} = D_{(3)}  = \frac{2}{N_c^2 T^2} \frac{(2-\kappa)^2 (1-\kappa)}{(1 + \kappa)} (\sigma_{11} - \sigma_{12})\,,
\end{align}
The conductivity matrix $\sigma_{ab}$ in eq.~\eqref{eq:sigma-3kappa} can be evaluated using the standard holographic methods (see Appendix \ref{conductivity-matrix-app}). Substituting $\sigma_{ab}$ in terms of $\kappa$, one finds
\begin{align}
\label{eq:Da-kappa3}
  D_{(1)} = \frac{1}{8 \pi T} \frac{4-\kappa^2}{1+\kappa} \,,\ \ \ \ 
  D_{(2)} = D_{(3)} = \frac{1}{4\pi T} \frac{(2-\kappa) (1-\kappa)}{(1+\kappa)}\,. 
\end{align}
As expected, the diffusion coefficients $D_{(2)}$,  $D_{(3)}$ become negative at $\kappa > 1$, signalling a dynamic instability of the state with $(\mu_1, \mu_2, \mu_3) = (\mu, \mu, \mu)$ at $\mu > \sqrt{2}\, 2\pi T$.

\subsection{The unstable quasinormal mode}
\label{instability-sound-channel}
So far, we have evaluated the diffusion coefficients $D_{(a)}$ in \eqref{eq:Da-kappa3} by combining the results for thermodynamic susceptibilities (evaluated from the equation of state) with the results for charge conductivities (evaluated using the Kubo formulas). Hydrodynamics predicts that dispersion relations such as \eqref{eq:w-shear}, \eqref{eq:wD}, \eqref{eq:wS} appear as singularities in real-time response functions of the corresponding operators, such as the energy-momentum tensor and R-currents. In the holographic description, such singularities in response functions manifest themselves as quasinormal modes of the corresponding bulk fields~\cite{Kovtun:2005ev}. 

We will focus on the diffusive mode with $\omega(q) = -i D_{(2)} q^2 + \dots$. Hydrodynamic expectation dictates that the STU background with $(\kappa_1, \kappa_2, \kappa_3) = (\kappa, \kappa, \kappa)$ has two unstable hydrodynamic quasinormal mode when $\kappa>1$, with ${\rm Im}(\omega) > 0$ at small real $q$. These quasinormal modes arise from fluctuations of $A_\mu^2 - A_\mu^1$, or $A_\mu^3 - A_\mu^1$, or $A_\mu^3 - A_\mu^2$ in the bulk.%
\footnote{
  The unstable modes we discuss here were also observed in \cite{Gentle:2012rg} at zero spatial momentum. }

We consider perturbations of the background \eqref{metric_u_3_diag_x}, \eqref{gauge-bg_x}, \eqref{scal_gauge_u_3_diag_x}, as in \eqref{fluct-metric}, \eqref{fluct-gauge}, and \eqref{fluct-scalar}, where the background values $g^{(0)}_{\mu\nu}$, $A_\mu^{a\; (0)}$ and $H_a^{(0)}$ now correspond to the STU background with $\kappa_a = \kappa$. We define electric field perturbations ${\bf E}_z = (E^1_z, E^2_z, E^3_z)$ as in eq.~\eqref{eq:Ez-def}, and scalar perturbations ${\bf s} = (s_1, s_2, s_3)$ as in \eqref{eq:sa-def}. We can work either with pair-wise differences of the perturbations, or with combinations
\begin{align}
   \mathfrak{E}^1_z & =\coeff13 (  2,-1,-1  ) {\cdot} {\bf E}_z\,, &
   \mathfrak{E}^2_z & =\coeff13 (  -1,2,-1  ) {\cdot} {\bf E}_z\,, &
   \mathfrak{E}^3_z & =\coeff13 (  -1,-1,2  ) {\cdot} {\bf E}_z\,,\\[5pt]
   \mathfrak{s}_1 & = \coeff13 (  2,-1,-1  ) {\cdot} {\bf s}\,, &
   \mathfrak{s}_2 & = \coeff13 (  -1,2,-1  ) {\cdot} {\bf s}\,, &
   \mathfrak{s}_3 & = \coeff13 (  -1,-1,2  ) {\cdot} {\bf s}\,, 
\end{align}
reflecting the eigenvectors \eqref{eigenvectors-hessian-1} - \eqref{eigenvectors-hessian-3} of the energy Hessian.

\begin{figure}
\centering
\includegraphics[width=0.46\textwidth]{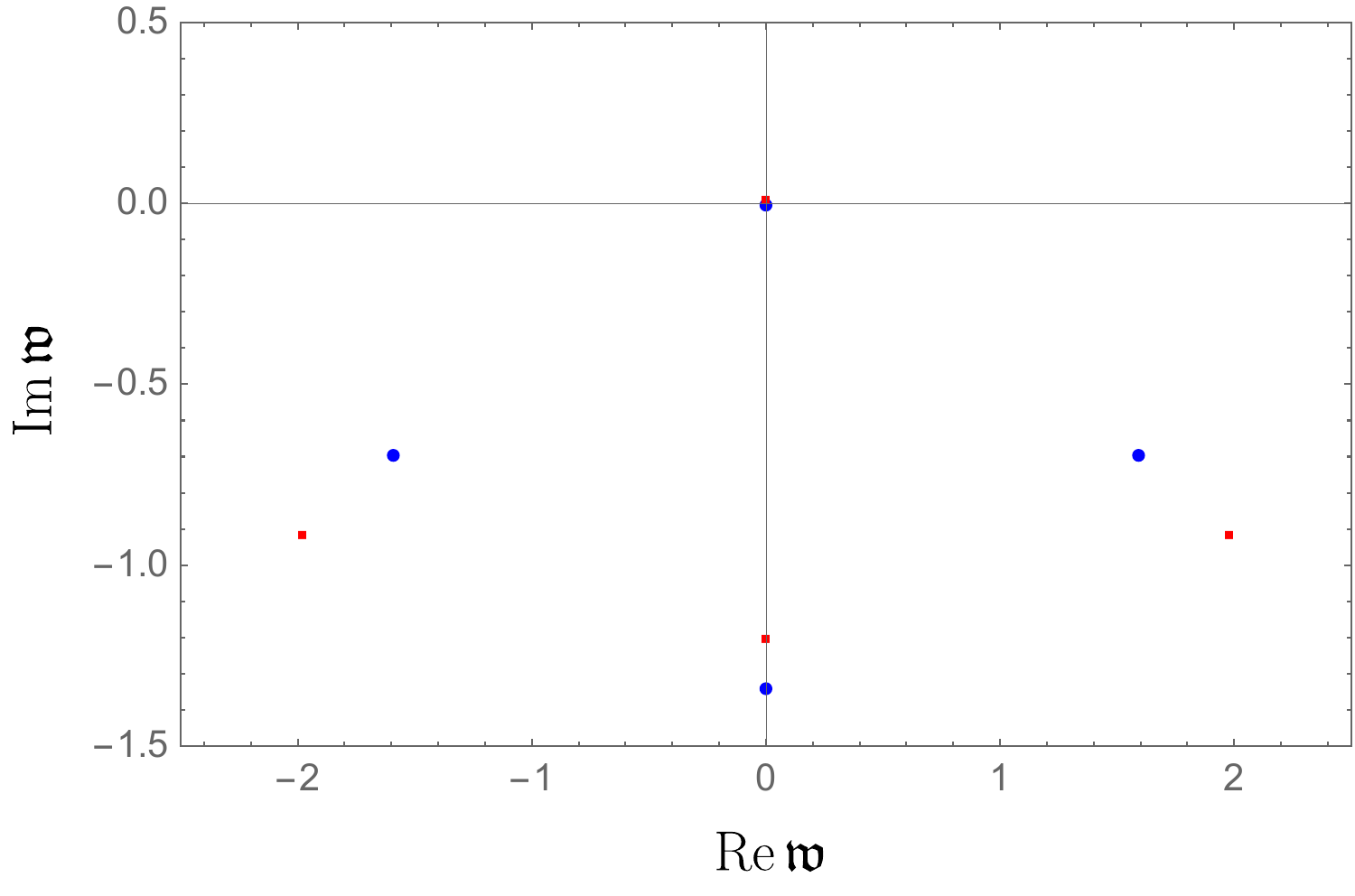}
%{figs/diffusion-mode-pkk.pdf}
\hspace{0.05\textwidth}
\includegraphics[width=0.46\textwidth]{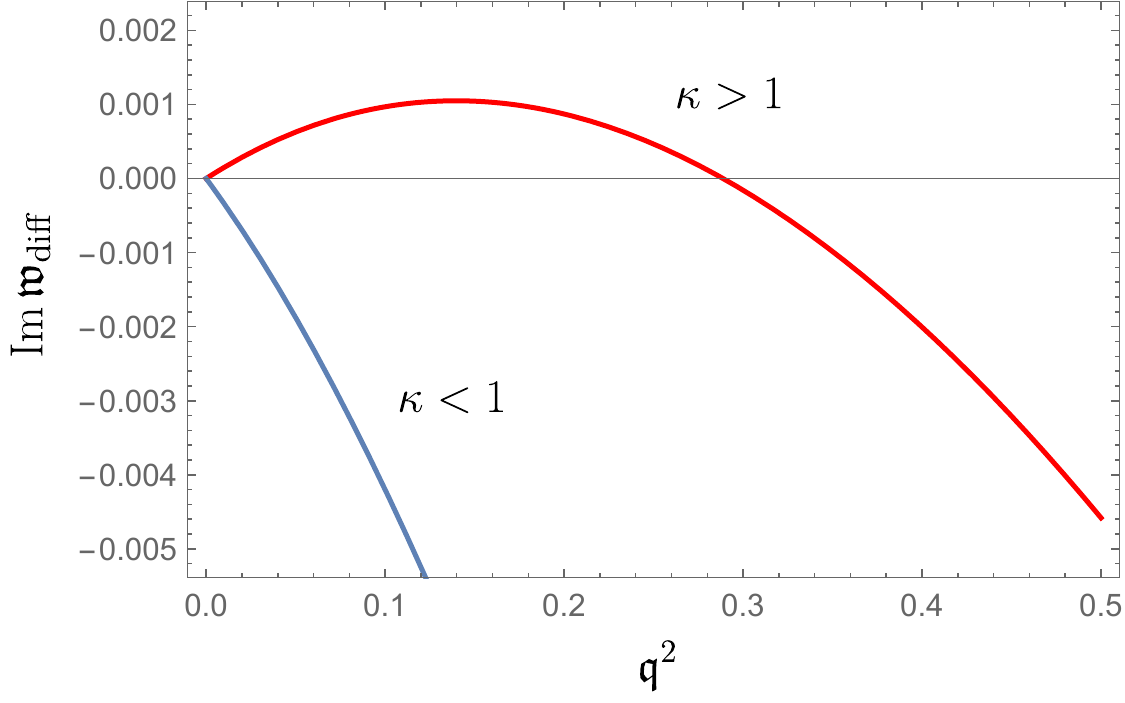}
\caption{
\label{fig:qnm-instability}
{\small   The case of $(\kappa_1,\kappa_2,\kappa_3)=(\kappa,\kappa,\kappa)$.  The closest to the origin quasinormal modes of the system \eqref{fluct-01}---\eqref{fluct-02} in the complex frequency plane for $\kappa = 0.9<\kappa_c=1$ (blue dots) and $\kappa =1.1 > \kappa_c = 1$ (red squares) at $\qfr^2 = 0.15$ --- left panel; the gapless diffusion mode on the imaginary axis crosses into the upper half-plane at $\kappa=\kappa_c=1$.  The unstable diffusion mode $\mbox{Im}\, \wfr_{\mbox{\scriptsize diff}}$ as a function of $\qfr^2$ for $\kappa=0.9$ (blue curve) and $\kappa =1.1$ (red curve) --- right panel.}
}
\end{figure}

The fluctuations $\mathfrak{E}^a_z$ and $\mathfrak{s}_a$ decouple from the general system of fluctuations: the metric fluctuations couple to $(1,1,1){\cdot}{\bf E}_z$ only,  whereas the fluctuations $\mathfrak{E}^a_z$ and $\mathfrak{s}_a$ obey the closed system of equations (see ref.~\cite{Gladden:2024ssb}):
\begin{eqnarray}
  &\,& {\mathfrak{E}^{a}_z}'' + \mathfrak{D}^{-1}\Biggl[\mathfrak{w}_0^2(1+\kappa u)^3\left( \frac{f'}{f} 
- \frac{\kappa}{1+\kappa u}\right) - \frac{2\kappa\mathfrak{q}_0^2f}{1+\kappa u}\Biggr] {\mathfrak{E}^{a}_z}'   
+ \frac{\mathfrak{D}}{uf^2} \mathfrak{E}^a_z \nonumber \\
 &+& \frac{2\sqrt{\kappa}\, \mathfrak{q}_0}{(1+\kappa u)^3}\mathfrak{s}_a'  
 + \frac{2\kappa^{1/2}\mathfrak{q}_0}{\mathfrak{D}}\cdot \Biggl[\mathfrak{w}_0^2\left(\frac{f'}{f} 
 -\frac{4\kappa}{1+\kappa u}\right) + \frac{\kappa\mathfrak{q}_0^2 f}{(1+\kappa u)^4}\Biggr]\mathfrak{s}_a  = 0\,,
 \label{fluct-01}
\end{eqnarray}
\begin{eqnarray}
&\,& \mathfrak{s}_a'' + \left(\frac{f'}{f} - \frac{1+3\kappa u}{u(1+\kappa u)}\right) \mathfrak{s}_a' 
+\Biggl[  \frac{\mathfrak{D}}{uf^2} + \frac{1+\kappa u}{u^2f} + \frac{2\kappa (1+\kappa)^3u}{(1+\kappa u)^2f} \nonumber \\
&-& \frac{\kappa}{1+\kappa u}\left(\frac{f'}{f} 
- \frac{1+3\kappa u}{u(1+\kappa u)}\right) -\frac{4\kappa (1+\kappa )^3(1+\kappa u)u \mathfrak{w}_0^2}{\mathfrak{D} f}\Biggr] \mathfrak{s}_a
 \nonumber \\ & - & \frac{2\kappa^{1/2}(1+\kappa)^3(1+\kappa u)u \mathfrak{q}_0}{\mathfrak{D}}  {\mathfrak{E}^{a}_z}'    = 0.
\label{fluct-02}
\end{eqnarray}
Here,  $f(u)= (1+\kappa u )^3 - (1+\kappa)^3 u^2$ and $\mathfrak{D} (u) = (1+\kappa u)^3\mathfrak{w}_0^2 - \mathfrak{q}_0^2f$.  
In the following, it will be convenient to normalise the frequency and momentum as
\begin{equation}
 \wfr = \frac{\omega}{2\pi T}\,,   \qquad \qquad  \qfr = \frac{q}{2\pi T}\,,
\end{equation}
where $T$ is the Hawking temperature of the background \eqref{T-H-diag_x}.

At the horizon ($u=1$), the exponents of the fluctuations $\mathfrak{E}^a_z$ and $\mathfrak{s}_a$  are $\nu_\pm =\pm i \wfr/2$, and we choose $\nu=\nu_-$ following the recipe of  ref.~\cite{Son:2002sd}.  At the boundary ($u=0$), the exponents 
of the scalar $\mathfrak{s}_i$ are $(1,1)$ and those of the $U(1)$ field  $\mathfrak{E}^a_z$  are $(0,1)$.  For the scalar, this corresponds to the conformal dimension   $\Delta_+=\Delta_-=2$ of a dual operator.\footnote{In ${\cal N}=4$ SYM, such (gauge-invariant) operators are the bilinears in scalar fields $\Phi^i$,  e.g. the operator
${\cal O}^{ij} = \mbox{Tr} \Phi^i \Phi^j - \frac{1}{6} \delta^{ij} \mbox{Tr} \, \Phi^k\Phi^k$,
where  $i=1,...6$ are the indices of  the ${\bf 6}$ representation of $SU(4)_R$.}
%\begin{figure}
%\includegraphics[width=0.45\textwidth]{figs/stu-sound-diffusion-mode-kappa=0-5.pdf}
%\hspace{0.01\textwidth}
%\includegraphics[width=0.45\textwidth]{figs/stu-sound-diffusion-mode-kappa=1-1.pdf}
%\caption{
%\label{fig:qnm-instability} The unstable diffusion mode 
%$\mbox{Im}\, \wfr$ as a function of $\qfr^2$ for $\kappa = 0.5<\kappa_c=1$ (left panel) and $\kappa =1.1 >\kappa_c = 1$ (right panel).
%}
%\end{figure}
%
At the boundary,  the scalar field fluctuations behave as $ \mathfrak{s}_a = {\cal A}_a u \log{u}+\cdots +{\cal B}_a u+\cdots$,  and the standard recipe to obtain the poles of the dual retarded correlators implies setting ${\cal A}_a=0$. Technically,  this may be inconvenient when solving the system  \eqref{fluct-01}---\eqref{fluct-02} numerically.  Instead, one can write an equivalent system for the coupled 
variables $\mathfrak{E}^a_z$ and $\varphi_a \equiv (1-u)^2\, u  \, \mathfrak{s}_a '' $ as
\begin{eqnarray}
  {\mathfrak{E}^{a}_z}''  &+&  A_E {\mathfrak{E}^{a}_z}' + B_E\, \mathfrak{E}^a_z + C_E \varphi_a ' + D_E \varphi_a =0\,,  \\
\varphi_a '' &+& A_h \varphi_a ' + B_h \,  \varphi_a + C_h  {\mathfrak{E}^{a}_z}'  + D_h \mathfrak{E}^a_z=0\,,  
\end{eqnarray}
where the coefficients are derived from the system   \eqref{fluct-01}---\eqref{fluct-02}. The field $\varphi_a$ has indices $(0,1)$ at $u=0$, and the standard Dirichlet condition $\varphi_a(0)=0$ gives ${\cal A}_a=0$. Alternatively, one can integrate the original system \eqref{fluct-01}---\eqref{fluct-02} from $u=0$ and from $u=1$ and match solutions at an intermediate point.

The behaviour of the quasinormal spectrum  has been studied numerically in ref. \cite{Gladden:2024ssb}.  It is qualitatively similar to the one discussed in section \ref{section-two-kappas} for the case of   $(\kappa_1,\kappa_2,\kappa_3)=(\kappa,\kappa,0)$ (see Fig.~\ref{fig:qnm-instability} and compare with Fig.~\ref{fig:2chg-qnm-instability}).  The hydrodynamic $R$-charge diffusion mode is unstable: it crosses into the upper half-plane for $\kappa>1$. The dispersion relations $\mbox{Im}\, \wfr_{\mbox{\scriptsize diff}} (\qfr^2)$  for $\kappa=0.9$ (blue curve) and $\kappa =1.1$ are shown in the right panel of Fig.~\ref{fig:qnm-instability}. For $|\qfr^2|\ll 1$, both curves are clearly linear in $\qfr^2$. The unstable mode initially moves up from the origin along the imaginary axis with $|\qfr^2|$ increasing,  but then moves down into the lower half-plane.  In this sense,  the instability is an infrared effect, affecting the long-distance scales. 

\begin{figure}
\centering
\includegraphics[width=0.46\textwidth]{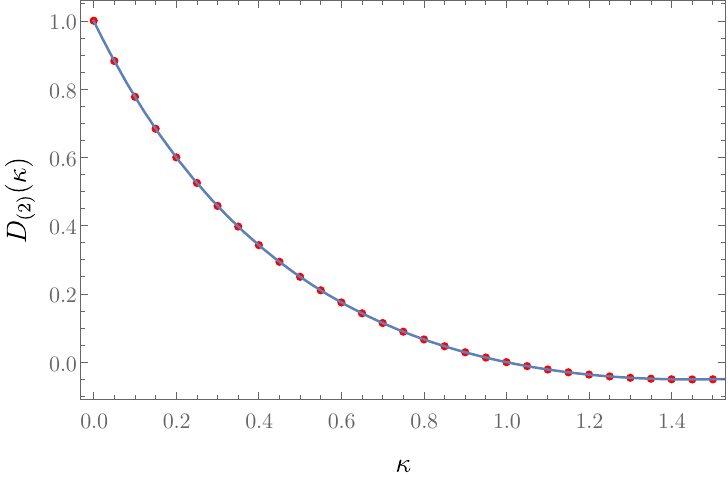}
\hspace{0.05\textwidth}
\includegraphics[width=0.46\textwidth]{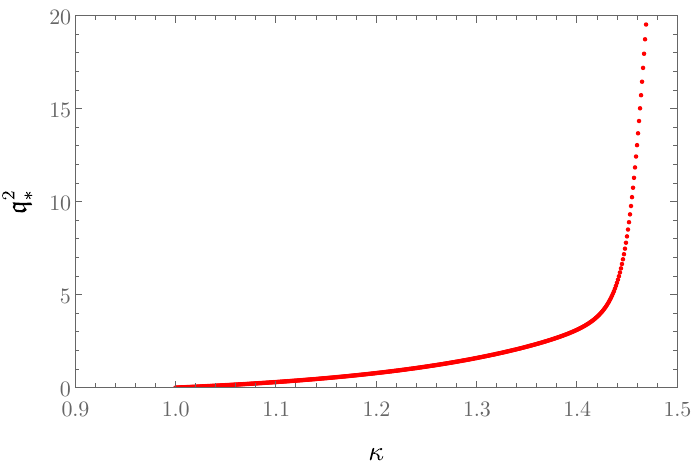}
\caption{ 
\label{fig:2chg-qnm-instability-2}
 The case of $(\kappa_1,\kappa_2,\kappa_3)=(\kappa,\kappa,\kappa)$.  Left panel: The analytic prediction for the diffusion coefficient $D_{(2)}=D_{(3)}$  (normalised by $1/2\pi T$) in Eq.~\eqref{diffusion-3-kappa} as a function of $\kappa$ is shown in blue. Numerical data are plotted in red, showing agreement.
Right panel: The endpoint $\qfr_*^2$ of the  instability domain as a function of $\kappa$.
}
\end{figure}

The dispersion relation of the unstable mode is shown in the right panel of Fig.~\ref{fig:qnm-instability}.  For sufficiently small $\mathfrak{q}^2$, the dispersion relation is linear in $\qfr^2$, and agrees with our analytic result from eq.~\eqref{eq:Da-kappa3},
\begin{equation}
  \wfr^{\rm diffusion}_{(2)}(\qfr) = -i \, \frac{(2-\kappa)(1-\kappa)}{2(1+\kappa)} \, \qfr^2 +\cdots\,,
\label{diffusion-3-kappa}
\end{equation}
see Fig.~\ref{fig:2chg-qnm-instability-2}, left panel.  The instability region $\qfr^2 \in [0,\qfr^2_*]$ in Fig.~\ref{fig:qnm-instability} corresponds to the red curve lying above the horizontal axis.  The value of $\qfr^2_*$ depends on $\kappa$: it increases with $\kappa$ increasing (see Fig.~\ref{fig:2chg-qnm-instability-2}, right panel).  
In addition to the hydrodynamic mode, other modes are also present on the imaginary frequency axis (see Fig.~\ref{fig:qnm-instability}, left panel). These modes move up the axis as $\kappa$ increases at fixed $\qfr$ and eventually cross into the upper half-plane for sufficiently large $\kappa$, within the interval $1 < \kappa < 2$, i.e.  in the thermodynamically unstable region.  Mode collisions occur in this domain, involving modes that move off the imaginary axis.

 \section{Discussion}
 \label{sec:discussion}
We now summarize our results. We have identified hydrodynamic modes in relativistic theories with $\Nf$ flavours of global $U(1)$ charges, in thermal equilibrium states with $\Nf$ chemical potentials $\mu_a$ for the corresponding charges. The first order hydrodynamic  transport coefficients include shear viscosity $\eta$, bulk viscosity $\zeta$, and a symmetric $\Nf \times \Nf$ conductivity matrix $\sigma_{ab}$. We have given explicit expressions for the speed of sound, eq.~\eqref{eq:cs2-1}, and for the sound damping coefficient, eq.~\eqref{eq:Gamma-2}. 

The diffusive modes at non-zero $\mu_a$ mix the fluctuations of temperature with the fluctuations of the charge densities. The diffusion coefficients $D_{(a)}$ are given by the eigenvalues of the $\Nf\times\Nf$ diffusion matrix~\eqref{eq:D-matr}. The expressions for $D_{(a)}$ as eigenvalues of \eqref{eq:D-matr} generalise the well-known expression $D = \sigma/\chi$ for $\Nf=1$, $\mu=0$, as well as the more complicated expression for $D$ for $\Nf=1$, $\mu\neq0$~\cite{Kovtun:2012rj}.

In hydrodynamics, diffusive fluctuations at $\mu_a \neq 0$ involve the fluctuations of temperature, and, as such, they belong to the ``sound channel'' of near-equilibrium perturbations. However, flavour symmetry may give rise to separate ``diffusion channels'' which decouple from the sound channel to all orders in the derivative expansion. For example, for $\Nf=2$, an extended flavour symmetry may give rise to such a decoupled diffusion channel in a state with $\mu_1 = \mu_2$, describing diffusion of charge density difference $\delta n_1 - \delta n_2$. 

As expected, the shear damping coefficient $\eta/(\epsilon{+}p)$ in \eqref{eq:w-shear}, the sound damping coefficient $\Gamma$ in \eqref{eq:wS}, and the diffusion coefficients $D_{(a)}$ in \eqref{eq:wD} all have the structure ``transport in the numerator, thermodynamics in the denominator'', analogous to the well-known $\Nf=1$, $\mu=0$ expression $D= \sigma/\chi$. Thus for positive non-equilibrium hydrodynamic entropy production, hydrodynamic stability of the equilibrium state is determined by thermodynamic susceptibilities. See~\cite{Gouteraux:2024adm} for other examples of thermodynamic vs hydrodynamic instabilities. 

We have applied the above hydrodynamic results to strongly coupled ${\cal N}=4$ supersymmetric $SU(N_c)$ Yang-Mills (SYM) theory in 3+1 dimensions. This theory has $\Nf=3$, for $U(1)^3 \subset SU(4)$ R-symmetry. In the limit $N_c\to\infty$, we used the dual classical gravitational description in terms of the STU supergravity truncation. 

The sound dispersion relation in SYM theory is fixed by conformal symmetry to be \eqref{eq:sound-cft}, hence the sound damping coefficient $\Gamma$ in \eqref{eq:wS} is determined by thermodynamics, by virtue of the general holographic result $\eta = s/4\pi$~\cite{Kovtun:2003wp,Buchel:2003tz,Kovtun:2004de}, which also holds at non-zero chemical potential~\cite{Son:2006em, Mas:2006dy}. It is straightforward to obtain $\Gamma$ as a function of $T$ and $\mu_a$, using eqs.~\eqref{eq:TH-kappa}, \eqref{eq:s-kappa}, \eqref{eq:e-kappa}, and  \eqref{eq:mu-kappa}.

On the other hand, obtaining the three diffusion coefficients $D_{(a)}$ requires doing new calculations, which we have performed in this paper. We have calculated the charge conductivity matrix $\sigma_{ab}$ in three illustrative cases: $(\mu_1, \mu_2, \mu_3) = (\mu, 0, 0)$,  $(\mu_1, \mu_2, \mu_3) = (\mu, \mu, 0)$, and $(\mu_1, \mu_2, \mu_3) = (\mu, \mu, \mu)$. Given the conductivity matrices, the three diffusion coefficients can be obtained using the formulas in sec.~\ref{hydro}, and expressed as functions of $T$ and $\mu_a$. The analytic expressions for the diffusion coefficients are given in eqs.~\eqref{eq:D-11}, \eqref{eq:D-21} (single non-vanishing chemical potential), \eqref{eq:Da-kappa2} (two equal non-vanishing chemical potentials), and \eqref{eq:Da-kappa3} (three equal non-vanishing chemical potentials).

The SYM theory equation of state which is obtained from the STU background has thermodynamic instabilities when $\mu_a > (\mu_{a})_{\rm crit.}$, where the critical value is equal to an $O(1)$ number times $2\pi T$. Thus the extremal $T\to0$ limit of the STU background is always thermodynamically unstable. The thermodynamic instability manifests through negative eigenvalues of the susceptibility matrix, which must be positive-definite in thermodynamically stable states. As expected from hydrodynamics, thermodynamic instabilities give rise to long-wavelength dynamic instabilities of the equilibrium state. These hydrodynamic instabilities arise through negative diffusion coefficients for the corresponding modes. For example, in a state with $(\mu_1, \mu_2, \mu_3) = (\mu, \mu, \mu)$, two eigenvalues of the thermodynamic Hessian matrix change sign at $\mu = \sqrt{2}\, 2\pi T$, leading to two negative diffusion coefficients for R-charge differences at low temperature~\cite{Gladden:2024ssb}. 

In the dual gravitational description, dynamical instabilities appear as unstable quasinormal modes of the charged STU black brane, located in the upper half of the complex frequency plane at small real spatial momentum~$q$. Consequently, charged five-dimensional STU black branes are always dynamically unstable as $T \to 0$. This includes the special case with three equal chemical potentials, whose dual background is the $AdS_5$–Reissner–Nordström black brane \cite{Gladden:2024ssb}.    While this brane is stable within Einstein–Maxwell theory, it becomes unstable at low temperature when viewed as a special case of the more general STU solution, in this case,  due to fluctuations of gauge fields and neutral scalar fields.  Interestingly, the quasinormal mode that is unstable at small $q$ becomes stable again at large $q$; see Figs.~\ref{fig:2chg-qnm-instability} and \ref{fig:qnm-instability}. 
%We plan to return to a more detailed investigation of this phenomenon in future work.

What do the above instabilities mean physically? On the field theory side, the low-temperature instability is the clumping of R-charges. The options seem to be: another (perhaps, inhomogeneous) phase of the SYM theory at low temperature, or the end of the phase diagram, signifying the breakdown of the grand canonical ensemble at $\mu_a > (\mu_{a})_{\rm crit.}$. The latter option is realized in both free, and in weakly coupled SYM theory~\cite{Yamada:2006rx}. The reason is that the scalar fields of the SYM theory are charged under R-symmetry, and the corresponding scalar potential has flat directions. A chemical potential for charged scalars amounts to a negative mass squared, and as a result the grand canonical partition function fails to converge. There's no {\em a priori} reason why an analogous breakdown of the grand canonical ensemble should not happen in strongly coupled SYM theory. 
On the dual holographic side, the fate of the instability may be determined by numerically following the growth of small perturbations of the STU background. The non-linear dynamical gravitational evolution may end up in either a new phase, or in a naked singularity. The latter would be interpreted as a breakdown of the grand canonical ensemble.  If the grand canonical description of the strongly coupled SYM theory indeed breaks down at $\mu_a > (\mu_{a})_{\rm crit.}$, the breakdown should be visible in the microcanonical ensemble, whose gravitational description is given by fixed-charge, fixed mass large AdS black holes (rather than branes) \cite{Chamblin:1999tk},\cite{Yaffe:2017axl}.   Another perspective on the instability was recently discussed in refs.~\cite{Buchel:2025cve,Buchel:2025tjq,Buchel:2025ves}. 
%\hl{[[say something about the black hole literature]]}. 
%
%\hl{[[We need to comment on Buchels' papers} 
%\cite{Buchel:2025cve}, \cite{Buchel:2025tjq}, 
%but this is perhaps better done below 
%eq.~\eqref{Gibbs-three-kappas} than in the Discussion.]]
%

Finally, it is important to note the five-dimensional truncation used to study the phase structure of SYM theory, and the associated instabilities. In this paper, we have used the STU consistent truncation whose field content includes the metric, three $U(1)$ gauge fields, and two real scalar fields. In the STU truncation, the stable part of the phase diagram is shown Fig.~\ref{fig-kappa-space}. One can ask what would happen in larger consistent truncations which include more fields, or in ten dimensions\footnote{Hydrodynamics of the spinning $Dp$ branes   was recently discussed in ref.~\cite{Armas:2025fvo}.} \cite{Choi:2024xnv}.  In particular, larger truncations which contain charged complex scalar fields may develop instabilities of the ``holographic superconductor'' type~\cite{Gubser:2008px}. This question was addressed in~\cite{Gentle:2012rg}, and more recently in~\cite{Buchel:2025ves}. 
The results of~\cite{Gentle:2012rg} indicate superconducting phases at somewhat higher temperatures than the STU  instability we discuss here. However, thermodynamics of these phases is not known, and it is not clear whether they have a lower or a higher free energy compared to the STU phase. Ref.~\cite{Buchel:2025ves} used a different consistent truncation, and found a superconducting instability below the STU critical temperature discussed here. The dynamic stability analysis of such superconducting phases is yet to be performed. Clearly, more work is needed in order to elucidate the low-temperature behaviour of ${\cal N}=4$ SYM theory at finite density.

\begin{acknowledgments}
We would like to thank Alex Buchel,  Blaise Goutéraux, Ben Withers and Laurence~G.~Yaffe for helpful discussions, and Andres Anabalon, Oscar Dias, Oscar Henriksson, Carlos Hoyos, 
Niko Jokela and Makoto Natsuume for useful correspondence. 
%P.K.\ and A.O.S.\ thank the participants of the Galileo Galilei Institute for Theoretical Physics program ``Foundations and Applications of Relativistic Hydrodynamics'' for discussions.  
P.K.~and A.O.S.\ thank the Galileo Galilei Institute for Theoretical Physics for the hospitality, and the INFN for partial support during the completion of this work. The work of P.K. was supported in part by the NSERC of Canada. P.K.\ would like to thank the Simons Center for Geometry and Physics for supporting the program ``Black hole physics from strongly coupled thermal dynamics'' where part of this work was completed. P.K.\ would like to thank the Kavli Institute for Theoretical Physics for supporting the program ``Relativistic Plasma Physics: From the Lab to the Cosmos'' during which part of this work was completed. P.K.'s research was supported in part by grant NSF PHY-2309135 to the Kavli Institute for Theoretical Physics (KITP). A.O.S.~is partially supported by the UK STFC grant ST/X000761/1. 
\end{acknowledgments}

 \appendix
 \section{The probability of fluctuations in variables $\varepsilon$ and $n_a$ }
 \label{einstein-formula-appendix}
The Einstein's formula \eqref{einstein-1} can be re-written for the internal energy density considered 
as a function of the entropy density and densities of charges \cite{zubarev-book,kvasnikov-book-3}.   
Consider a fixed volume system   embedded in a large thermostat.  The system is allowed to 
exchange the energy ${\cal E}$ and the charges $Q_a$  (whose volume densities we will denote by $\varepsilon$ 
and $n_a$) with the thermostat,  the exchange being regulated by the temperature $T$ and the chemical 
potentials $\mu_a$.  Since the global charges in the full system are conserved, for fluctuations $\Delta \varepsilon$ 
and $\Delta n_a$ in the small system we have 
\begin{equation}
\Delta \varepsilon = - \Delta \varepsilon^T\,, \qquad    \Delta n_a = - \Delta n_{a}^T\,,
\label{thermostat}
\end{equation}
where the index $T$  denotes the quantities associated with the thermostat.  The probability of a fluctuation in the total system (the small system plus the thermostat) is given by 
\begin{equation}
w_\Delta = e^{\Delta s_{ \text{\tiny TOT} }}\,,  
\label{einstein-1-tot}
\end{equation}
where $\Delta s_{ \text{\tiny TOT}} = \Delta s_T + \Delta s$ is the total entropy density.  In the thermodynamic limit, the intensive parameters of the thermostat such as temperature, pressure, chemical potentials retain their equilibrium values in the presence of fluctuations.  Indeed,  for example, for pressure we have
\begin{equation}
p_T' =p_T \left( T, n_a^T +\Delta n_a^T\right) = p (T,n_a) +\left( \frac{\partial p}{\partial n_a} \right)_T \Delta n_a^T +\cdots\,,
\label{einstein-1-tot-p}
\end{equation}
where, in view of \eqref{thermostat},
\begin{equation}
\bigg|  \Delta n_a^T \bigg|  = \bigg| n_a^T \frac{\Delta Q_a^T}{Q_a^T} \bigg|  =
 \bigg| n_a^T \frac{\Delta Q_a}{Q_a^T} \bigg|  = \bigg| n_a^T \frac{\Delta Q_a}{Q_a} \bigg| \bigg| \frac{Q_a}{Q_a^T}\bigg| \to 0
\end{equation}
in the  thermodynamic limit $|Q_a^T/Q_a|\to \infty$.   This allows us to write
\begin{equation}
\Delta s_{ \text{\tiny TOT}} = \frac{1}{T_T} \left( \Delta \varepsilon^T - \mu_a^T \Delta n_a^T\right) + \Delta s= 
 \frac{1}{T} \left( - \Delta \varepsilon + \mu_a \Delta n_a\right) + \Delta s
\end{equation}
and thus 
\begin{equation}
w_\Delta = e^{ - \frac{1}{T} \left( \Delta \varepsilon -\mu_a \Delta n_a - T \Delta s\right) }\,.
\label{einstein-1-tot-z}
\end{equation}
Choosing $y_i = (s, n_a)$ as independent thermodynamic parameters, we have
\begin{equation}
\Delta \varepsilon = \sum\limits_{i} \frac{\partial \varepsilon}{\partial y_i} \Delta y_i + \frac 12 \sum\limits_{ij} \frac{\partial^2 \varepsilon}{\partial y_i \partial y_j} \Delta y_i \Delta y_j + \cdots\,,
\end{equation}
where $(\partial \varepsilon/\partial s)_{n_a} = T$ and  $(\partial \varepsilon/\partial n_a)_{s} = \mu_a$.  Substituting this into Eq.~\eqref{einstein-1-tot-z}, we find the probability of fluctuations 
\begin{equation}
w_\Delta \sim \exp{\left\{ - \frac{1}{2 T} \sum\limits_{ij} \frac{\partial^2 \varepsilon}{\partial y_i \partial y_j} \Delta y_i \Delta y_j \right\}}\,.
\label{fluctuation-02x}
\end{equation}
The condition of stable equilibrium requires the quadratic form in the exponent to be positive definite.

\section{Cayley-Hamilton theorem}
\label{cayley-hamilton}
Thermodynamic quantities, such as free energy and its derivatives, will generically depend on the three chemical potentials in the STU model. The charge operators, $Q^a$, are elements of the Cartan subalgebra of the R-symmetry group, and so chemical potentials are elements of the dual space. Since free energy must be a singlet, its dependence on chemical potentials must come only through symmetric group invariants of $SU(4)$. The purpose of this Appendix is to show that it can be a function only of three invariants $m_2$, $m_3$, and $m_4$ defined by Eqs.~\eqref{eq:invariants} in the case of $SU(4)$, or, more generally,  a function of $N_f$ invariants in the case of $SU(N_f+1)$ global symmetry group.

Let $G=SU(N)$, and let $\{T^a\}_{a=1}^r$, where $r=\text{rank}(G)$, be a basis for the corresponding Lie algebra, $\mathfrak{g}=\mathfrak{su}(N)$. A symmetric invariant tensor, $\kappa^{i_1\hdots i_m}_{(m)}$, of rank $m$ may be constructed as the symmetric trace of products of generators.
\begin{equation}
\kappa^{i_1 \hdots i_m}_{(m)} :=\text{Tr}\left(T^{(i_1}\hdots T^{i_m)}\right)\label{eq:InvTensorM}\,.
\end{equation}
Such invariant tensors are in one-to-one correspondence with Casimir operators, defined by
\begin{equation}
C_{(m)} = \kappa_{i_1\hdots i_m}^{(m)} T^{i_1}\hdots T^{i_m},
\end{equation}
which obey
\begin{equation}
\left[ C_{(m)}, T^a \right] = 0,    \qquad \forall a\,.
\end{equation}
That is, they are invariant under the action of all generators.

We shall now prove that there are only $N-1$ independent symmetric invariants. Consider a generic element of $\mathfrak{g}$, $X=X_aT^a$. Here, the $T^a$ are thought of as being $N\times N$ matrices in the defining representation of $\mathfrak{su}(N)$, i.e. they are anti-Hermitian and traceless, and $X_a\in\mathbb{C}$. The Cayley-Hamilton theorem \cite{Frobenius1877} states that there exist $\{c_i\}_{i=0}^{N-1}\in\mathbb{C}$ such that the characteristic matrix polynomial obeys
\begin{equation}
p_X := X^N + c_{N-1}X^{N-1} \hdots + c_1 X + c_0 \mathbf{1}_N = 0,
\end{equation}
where, for concreteness, $X^n = X_{a_1}\hdots X_{a_n}\cdot T^{a_1}\hdots T^{a_n}$. Multiplying this by $X$, and noticing that $\text{Tr}(X)=0$, it is clear that there exists $f$ such that 
\begin{equation}
\text{Tr}\left(X^{N+1}\right) = f\biggl(\text{Tr}\left(X^N\right), \hdots, \text{Tr}\left(X^2\right)\biggr).\label{eq:TrXN+1}
\end{equation}
Since multiplication over $\mathbb{C}$ is commutative, 
\begin{equation}
X_{a_1}\dots X_{a_n} \cdot T^{a_1}\hdots T^{a_n} = n!\cdot X_{a_1}\hdots X_{a_n} \cdot T^{(a_1}\hdots T^{a_n)}.
\end{equation} 
In other words, the symmetric prefactor projects out the antisymmetric part. Taking the trace of this equation,  we find
\begin{equation}
\text{Tr}\left(X^n\right) = n!\cdot \kappa_{(n)}^{a_1\hdots a_n}\cdot X_{a_1}\hdots X_{a_n}.
\end{equation}
Eq.~(\ref{eq:TrXN+1}) may therefore be understood as stating that an arbitrary linear combination of symmetric invariant tensors of order $n>N$ may always be expressed as a function of the `primitive' symmetric invariants of lower order, of which there are $N-1=r$. Since the symmetric invariants are in one-to-one correspondence with Casimirs, the same applies to these. This is shown explicitly in refs.~\cite{BiedenharnL.C.1963OtRo}, \cite{KleinAbraham1963IOot}. It is a general result that for semi-simple Lie algebras, the number of independent Casimirs is equal to the rank of the group. The invariants of $m_2$, $m_3$, and $m_4$ of eq.(\ref{eq:invariants}) are of precisely the form above, namely the Sudbery basis form for symmetric invariant tensors \cite{deAzcarraga:1997ya}. For the case of $SU(4)$, we have shown that there exist only these three primitive invariants, and higher-order symmetric tensors will be built out of them.

\section{The scalar and redundancy}
\label{scalar-and-redundancy}
%
%The scalars in the STU model are denoted $X_i$, $i\in\{1,2,3\}$. It is convenient to introduce 
%new variables, $H_i$, such that 
%\begin{equation}
%X_i = \frac{\mathcal{H}^{1/3}}{H_i}, \qquad \text{where }\mathcal{H}=\prod_{i=1}^3 H_i.
%\end{equation}
In the STU model, the scalars  $X_a$ satisfy the constraint
\begin{equation}
\prod_{a=1}^3 X_a= 1.\label{eq:const}
\end{equation}
We are interested in computing the equations of motion of perturbations about the STU background, and it will be convenient to express these in terms of the variations $\delta H_a$ of the fields $H_a$  related to $X_a$ via Eq.~\eqref{fields-H}. We seek to understand the implications of the constraint above on this variation.

To linear order in perturbations, the constraint becomes
\begin{equation}
\delta\left(\prod_{a=1}^n X_a\right) = \sum_{i=1}^3 \left( \delta X_a   \prod_{b\neq a} X_b\right) = 0.\label{eq:PertConst}
\end{equation}
In terms of the new variables, 
\begin{align}
\delta X_a = \delta\left( \frac{H^{1/3}}{H_a}\right) &= \frac{H^{1/3}}{H_a}\left(\frac{1}{3}\frac{\delta H}{H} - \frac{\delta H_a}{H_a}\right)\\
&= X_a\left(\frac{1}{3}\frac{\delta H}{H} -  \frac{\delta H_a}{H_a}\right)\,.
\end{align}
Thus, 
\begin{equation}
\delta X_a \cdot  \prod_{b\neq a} X_b = \left(\frac{1}{3}\frac{\delta H}{H} - \frac{\delta H_a}{H_a}\right),
\end{equation}
where we used Eq.(\ref{eq:const}). From Eq.(\ref{eq:PertConst}), therefore, one finds
\begin{equation}
\delta H= H \sum_{a=1}^3 \frac{\delta H_a}{H_a}.\label{eq:varCalH}
\end{equation}
%The constraint implies that, though there are three variables $X_a$, they are not independent. The same
% is true for the $H_a$, but naive attempts to use eq.(\ref{eq:const}) to express some $H_i$ in terms of the other 
%two will merely yield a tautology. Instead, the redundancy manifests through the fact that 
Note that one may perform a rescaling $H_a \to \lambda H_a$, for $\lambda\in \mathbb{C}$, and the physical variable $X_a$ remains unchanged.  This can be leveraged to simplify the variational problem. Let $H_a\to \lambda H_a$ for all $a$, whereby $H\to \lambda^3 H$. The expression above becomes
\begin{equation}
\delta H = \lambda^2 H\sum_{a=1}^3 \frac{\delta H_a}{H_a}.
\end{equation}
Since $\lambda$ is arbitrary, one can select $\lambda = 0$, thereby fixing the redundancy with the constraint $\delta H=0$.  By the same token, this can be used to express the variation of one scalar in terms of the other two. From eq.(\ref{eq:varCalH}), it is clear that 
\begin{equation}
\delta H_3 = -H_3\left(\frac{\delta H_1}{H_1}+\frac{\delta H_2}{H_2}\right),
\end{equation}
since there are really only two scalars in the theory.

\section{Thermodynamic parameters vs black brane parameters}
\label{app:tbb}

\noindent
In this Appendix we consider thermodynamics of STU black brane solutions, as described in sec.~\ref{sec:STU-background}. The black brane solutions are characterized by four integration constants $(T_0, \kappa_1, \kappa_2, \kappa_3)$. The constant $T_0$ has units of temperature, while $\kappa_a$ are dimensionless. The temperature and the chemical potentials of the dual field theory are expressed in terms of the four integration constants as
\begin{align}
  & T = \frac{2 + \kappa_1 + \kappa_2 + \kappa_3 - \kappa_1 \kappa_2 \kappa_3 }{2 \sqrt{(1+\kappa_1) (1+\kappa_2) (1+\kappa_3)}} T_0 \,,\\[5pt]
  & \mu_1 = \sqrt{2} \pi T_0  \frac{ \sqrt{ \kappa_1 (1+\kappa_2) (1+\kappa_3) } }{ \sqrt{ 1+ \kappa_1} } \,,\\[5pt]
  & \mu_2 = \sqrt{2} \pi T_0  \frac{ \sqrt{ \kappa_2 (1+\kappa_1) (1+\kappa_3) } }{ \sqrt{ 1+ \kappa_2} } \,,\\[5pt]
\label{eq:mu3}
  & \mu_3 = \sqrt{2} \pi T_0  \frac{ \sqrt{ \kappa_3 (1+\kappa_1) (1+\kappa_2) } }{ \sqrt{ 1+ \kappa_3} } \,.
\end{align}
The integration constants are sign-definite: $T_0 > 0$, $\kappa_a \geq 0$. They must satisfy
\begin{align}
  2 + \kappa_1 + \kappa_2 + \kappa_3 - \kappa_1 \kappa_2 \kappa_3  > 0 \,,
\end{align}
by positivity of temperature, and 
\begin{align}
  & 2 - \kappa_1 - \kappa_2 - \kappa_3 + \kappa_1 \kappa_2 \kappa_3 > 0 \,,\\
  & \kappa_1 + \kappa_2 + \kappa_3 < 3\,,
\end{align}
by thermodynamic stability~\cite{Gladden:2024ssb}. The pressure is 
\begin{align}
  p = \frac{\pi^2 N_c^2}{8}  T_0^4 (1+\kappa_1) (1+\kappa_2) (1+\kappa_3) \,.
\end{align}
The grand canonical free energy $\Omega = -pV$, where $V$ is the field theory volume, $V\to\infty$. Let us also introduce the dimensionless free energy density
\begin{align}
\label{eq:Omega-bar}
  \bar\Omega(\mu_a/T) \equiv \frac{8}{\pi^2 N_c^2} \frac{\Omega(T, \mu_a)}{ V T^4 } 
  =
  -\frac{16 (1+\kappa_1^2) (1+\kappa_2^2) (1+\kappa_3^2)}{(2 + \kappa_1 + \kappa_2 + \kappa_3 - \kappa_1 \kappa_2 \kappa_3)^4}\,.
\end{align}
The map from $(T_0, \kappa_1, \kappa_2, \kappa_3)$ to $(T, \mu_1, \mu_2, \mu_3)$ is not one-to-one: for a given set of physical thermodynamic parameters $(T, \mu_1, \mu_2, \mu_3)$, there can be several different sets of $(T_0, \kappa_1, \kappa_2, \kappa_3)$ which give rise to the same $(T, \mu_1, \mu_2, \mu_3)$. To see this explicitly, let us look at examples.

\subsubsection*{A single non-zero chemical potential}
Consider a state with temperature $T$, and $(\mu_1, \mu_2, \mu_3) = (\mu, 0, 0)$.  By the above formulas, this must correspond to a geomerty with $\kappa_2 = \kappa_3 = 0$. Next, consider a black brane solution with $(T_0, \kappa_1, \kappa_2, \kappa_3) = (X_0, x, 0, 0)$, and another black brane solution with $(T_0, \kappa_1, \kappa_2, \kappa_3) = (Y_0, y, 0, 0)$. In order for these two black brane solutions to give rise to the same $T$ and $\mu$, we must have 
\begin{align}
  \frac{(2+x) X_0}{2\sqrt{1+x} } = \frac{(2+y) Y_0}{2\sqrt{1+y} } \,,\ \ \ \ 
  \frac{\sqrt{x} \, X_0}{\sqrt{1+x}} = \frac{\sqrt{y}\, Y_0}{\sqrt{1+y}} \,.
\end{align}
For given $X_0 > 0$ and $x \geq0$, these are two equations for $Y_0$ and $y$. Clearly, there is a solution $Y_0 = X_0$, $y=x$. But there is one more: 
$  Y_0 = \frac12 \sqrt{x}(x+4) X_0 / \sqrt{x^2 + 5x + 4}$, $y=4/x$. 
Note that both $Y_0$ and $y$ are positive for $X_0 >0$, $x>0$, and so $Y_0$ and $y$ represent a valid black brane background. Thus, two black brane backgrounds,
\begin{align}
\label{eq:bg1-I}
  & \textrm{Background I:}\ \  (T_0, \kappa_1, \kappa_2, \kappa_3) = (\tau_0, \kappa, 0, 0) \,, \\[5pt]
\label{eq:bg1-II}
  & \textrm{Background II:}\ \  (T_0, \kappa_1, \kappa_2, \kappa_3) = \left( \frac{ \sqrt{\kappa}(\kappa+4) \tau_0 }{2 \sqrt{\kappa^2 + 5\kappa + 4} }, \frac{4}{\kappa}, 0, 0 \right) , 
\end{align}
both give rise to the same $T$, and to the same  $(\mu_1, \mu_2, \mu_3) = (\mu, 0, 0)$. 

For $\kappa<2$, Background~I is thermodynamically stable and Background~II is thermodynamically unstable. For $\kappa>2$, it's the other way: Background~I is thermodynamically unstable, while Background~II is thermodynamically stable. 

\begin{figure}[t]
\centering
\includegraphics[width=0.45\textwidth]{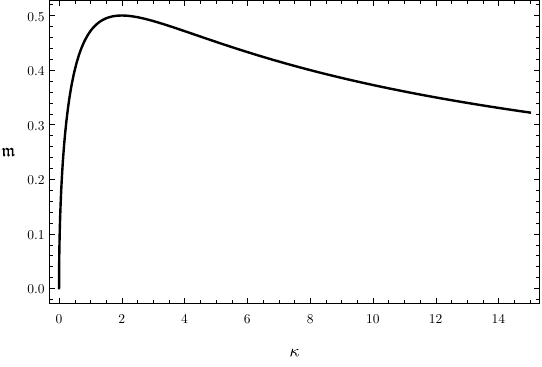}
\hspace{0.05\textwidth}
\includegraphics[width=0.45\textwidth]{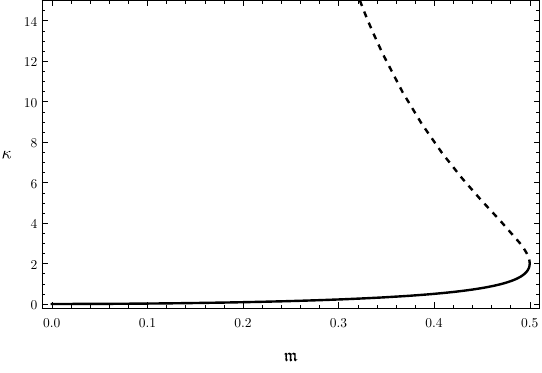}
\caption{
  Left: The dependence of $\m\equiv \mu/(2\pi T)$ on the parameter $\kappa$ in $(\kappa_1, \kappa_2, \kappa_3) = (\kappa, 0, 0)$. Right: The dependence of $\kappa$ on $\m$ is a double-valued function, with the branch corresponding to $\kappa>2$ (or to $4/\kappa < 2$) shown by a dashed line. 
}
\label{fig:mkm-1}
\end{figure}

\begin{figure}[t]
\centering
\includegraphics[width=0.45\textwidth]{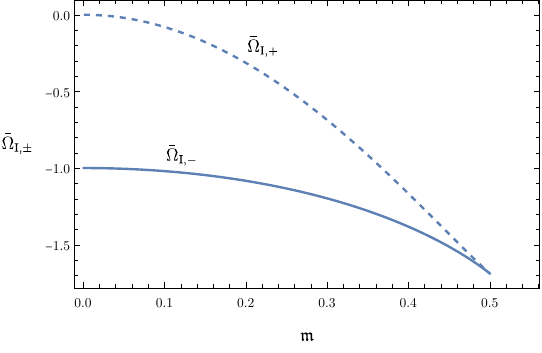}
\hspace{0.05\textwidth}
\includegraphics[width=0.45\textwidth]{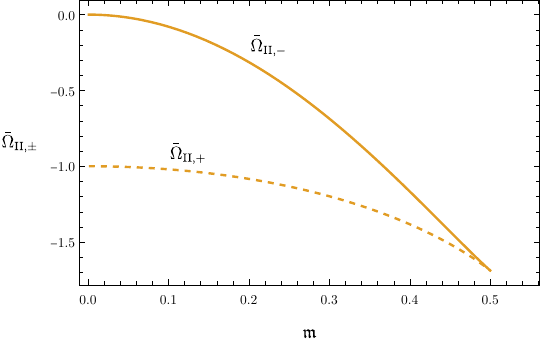}
\caption{
  The dependence of the free energy $\bar\Omega$ on $\m\equiv \mu/(2\pi T)$ for $(\mu_1, \mu_2, \mu_3) = (\mu,0,0)$. Left: $\bar\Omega_{\rm I, \pm}(\m)$, right: $\bar\Omega_{\rm II, \pm}(\m)$. The solid and dashed lines correspond to the solid and dashed lines in the right panel of Fig.~\ref{fig:mkm-1}. 
}
\label{fig:Om-1}
\end{figure}

The free energy density \eqref{eq:Omega-bar} for the two backgrounds in \eqref{eq:bg1-I}, \eqref{eq:bg1-II} is
\begin{align}
\label{eq:Omegabar-1}
    \bar\Omega_{\rm I} = -\frac{16(1+\kappa)^3}{(2+\kappa)^4}\,,\ \ \ \ 
    \bar\Omega_{\rm II} = -\frac{\kappa(4+\kappa)^3}{(2+\kappa)^4}\,.
\end{align}
In order to compare $\bar\Omega_{\rm I}$ to $\bar\Omega_{\rm II}$, we will express them as functions of $\m \equiv \mu/(2\pi T)$. Noting that $\m = \sqrt{2\kappa}/(2+\kappa)$, we have $\kappa = (1-2\m^2 \pm \sqrt{1-4\m^2})/\m^2$. This dependence is illustrated in Fig.~\ref{fig:mkm-1}. Substituting $\kappa(\m)$ into $\bar\Omega$, one finds two branches for each free energy in eq.~\eqref{eq:Omegabar-1}, $\bar\Omega_{\rm I, \pm}(\m)$, and $\bar\Omega_{\rm II, \pm}(\m)$; the corresponding functions are shown in Fig.~\ref{fig:Om-1}.
For each $0<\m<\frac12$, we have four phases, corresponding to two branches $\kappa(\m)$ for each background I and II. Note that 
\begin{align}
\label{eq:Opm12}
  \bar \Omega_{\rm I, \pm} = \bar\Omega_{\rm II, \mp} \,,
\end{align}
and the two phases with the lowest and equal free energy are $\bar\Omega_{\rm I, -}$ and $\bar\Omega_{\rm II, +}$. These two phases are the two STU backgounds with:
\begin{align}
\label{eq:Tk-11}
  & (T_0, \kappa_1, \kappa_2, \kappa_3) = (\tau_0, \kappa, 0, 0) , \ \textrm{for}\  \kappa<2 , \\[5pt]
\label{eq:Tk-12}
  & (T_0, \kappa_1, \kappa_2, \kappa_3) = \left( \frac{ \sqrt{\kappa}(\kappa+4) \tau_0 }{2 \sqrt{\kappa^2 + 5\kappa + 4} }, \frac{4}{\kappa}, 0, 0 \right) , \ \textrm{for}\ \kappa > 2\,. 
\end{align}
Both of the phases in eqs.~\eqref{eq:Tk-11} and \eqref{eq:Tk-12} are thermodynamically stable. 
The equality~\eqref{eq:Opm12} is not a coincidence: note that $T$ and $\m$ are invariant under 
\begin{align}
  T_0 \to \frac{ \sqrt{\kappa}(\kappa+4) }{2 \sqrt{\kappa^2 + 5\kappa + 4} } T_0\,,\ \ \ \ 
  \kappa \to \frac{4}{\kappa}\,,
\end{align}
hence any physical quantity in Background~I which is a (potentially multivalued) function of $T$ and $\m$ only, when evaluated for Background~II, gives the same answer.

\subsubsection*{Two equal non-zero chemical potentials}
Next, consider states with temperature $T$, and $(\mu_1, \mu_2, \mu_3) = (\mu, \mu, 0)$.  By eq.~\eqref{eq:mu3}, this must correspond to a geomerty with $\kappa_3 = 0$. Following the previous discussion of a single non-zero chemical potential, consider a black brane solution with $(T_0, \kappa_1, \kappa_2, \kappa_3) = (X_0, x_1, x_2, 0)$, and another solution with $(T_0, \kappa_1, \kappa_2, \kappa_3) = (Y_0, y_1, y_2, 0)$. In order for these two black brane solutions to give rise to the same $T$ and $\mu$, we must have 
\begin{align}
\label{eq:mu1mu2-12}
  & \frac{\sqrt{x_1 (1 + x_2)}}{\sqrt{1+x_1}} = \frac{\sqrt{x_2 (1 + x_1)}}{\sqrt{1+x_2}} \,,\\[5pt]
\label{eq:mu1mu2-22}
  & \frac{\sqrt{y_1 (1 + y_2)}}{\sqrt{1+y_1}} = \frac{\sqrt{y_2 (1 + y_1)}}{\sqrt{1+y_2}} \,,\\[5pt]
\label{eq:mumu-11}
  & \frac{\sqrt{x_1 (1 + x_2)} \, X_0 }{\sqrt{1+x_1}} = \frac{\sqrt{y_2 (1 + y_1)} \, Y_0}{\sqrt{1+y_2}} \,,\\[5pt]
\label{eq:TT-11}
  & \frac{(2 + x_1 + x_2) X_0}{\sqrt{(1+x_1) (1+x_2)}} =  \frac{(2 + y_1 + y_2) Y_0}{\sqrt{(1+y_1) (1+y_2)}} \,.
\end{align}
The first equation is the condition that $\mu_1 = \mu_2$ for the first solution. The second equation is the condition that $\mu_1 = \mu_2$ for the second solution. The third equation is the condition that the chemical potentials are the same between the two backgrounds. The fourth equation is the condition that the temperatures are the same. 

Equation~\eqref{eq:mu1mu2-12} has two solutions: $x_2 = x_1$, and $x_2 = 1/x_1$. Similarly, eq.~\eqref{eq:mu1mu2-22} has two solutions: $y_2 = y_1$, and $y_2 = 1/y_1$. Thus we have four cases: 
\begin{align}
\label{eq:bg2-Ia}
  & \textrm{Background Ia:}\ \  (T_0, \kappa_1, \kappa_2, \kappa_3) = (X_0, x, x, 0) \,, \\[5pt]
\label{eq:bg2-Ib}
  & \textrm{Background Ib:}\ \  (T_0, \kappa_1, \kappa_2, \kappa_3) = (X_0, x, 1/x, 0) , \\[5pt]
\label{eq:bg2-IIa}
  & \textrm{Background IIa:}\ \  (T_0, \kappa_1, \kappa_2, \kappa_3) = (Y_0, y, y, 0) \,, \\[5pt]
\label{eq:bg2-IIb}
  & \textrm{Background IIb:}\ \  (T_0, \kappa_1, \kappa_2, \kappa_3) = (Y_0, y, 1/y, 0) .
\end{align}
Backgrounds Ib and IIb are thermodynamically unstable because the corresponding $\kappa_{1,2}$ are outside the stability region, which is the orange region in Fig.~\ref{fig-kappa-space}. 

Further, note that the pair Ia-Ib can not have the same $T$ and $\mu$ (except for $x=1$), and similary IIa-IIb can not have the same $T$ and $\mu$ (except for $y=1$). The pair Ia-IIa can have the same $T$ and $\mu$ only for $X_0 = Y_0$, $x=y$, thus representing the same background. 

Other pairs of backgrounds can have the same $T$ and $\mu$ for a range of $\kappa$'s. 
In particular, backgrounds Ia and IIb (or Ib and IIa) can have the same $T$ and $\mu$ if eqs.~\eqref{eq:mumu-11} and \eqref{eq:TT-11} hold:
\begin{align}
  & \sqrt{x} X_0 = Y_0 \,,\\[5pt]
  & 2 X_0 = \frac{1+y}{\sqrt{y}} Y_0 \,.
\end{align}
The solutions are $Y_0 = \sqrt{x} X_0$, and $y = (2\pm 2\sqrt{1-x} - x)/x$. The product of the two solutions for $y$ is 1. Thus, the following backgrounds have the same $T$ and $\mu$: 
\begin{align}
\label{eq:Ia-2}
    & \textrm{Background Ia:}\ \  (T_0, \kappa_1, \kappa_2, \kappa_3) = (\tau_0, \kappa, \kappa, 0) \,, \\[5pt]
\label{eq:IIbp}
    & \textrm{Background IIb$'$:}\ \  (T_0, \kappa_1, \kappa_2, \kappa_3) = (\sqrt{\kappa}\,\tau_0, \frac{ 2 + 2\sqrt{1{-}\kappa} - \kappa)}{\kappa}, \frac{ 2 - 2\sqrt{1{-}\kappa} - \kappa)}{\kappa}, 0) \,, \\[5pt]
\label{eq:IIbpp}
    & \textrm{Background IIb$''$:}\ \  (T_0, \kappa_1, \kappa_2, \kappa_3) = (\sqrt{\kappa}\,\tau_0, \frac{ 2 - 2\sqrt{1{-}\kappa} - \kappa)}{\kappa}, \frac{ 2 + 2\sqrt{1{-}\kappa} - \kappa)}{\kappa}, 0) \,.
\end{align}
Analogously, Ib and IIb will have the same $T$ and $\mu$ if \eqref{eq:mumu-11} and \eqref{eq:TT-11} hold: 
\begin{align}
  & X_0 =  Y_0 \,,\\[5pt]
  & \frac{ \sqrt{1+x} }{\sqrt{x}}\, X_0 = \frac{ \sqrt{1+y} }{\sqrt{y}}\, Y_0 \,.
\end{align}
The solutions are $Y_0 = X_0$, $y = x$, or $y=1/x$. The first of these gives IIb = Ib. The second one gives the pair of backgrounds with the same $T$ and $\mu$:
\begin{align}
\label{eq:Ib-2}
    & \textrm{Background Ib:}\ \  (T_0, \kappa_1, \kappa_2, \kappa_3) = (\tau_0, \kappa, \frac{1}{\kappa}, 0) \,, \\[5pt]
\label{eq:IIb-2}
    & \textrm{Background IIb:}\ \  (T_0, \kappa_1, \kappa_2, \kappa_3) = (\tau_0, \frac{1}{\kappa}, \kappa, 0) \,.\end{align}

\begin{figure}[t]
\centering
\includegraphics[width=0.45\textwidth]{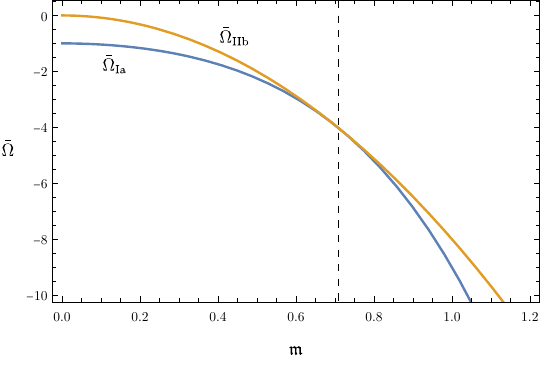}
\hspace{0.05\textwidth}
\includegraphics[width=0.45\textwidth]{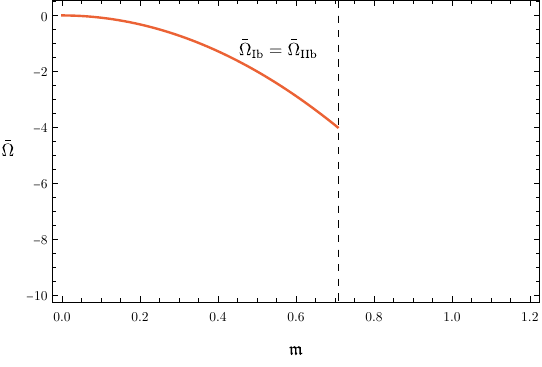}
\caption{
   The dependence of the free energy $\bar\Omega$ on $\m\equiv \mu/(2\pi T)$ for $(\mu_1, \mu_2, \mu_3) = (\mu, \mu, 0)$. Left: The blue line corresponds to the background \eqref{eq:Ia-2}, the orange line corresponds to the backgrounds \eqref{eq:IIbp}, \eqref{eq:IIbpp} which have the same free energy. The curves touch at $\m_c = 1/\sqrt{2}$, corresponding to $\kappa=1$, indicated by a dashed line. The thermodynamically unstable background IIb has a higher free energy that Ia for both $\m<\m_c$, and for $\m>\m_c$. 
  Right: The free energies for the thermodynamically unstable backgrounds \eqref{eq:Ib-2}, \eqref{eq:IIb-2}.  The plot is analogous to a degenerate version of Fig.~\ref{fig:Om-1}, in which solid and dashed lines are on top of each other. 
}
\label{fig:Om-2}
\end{figure}

Now let's plot the free energy $\bar\Omega$ as a function of $\m \equiv \mu/(2\pi T)$. For the backgrounds \eqref{eq:Ia-2}, \eqref{eq:IIbp}, \eqref{eq:IIbpp}, we have $\m = \sqrt{\kappa/2}$, which is a monotonic function of $\kappa$, with $\m>0$. Substituting $\kappa=2\m^2$ into \eqref{eq:Ia-2}, \eqref{eq:IIbp}, \eqref{eq:IIbpp}, we find $(\kappa_1, \kappa_2, \kappa_3)$ as functions of $\m$,
\begin{align}
  & \bar\Omega_{\rm Ia} = -(1+2\m^2)^2\,,\\
  & \bar\Omega_{\rm IIb} = -8\m^2 \,.
\end{align}
For the backgrounds \eqref{eq:Ib-2} and \eqref{eq:IIb-2}, we have $\m = \sqrt{2\kappa}/(1+\kappa)$ (hence $\m$ is restricted to be less than $\m_c = 1/\sqrt{2}$), and the free energies are $\bar\Omega_{\rm Ib} = \bar\Omega_{\rm IIb} = -8\m^2$. The free energy $\bar\Omega$ is plotted in Fig.~\ref{fig:Om-2}. The background Ia has a lower free energy than IIb, except for one point $\m=\m_c=1/\sqrt{2}$ when the two free energies are equal. The background Ia is thermodynamically stable for $\m<\m_c$. For $\m>\m_c$, both Ia and IIb are thermodynamically unstable. 

\subsubsection*{Three equal non-zero chemical potentials}

\begin{figure}[t]
\centering
\includegraphics[width=0.45\textwidth]{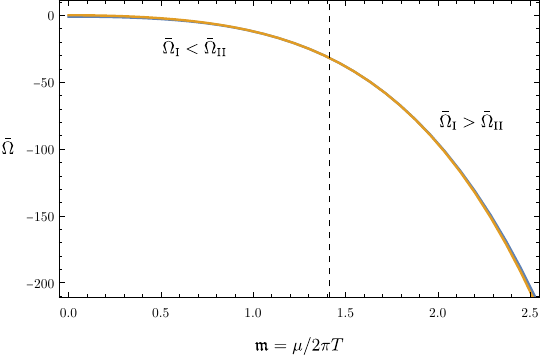}
\hspace{0.05\textwidth}
\includegraphics[width=0.46\textwidth]{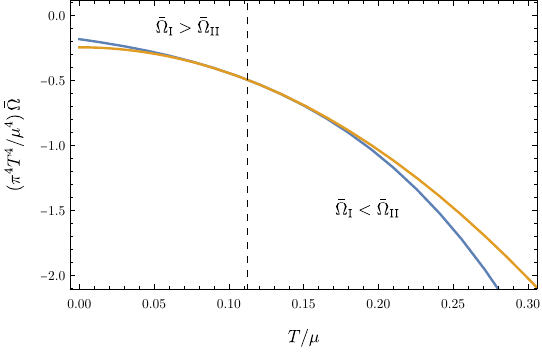}
\caption{
   Left: The dependence of the free energy $\bar\Omega$ on $\m\equiv \mu/(2\pi T)$ for $(\mu_1, \mu_2, \mu_3) = (\mu, \mu, \mu)$. Left: The blue line corresponds to the background \eqref{eq:bg3-I}, the orange line corresponds to the backgrounds \eqref{eq:bg3-II1}, \eqref{eq:bg3-II2}, \eqref{eq:bg3-II3} which have the same free energy. The curves intersect at $\m_c = \sqrt{2}$, corresponding to $\kappa=1$, indicated by a dashed line. 
     Right: The same plot presented in different variables, showing $(\pi^4 T^4/\mu^4)\bar\Omega$ as a function of $T/\mu$, with two branches crossing at $T/\mu = 1/(2\pi\sqrt{2})$, see ref.~\cite{Buchel:2025tjq}. 
}
\label{fig:Om-3}
\end{figure}

Finally, consider states with temperature $T$, and $(\mu_1, \mu_2, \mu_3) = (\mu, \mu, \mu)$. As before, consider a black brane solution with $(T_0, \kappa_1, \kappa_2, \kappa_3) = (X_0, x_1, x_2, x_3)$, and another black brane solution with $(T_0, \kappa_1, \kappa_2, \kappa_3) = (Y_0, y_1, y_2, y_3)$. Analogous to the previous discussion, there are several backgrounds which give rise to the same $T$ and $\mu$, studied recently in ref.~\cite{Buchel:2025tjq}.
An exercise show that the following backgrounds:
\begin{align}
\label{eq:bg3-I}
  & \textrm{Background I:}\ \  (T_0, \kappa_1, \kappa_2, \kappa_3) = (\tau_0, \kappa, \kappa, \kappa) \,, \\[5pt]
\label{eq:bg3-II1}
  & \textrm{Background II$'$:}\ \  \left( T_0 , \kappa_1, \kappa_2, \kappa_3 \right) = \left( \tau_0 \frac{(2{-}\kappa) \sqrt{\kappa (1{+}\kappa)} }{\sqrt{\kappa^2 - 3\kappa + 4}}  , \frac{\kappa}{(2{-}\kappa)^2},  \frac{\kappa}{(2{-}\kappa)^2},  \frac{(2{-}\kappa)^2}{\kappa} \right) , \\[5pt]
\label{eq:bg3-II2}
  & \textrm{Background II$''$:}\ \  \left( T_0 , \kappa_1, \kappa_2, \kappa_3 \right) = \left( \tau_0 \frac{(2{-}\kappa) \sqrt{\kappa (1{+}\kappa)} }{\sqrt{\kappa^2 - 3\kappa + 4}}  , \frac{\kappa}{(2{-}\kappa)^2},  \frac{(2{-}\kappa)^2}{\kappa} ,  \frac{\kappa}{(2{-}\kappa)^2} \right) , \\[5pt]
\label{eq:bg3-II3}
  & \textrm{Background II$'''$:}\ \  \left( T_0 , \kappa_1, \kappa_2, \kappa_3 \right) = \left( \tau_0 \frac{(2{-}\kappa) \sqrt{\kappa (1{+}\kappa)} }{\sqrt{\kappa^2 - 3\kappa + 4}}  , \frac{\kappa}{(2{-}\kappa)^2},  \frac{(2{-}\kappa)^2}{\kappa} ,  \frac{\kappa}{(2{-}\kappa)^2} \right) , 
\end{align}
have the same $T$ and the same $\mu_a = \mu$. Their free energies are shown in Fig.~\ref{fig:Om-3}. The right panel in Fig.~\ref{fig:Om-3} corresponds to Fig.~1 in ref.~\cite{Buchel:2025tjq}, where thermodynamics of STU phases with $(\kappa_1, \kappa_2, \kappa_3) = (x,x,x)$ was compared to thermodynamics of STU phases with  $(\kappa_1, \kappa_2, \kappa_3) = (y, y, 1/y)$. Background I is thermodynamically stable for $\m < \m_c = \sqrt{2}$ (corresponding to $\kappa=1$), and thermodynamically unstable for $\m > \m_c$. Background II is thermodynamically unstable for all~$\m>0$.

 \section{The conductivity matrix}
 \label{conductivity-matrix-app}
 The conductivity matrix $\sigma_{ab}$ can be computed by applying Kubo formula to the two-point functions of the spatial components of currents $J_i^a$ at zero spatial momentum,
 \begin{equation}
 \sigma_{ab} = - \frac{1}{\omega}\; \mbox{Im}\, G_{J^a_i J^b_i}\left(\omega,  {\bf k}=0\right)\,,
 \label{Kubo-sigma}
 \end{equation}
 where $i=x,y,z$.  The correlators in the hydrodynamic approximation are determined by solving the fluctuation equations for gauge fields at zero spatial momentum and using the relevant part of the boundary action.  
 
 Note that at zero spatial momentum the difference between the sound and shear channels disappear, and $G_{J^a_x J^b_x}\left(\omega,  {\bf k}=0\right)=G_{J^a_y J^b_y}\left(\omega,  {\bf k}=0\right)=G_{J^a_z J^b_z}\left(\omega,  {\bf k}=0\right)$.  We borrow the shear channel
  equations of motion for the ``electric'' fields 
 $$
 \tilde{E}_{\text{\tiny\itshape L/R}}^{a}= \wfr_0 \left( \tilde{A}_{x}^{a}\pm i \tilde{A}_{y}^{a}\right),
 $$
 where
 \begin{equation}
\tilde{A}_{x,y}^{a}=\frac{\sqrt{\kappa_{a}}}{\mu_{a}(1+\kappa_{a})}\, A_{x,y}^{a}\,,
\end{equation}
  from ref.~\cite{chiral-life}, setting $\qfr_0=0$ there (alternatively, they can be obtained from the sound channel equations of motion with $\qfr_0=0$).   At vanishing spatial momentum,  $\tilde{E}_{\text{\tiny\itshape L}}^{a}$ and $\tilde{E}_{\text{\tiny\itshape R}}^{a}$ satisfy the same equations,  and we use the notation $\tilde{E}^{a}$ for either of the fields.  The equations are
\begin{eqnarray}
&\,&\tilde{E}^{1''}+\bigg(\frac{2 H_{1}'}{H_{1}}-\frac{H'}{H}+\frac{f'}{f}\bigg)\tilde{E}^{1 '}+\frac{\wfr_0^{2}H}{ u f^{2}}\tilde{E}^{1}\nonumber \\&-& \frac{\sqrt{\kappa_{1}} u }{H_{1}^{2}f}(1+\kappa_{1})(1+\kappa_{2})(1+\kappa_{3})(\sqrt{\kappa_{1}}\tilde{E}^{1}+\sqrt{\kappa_{2}}\tilde{E}^{2}+\sqrt{\kappa_{3}}\tilde{E}^{3})=0\,, 
\label{ginv-eq-3-e1}\\
%\end{gathered}
%\end{equation}
%
%\begin{equation}
%\begin{gathered}
&\,& \tilde{E}^{2''}+\bigg(\frac{2 H_{2}'}{H_{2}}-\frac{H'}{H}+\frac{f'}{f}\bigg)\tilde{E}^{2 '}+\frac{\wfr_0^{2}H}{ u f^{2}}\tilde{E}^{2} \nonumber  \\&-&\frac{\sqrt{\kappa_{2}} u }{H_{2}^{2}f}(1+\kappa_{1})(1+\kappa_{2})(1+\kappa_{3})(\sqrt{\kappa_{1}}\tilde{E}^{1}+\sqrt{\kappa_{2}}\tilde{E}^{2}+\sqrt{\kappa_{3}}\tilde{E}^{3}) =0\,, \\
\label{ginv-eq-3-e2}
%\end{gathered}
%\end{equation}
%
%\begin{equation}
%\begin{gathered}
&\,& \tilde{E}^{3''}+\bigg(\frac{2 H_{3}'}{H_{3}}-\frac{H'}{H}+\frac{f'}{f}\bigg)\tilde{E}^{3 '}+\frac{\wfr_0^{2}H}{ u f^{2}}\tilde{E}^{3}\nonumber \\&-&\frac{\sqrt{\kappa_{3}} u }{H_{3}^{2}f}(1+\kappa_{1})(1+\kappa_{2})(1+\kappa_{3})(\sqrt{\kappa_{1}}\tilde{E}^{1}+\sqrt{\kappa_{2}}\tilde{E}^{2}+\sqrt{\kappa_{3}}\tilde{E}^{3}) =0\,.
\label{ginv-eq-3-e3}
%\end{gathered}
\end{eqnarray}
 The equations \eqref{ginv-eq-3-e1}---\eqref{ginv-eq-3-e3} are coupled,  and we were not able to solve them analytically even in the hydrodynamic approximation. However, in some special cases, an analytic solution can be found.
 
 The part of the boundary action involving the gauge field fluctuations only is given by \cite{chiral-life}
\begin{equation}
%\begin{gathered}
S_{\text{reg.  shear}}=\frac{\pi^{2} N^{2}_c T_{0}^{4}}{8}(1+\kappa_{1})(1+\kappa_{2})(1+\kappa_{3})\sum_{a=1}^{3}\int d^{4}x  \Biggl( {\Tilde{A}}_{x}^a{\Tilde{A}_{x}}^{a'}+{\Tilde{A}}_{y}^a{\Tilde{A}_{y}}^{a '} \Biggr)\,. \label{action-shear}
%\end{gathered}
\end{equation}
 %where
% \begin{equation}
%\tilde{A}_{x,y}^{a}=\frac{\sqrt{\kappa_{a}}}{\mu_{a}(1+\kappa_{a})}\, A_{x,y}^{a}\,.
%\end{equation}
Expressing the chemical potentials via $\kappa_a$, we find
\begin{equation}
%\begin{gathered}
S_{\text{reg.  shear}}=\frac{N^{2}_c T_{0}^{2}}{16}\sum_{a=1}^{3}\int d^{4}x  \Biggl( A_{x}^a A_{x}^{a'}+A_{y}^a A_{y}^{a '} \Biggr)\,.\label{action-shear-simple}
%\end{gathered}
\end{equation}
 \subsection{The case of $(\kappa_1,\kappa_2,\kappa_3) = (\kappa,0,0)$}
 In this case, the equations  \eqref{ginv-eq-3-e1}---\eqref{ginv-eq-3-e3}  read
\begin{eqnarray}
&\,&\tilde{E}^{1''}+\bigg(\frac{f'}{f}+ \frac{ H'}{H}   \bigg)\tilde{E}^{1 '}+\frac{\wfr_0^{2}H}{ u f^{2}}\tilde{E}^{1} -\frac{\kappa (1+\kappa)  u }{H^{2}f}\tilde{E}^{1}=0\,, 
\label{ginv-eq-3-e1-0} \\
&\,&\tilde{E}^{a''}+\bigg(\frac{f'}{f}- \frac{ H'}{H}   \bigg)\tilde{E}^{a '}+\frac{\wfr_0^{2}H}{ u f^{2}}\tilde{E}^{a} =0\,, 
\label{ginv-eq-3-e1-00}
\end{eqnarray}
 where $a=1,2$, $H(u)=1+\kappa u$ and $f(u)=(1-u)(1+u+\kappa u)$.  Their solution to leading non-trivial  order the hydrodynamic approximation can be found 
 following the approach introduced in refs.~\cite{Policastro:2002se,Kovtun:2005ev} (see also Appendix C in  ref.~\cite{Grozdanov:2016fkt}):
\begin{eqnarray}
\tilde{E}^1 (u) &=&  \frac{C_1 (1 - u)^{-\frac{i \wfr_0 \sqrt{1+k}}{2+\kappa}}}{1+ \kappa u} \Biggl[ 1 + \frac{u \kappa}{2}  \nonumber \\
&+& \frac{i \wfr_0 \sqrt{1+\kappa} \left[ - (1 - u)\kappa^2 + (1 + \kappa)(2 + u \kappa) \ln\left( \frac{1 + u + u \kappa}{2 + \kappa} \right) \right]}{2 (2+\kappa) (1 + \kappa)}  \Biggr]\,,  \nonumber  \\
\tilde{E}^{2,3} (u) &=& C_{2,3} (1 - u)^{-\frac{i \wfr_0\sqrt{1+\kappa}}{2+\kappa}}     \left[ 1 + \frac{i \wfr_0 \sqrt{1+\kappa}}{(2+\kappa) (1+\kappa)^2}  \ln\left( \frac{1 + u + u \kappa}{2 + \kappa} \right) \right]\,. \nonumber
\end{eqnarray}
It is worth emphasising that the equations \eqref{ginv-eq-3-e1-0}---\eqref{ginv-eq-3-e1-00} decouple, so the integration constant in the solution for each field depends solely on the boundary value of that field.

To compute the correlators using the bulk solutions $\tilde{E}^a(u)$,  we write these solutions in the form
\begin{equation}
A_{x,y}^{a}(u) =  A_{x,y}^{a} (\epsilon)\; \frac{ \tilde{E}^{a}(u)}{\tilde{E}^{a}(\epsilon)}\,,
\end{equation}
where  $A_{x,y}^{a} (\epsilon)$ are the boundary values of the fields and the normalisation of $\tilde{E}^a(u)$ in the ratio is irrelevant. Then,  using the boundary action \eqref{action-shear-simple} and the recipe of ref.~\cite{Son:2002sd},  for any $i=x,y,z$, we have
\begin{equation}
G_{J^a_i J^b_i}\left(\omega,  {\bf k}=0\right) = - \frac{N^{2}_c T_{0}^{2}}{8} \left[ \frac{ \tilde{E}^{a '}(\epsilon)}{\tilde{E}^{a}(\epsilon)}\right]\,,
\end{equation}
where the square bracket means removing contact terms.   We find
\begin{eqnarray}
&\,& G_{J^1_i J^1_i}\left(\omega,  {\bf k}=0\right) = - \frac{N^{2}_c T_{0}^{2}}{32} \frac{i \wfr_0 (2+\kappa)^2}{\sqrt{1+\kappa}}\,, \\
&\,& G_{J^2_i J^2_i}\left(\omega,  {\bf k}=0\right) =  G_{J^3_i J^3_i}\left(\omega,  {\bf k}=0\right) = - \frac{N^{2}_c T_{0}^{2}}{8} \frac{i \wfr_0 }{\sqrt{1+\kappa}}\,, \\
&\,& G_{J^a_i J^b_i}\left(\omega,  {\bf k}=0\right) = 0\,, \qquad a\neq b\,.
\end{eqnarray}
The Kubo formula \eqref{Kubo-sigma} gives the matrix of conductivities for the equilibrium state with $(\kappa_1,\kappa_2,\kappa_3) = (\kappa,0,0)$:
\begin{align}
  \sigma_{ab} = \frac{N^2_c T}{8\pi}\, \begin{pmatrix}
 \frac{2+\kappa}{4} & 0 &  0\\
  0 & \frac{1}{2+\kappa} & 0 \\
  0 & 0 & \frac{1}{2+\kappa}
  \end{pmatrix} \,,
\label{susco-matrix}
\end{align}
 where we used the relation $T_0 = 2 \sqrt{1+\kappa} T/(2+\kappa)$ between the parameter $T_0$ and the physical temperature $T$.

 \subsection{The case of $(\kappa_1,\kappa_2,\kappa_3) = (\kappa,\kappa,0)$}
 For two equal chemical potentials, the equations  \eqref{ginv-eq-3-e1}---\eqref{ginv-eq-3-e3} are
 \begin{eqnarray}
\tilde{E}_1'' 
&+& \frac{2(u - \kappa + 2u \kappa)}{(1 - u)(1 + u + 2u \kappa)} \tilde{E}_1' 
+ \frac{\wfr_0^2 (1 + u  \kappa)^2}{(1 - u)^2 u (1 + u + 2u \kappa)^2} \tilde{E}_1 \nonumber \\
&+& \frac{u \kappa (1 + \kappa)^2}{(1 - u)(1 + u \kappa)^2 (1 + u + 2u \kappa)} (\tilde{E}_1 + \tilde{E}_2) = 0\,,   \label{eqmx1}\\
\tilde{E}_2'' 
&+& \frac{2(u - \kappa + 2u \kappa)}{(1 - u)(1 + u + 2u \kappa)} \tilde{E}_2' 
+ \frac{\wfr_0^2 (1 + u  \kappa)^2}{(1 - u)^2 u (1 + u + 2u \kappa)^2} \tilde{E}_2 \nonumber \\
&+& \frac{u \kappa (1 + \kappa)^2}{(1 - u)(1 + u \kappa)^2 (1 + u + 2u \kappa)} (\tilde{E}_1 + \tilde{E}_2) = 0\,,  \label{eqmx2}\\
\tilde{E}_3'' 
&+& \frac{2 u (1 + \kappa)^2}{(1 - u)(1 + u \kappa)(1 + u + 2 u \kappa)} \tilde{E}_3' \nonumber \\
&+& \frac{\wfr_0^2 (1 + u  \kappa)^2}{(1 - u)^2 u (1 + u + 2 u \kappa)^2} \tilde{E}_3 = 0 \label{eqmx3}\,.
\end{eqnarray}
The equations \eqref{eqmx1} and \eqref{eqmx2} are coupled.  The perturbative solution to \eqref{eqmx1} and \eqref{eqmx2} is given by 
 \begin{eqnarray}
 \tilde{E}_1 &=& C_1 \tilde{E}_1^{(1)} +C_2 \tilde{E}_1^{(2)} \,,   \label{magnumx1}\\
 \tilde{E}_2 &=& C_1 \tilde{E}_2^{(1)} + C_2 \tilde{E}_2^{(2)} \label{magnumx2} \,,
 \end{eqnarray}
 where, to linear order in $\wfr_0$,
\begin{eqnarray}
\tilde{E}_1^{(1)} = \tilde{E}_2^{(1)}  &=& - \frac{(1-u)^{- \frac{i\wfr_0}{2}}}{\kappa (3 + 4\kappa)(1 + u\kappa)}\Biggl[  1+ 
\nonumber  \\
&+&  \frac{i\wfr_0 \left(
  2 (1 - u) \kappa^2 (1 + 2 \kappa)
  + (1 + \kappa) \ln \left[ \frac{1 + u + 2 u \kappa}{2 (1 + \kappa)}\right]
\right)}{
  2  (1 + \kappa) (1 + 2 \kappa)^2 
}
\Biggr]
\end{eqnarray}
  and
\begin{equation}
\tilde{E}_1^{(2)} = - \tilde{E}_2^{(2)} =  (1-u)^{- \frac{i\wfr_0}{2}}\Biggl[  1+  
\frac{i \wfr_0}{2}  \ln \left( \frac{1 + u + 2 u \kappa}{2 (1 + \kappa)} \right)\,
\Biggr]\,.
\end{equation}
 At the boundary at $u=\epsilon \to 0$,  we demand  $\tilde{E}_1(u)\to \tilde{E}_1^{(0)}$ and $\tilde{E}_2(u)\to \tilde{E}_2^{(0)}$.  This implies that the integration constants $C_1$ and $C_2$ in Eqs.~\eqref{magnumx1}---\eqref{magnumx2} are related to the boundary values of the fields as
 \begin{eqnarray}
 C_1 &=& \alpha  \tilde{E}_1^{(0)} +\gamma  \tilde{E}_2^{(0)}\,, \nonumber \\
 C_2 &=& \beta  \tilde{E}_1^{(0)} +\delta  \tilde{E}_2^{(0)}\,, 
 \end{eqnarray}
 where the coefficients can be found from 
\begin{equation}
\begin{pmatrix} 
  \alpha   &  \beta\\ 
  \gamma & \delta
\end{pmatrix}
=
\frac{1}{\Delta} 
\begin{pmatrix} 
  \tilde{E}_2^{(2)} (\epsilon)   & - \tilde{E}_2^{(1)} (\epsilon)\\ 
  - \tilde{E}_1^{(2)} (\epsilon) & \tilde{E}_1^{(1)} (\epsilon)
\end{pmatrix}\,,
\label{connection-coeffi}
\end{equation}
with
\begin{equation}
\Delta = \mbox{det}\,
\begin{pmatrix} 
  \tilde{E}_1^{(1)} (\epsilon)   & \tilde{E}_1^{(2)} (\epsilon)\\ 
  \tilde{E}_2^{(1)} (\epsilon) & \tilde{E}_2^{(2)} (\epsilon)
\end{pmatrix}\, \to \frac{2}{\kappa (3+4\kappa)} + O(\wfr_0)\,.
\label{determinant-x}
\end{equation}
 To leading order in $\wfr_0$,  the relevant part of the action \eqref{action-shear-simple} is given by
 \begin{eqnarray}
S_{\text{reg.  shear}} &=& \frac{i \wfr_0 N_c^2 T_0^2}{32 (1+\kappa)}
\Biggl[
  -2 \tilde{E}_1^{(0)} \tilde{E}_2^{(0)} \kappa (2 + \kappa)
  + \tilde{E}_1^{(0)} \tilde{E}_1^{(0)}(2 + 2 \kappa + \kappa^2) \nonumber \\
 & +&  \tilde{E}_2^{(0)}\tilde{E}_2^{(0)}  (2 + 2 \kappa + \kappa^2)
\Biggr]\,.
\end{eqnarray}
 We find the correlators to leading order in $\wfr_0$:
\begin{eqnarray}
&\,& G_{J^1_i J^1_i}\left(\omega,  {\bf k}=0\right) = G_{J^2_i J^2_i}\left(\omega,  {\bf k}=0\right) = - \frac{N^{2}_c T_{0}^{2}}{16} \, \frac{i \wfr_0 (2+2 \kappa +\kappa^2)}{ (1+\kappa)}\,, \\
&\,& G_{J^1_i J^2_i}\left(\omega,  {\bf k}=0\right) =  G_{J^2_i J^1_i}\left(\omega,  {\bf k}=0\right) = \frac{N^{2}_c T_{0}^{2}}{16} \frac{i \wfr_0 \kappa (2+\kappa)}{1+\kappa}\,.
\end{eqnarray}
 The perturbative solution to  \eqref{eqmx3} is given by
 \[
\tilde{E}_3 (u) = C (1 - u)^{-i \omega_0 / 2} \left( 
1 + \frac{
  i \wfr_0 \left[ 
    2 (1 - u) \kappa^2 (1 + 2 \kappa) + 
    (1 + \kappa) \ln \left( \frac{1 + u + 2 u \kappa}{2 (1 + \kappa)} \right)
  \right]
}{
  2 (1 + \kappa)(1 + 2 \kappa)^2
}
\right)
\]
 which gives the correlator
 \begin{equation}
  G_{J^3_i J^3_i}\left(\omega,  {\bf k}=0\right) =  - \frac{N^{2}_c T_{0}^{2}}{8} \, \frac{i \wfr_0}{ 1+\kappa}\,, 
\end{equation}
 with $G_{J^3_i J^1_i}\left(\omega,  {\bf k}\right) =G_{J^1_i J^3_i}\left(\omega,  {\bf k}=0\right) =0$ and 
 $G_{J^3_i J^2_i}\left(\omega,  {\bf k}\right) =G_{J^2_i J^3_i}\left(\omega,  {\bf k}=0\right) =0$.

The Kubo formula \eqref{Kubo-sigma} then leads to the matrix of conductivities for the equilibrium state with $(\kappa_1,\kappa_2,\kappa_3) = (\kappa,\kappa,0)$:
\begin{align}
  \sigma_{ab} = \frac{N^2_c T}{16\pi}\, \begin{pmatrix}
 \frac{(2+2 \kappa +\kappa^2)}{ 2(1+\kappa)}& -\frac{\kappa (2+\kappa)}{2(1+\kappa)}\ &  0\\
  -\frac{\kappa (2+\kappa)}{2(1+\kappa)} &  \frac{(2+2 \kappa +\kappa^2)}{ 2(1+\kappa)} & 0 \\
  0 & 0 & \frac{1}{1+\kappa}
  \end{pmatrix} \,,
\label{susco-matrix-2-kappas}
\end{align}
 where we used the relation $T_0 = T$   between the parameter $T_0$ and the physical temperature $T$ in this case.

 \subsection{The case of $(\kappa_1,\kappa_2,\kappa_3) = (\kappa,\kappa,\kappa)$}
In this case,  the equations  \eqref{ginv-eq-3-e1}---\eqref{ginv-eq-3-e3}  read
 \begin{eqnarray}
\tilde{E}_1''(u) 
&+& \left(
-\frac{\kappa}{1 + u \kappa}
+ \frac{3 \kappa + 3 u^2 \kappa^3 - 2 u (1 + 3 \kappa + \kappa^3)}{(1 - u)(-1 + u^2 \kappa^3 - u (1 + 3 \kappa))}
\right) \tilde{E}_1'(u) \nonumber \\
&+& \frac{
\wfr_0^2 (1 + u \kappa)^3 \tilde{E}_1(u)
}{
(1 - u)^2 u (1 + u + 3 u \kappa - u^2 \kappa^3)^2
} \nonumber \\
&-& \frac{
u \kappa (1 + \kappa)^3 (\tilde{E}_1(u) + \tilde{E}_2(u) + \tilde{E}_3(u))
}{
(1 - u)(1 + u \kappa)^2 (-1 + u^2 \kappa^3 - u (1 + 3 \kappa))
} = 0\,, \\
\nonumber \\
\tilde{E}_2''(u) 
&+& \left(
-\frac{\kappa}{1 + u \kappa}
+ \frac{3 \kappa + 3 u^2 \kappa^3 - 2 u (1 + 3 \kappa + \kappa^3)}{(1 - u)(-1 + u^2 \kappa^3 - u (1 + 3 \kappa))}
\right) \tilde{E}_2'(u) \nonumber \\
&+& \frac{
\wfr_0^2 (1 + u \kappa)^3 \tilde{E}_2(u)
}{
(1 - u)^2 u (1 + u + 3 u \kappa - u^2 \kappa^3)^2
} \nonumber \\
&-& \frac{
u \kappa (1 + \kappa)^3 (\tilde{E}_1(u) + \tilde{E}_2(u) + \tilde{E}_3(u))
}{
(1 - u)(1 + u \kappa)^2 (-1 + u^2 \kappa^3 - u (1 + 3 \kappa))
} = 0\,, \\
\nonumber \\
\tilde{E}_3''(u) 
&+& \left(
-\frac{\kappa}{1 + u \kappa}
+ \frac{3 \kappa + 3 u^2 \kappa^3 - 2 u (1 + 3 \kappa + \kappa^3)}{(1 - u)(-1 + u^2 \kappa^3 - u (1 + 3 \kappa))}
\right) \tilde{E}_3'(u) \nonumber \\
&+& \frac{
\wfr_0^2 (1 + u \kappa)^3 \tilde{E}_3(u)
}{
(1 - u)^2 u (1 + u + 3 u \kappa - u^2 \kappa^3)^2
} \nonumber \\
&-& \frac{
u \kappa (1 + \kappa)^3 (\tilde{E}_1(u) + \tilde{E}_2(u) + \tilde{E}_3(u))
}{
(1 - u)(1 + u \kappa)^2 (-1 + u^2 \kappa^3 - u (1 + 3 \kappa))
} = 0\,.
\end{eqnarray}
 These equations are coupled.   To effectively decouple them,  we can write equations satisfied by ${\cal \tilde{E}}_1 = 2 \tilde{E}_1-\tilde{E}_2-\tilde{E}_3$,  ${\cal \tilde{E}}_2= 2 \tilde{E}_2-\tilde{E}_1-\tilde{E}_3$ and ${\cal \tilde{E}}_3 = 2 \tilde{E}_3-\tilde{E}_1-\tilde{E}_2$:
\begin{eqnarray}
&& {\cal \tilde{E}}_i'' + \left(-\frac{\kappa}{1 + u \kappa} + \frac{3 \kappa + 3 u^2 \kappa^3 - 2 u (1 + 3 \kappa + \kappa^3)}{(1-u)(-1 + u^2 \kappa^3 - u (1 + 3 \kappa))} \right) {\cal \tilde{E}}_i' \nonumber \\
&& + \frac{\wfr_0^2 (1 + u \kappa)^3}{(1-u)^2 u (1 + u + 3 u \kappa - u^2 \kappa^3)^2} {\cal \tilde{E}}_i= 0\,,
\end{eqnarray}
where $i=1,2,3$, as well as the equation for  ${\cal E}=\tilde{E}_1+\tilde{E}_2+\tilde{E}_3$:
 \begin{eqnarray}
&& {\cal E}'' + \left(-\frac{\kappa}{1 + u \kappa} + \frac{-1 + u^2 \kappa^3 - u (1 + 3 \kappa) + (1 - u) \left(1 + 3 \kappa - 2 u \kappa^3\right)}{(1 - u) \left(1 - u^2 \kappa^3 + u (1 + 3 \kappa)\right)} \right) {\cal E}' \nonumber \\
&& + \frac{\wfr_0^2 (1 + u \kappa)^3}{(1 - u)^2 u \left(1 - u^2 \kappa^3 + u (1 + 3 \kappa)\right)^2} {\cal E} \nonumber \\
&& - \frac{3 u \kappa (1 + \kappa)^3}{(1 - u) (1 + u \kappa)^2 \left(1 - u^2 \kappa^3 + u (1 + 3 \kappa)\right)} {\cal E} = 0\,.
\end{eqnarray}
 The perturbative solutions for  ${\cal \tilde{E}}_i$ and ${\cal E}$ are given by\footnote{Note that the variables ${\cal E}$ and ${\cal \tilde{E}}_i$  
 correspond, accordingly,  to the ``centre of mass'' and fluctuations around it described by the eigenvectors \eqref{eigenvectors-hessian-1}---\eqref{eigenvectors-hessian-3}.  See also ref.~\cite{Gladden:2024ssb}.}
\begin{equation}
{\cal E} (u) = \left(\tilde{E}_1^0+\tilde{E}_2^0+\tilde{E}_3^0\right) \frac{F(u)}{F(\epsilon)}\,,
\end{equation}
\begin{equation}
{\cal \tilde{E}}_i (u) =   \left(2 \tilde{E}_i^0-\tilde{E}_j^0-\tilde{E}_k^0\right)  \frac{F_1(u)}{F_1(\epsilon)}\,,  \qquad i,j,k = 1,2,3\,,  i\neq j\neq k\,, 
\end{equation}
%
% \begin{eqnarray}
% {\cal E} (u) &=& \left(\tilde{E}_1^0+\tilde{E}_2^0+\tilde{E}_3^0\right) \frac{F(u)}{F(\epsilon)}\\
% {\cal \tilde{E}}_1 (u) &=&   \left(2 \tilde{E}_1^0-\tilde{E}_2^0-\tilde{E}_3^0\right)  \frac{F_1(u)}{F_1(\epsilon)}\,,  \\
 %{\cal \tilde{E}}_2 (u) &=&   \left(2 \tilde{E}_2^0-\tilde{E}_1^0-\tilde{E}_3^0\right)  \frac{F_2(u)}{F_1(\epsilon)}\,,  \\
 %{\cal \tilde{E}}_3 (u)&=&   \left(2 \tilde{E}_3^0-\tilde{E}_1^0-\tilde{E}_2^0\right)  \frac{F_3(u)}{F_1(\epsilon)}\,,  
 %\end{eqnarray}
 where
\begin{eqnarray}
F_i(u) &=& (1 - u)^{- \frac{i \wfr_0}{(2 - \kappa) \sqrt{1 + \kappa}}} \Bigg(
1 + \frac{i \wfr_0}{2 (-2 + \kappa) \sqrt{1 + \kappa} \sqrt{1 + 4 \kappa}} \times \nonumber \\
&& \Bigg[
(-2 + 4 \kappa) \, \operatorname{ArcCoth} \left( 
\frac{(1 + \kappa) \sqrt{1 + 4 \kappa}}{-1 - 3 \kappa + 2 u \kappa^3} 
\right) \nonumber \\
&& + (2 - 4 \kappa) \, \operatorname{ArcTanh} \left( 
\frac{-1 + 2 (-1 + \kappa) \kappa}{\sqrt{1 + 4 \kappa}} 
\right) \nonumber \\
&& + \sqrt{1 + 4 \kappa} \left( 
\ln(2 + 3 \kappa - \kappa^3) - \ln(1 + u + 3 u \kappa - u^2 \kappa^3) 
\right)
\Bigg]
\Bigg)\,
\end{eqnarray}
 and
\begin{eqnarray}
F(u) &=& (1 - u)^{- \frac{i \wfr_0}{(2 - \kappa) \sqrt{1 + \kappa}}} \Bigg\{
\frac{ 2 - u \kappa}{2 (1 + u \kappa)} 
+ \frac{\wfr_0}{4 (-2 + \kappa)^4 \kappa (1 + \kappa)^{3/2} (1 + 4 \kappa)^{3/2} (1 + u \kappa)} 
\times \nonumber \\
&& \Bigg[
-2 i  (-2 + \kappa)^3 \kappa (-2 + u \kappa) ( -1 + 9 \kappa + 12 \kappa^2 + 2 \kappa^3 )
\operatorname{ArcTanh} \left( \frac{-1 - 3 \kappa + 2 u \kappa^3}{(1 + \kappa) \sqrt{1 + 4 \kappa}} \right) \nonumber \\
&& +  \sqrt{1 + 4 \kappa} \Bigg(
54 i (-2 + \kappa) \kappa^2 (8 - 14 \kappa + (4 + u) \kappa^2 + (-1 + u) \kappa^3) \nonumber \\
&& + 2 \pi (-2 + u \kappa)(1 + \kappa + 6 \kappa^2 + 95 \kappa^3 + 94 \kappa^4 
+ 9 \kappa^5 + 4 \kappa^6) \nonumber \\
&& + (-2 + \kappa)^3 \kappa (1 + 4 \kappa) (-2 + u \kappa) \Bigg(
- \frac{54 i \kappa (-4 + 5 \kappa)}{(-2 + \kappa)^2 (1 + 4 \kappa)} \nonumber \\
&& - \frac{2 \pi (1 - 3 \kappa + 18 \kappa^2 + 23 \kappa^3 + 2 \kappa^4 + \kappa^5)}{(-2 + \kappa)^3 \kappa} \nonumber \\
&& + \frac{2 i (-1 + 9 \kappa + 12 \kappa^2 + 2 \kappa^3)}{(1 + 4 \kappa)^{3/2}} 
\operatorname{ArcTanh} \left( \frac{-1 - 2 \kappa + 2 \kappa^2}{\sqrt{1 + 4 \kappa}} \right) \nonumber \\
&& - i (1 + \kappa) \ln(2 + 3 \kappa - \kappa^3)
\Bigg) \nonumber \\
&& + i (-2 + \kappa)^3 \kappa (-2 + u \kappa) (1 + 5 \kappa + 4 \kappa^2) 
\ln(1 + u + 3 u \kappa - u^2 \kappa^3) 
\Bigg)
\Bigg]
\Bigg\}\,.
\end{eqnarray}
 Then,  the three components of the ``electric'' field can be found as
 \begin{equation}
 \tilde{E}_i (u) = \frac 13 \left( {\cal E} + {\cal \tilde{E}}_i \right)\,.
 \end{equation}
 Subsituting the solution  $\tilde{E}_1 (u)$ into the boundary action \eqref{action-shear-simple},
  we find the term
 \begin{eqnarray}
S^{(1)} &=& 
\frac{i N_c^2 T_0^2 \wfr_0}{64 (1 + \kappa)^{3/2} } \left( 
- (\tilde{E}_{2}^0 + \tilde{E}_{3}^0) \kappa (4 + \kappa) 
+ \tilde{E}_{1}^0 (4 + 4 \kappa + 3 \kappa^2) 
\right)\tilde{E}_1^0\,,
\end{eqnarray}
and similarly for $\tilde{E}_2 (u)$ and $\tilde{E}_3 (u)$.  Correspondingly, the retarded correlators 
to leading order in $\wfr_0$ are given by
\begin{eqnarray}
&\,& G_{J^a_i J^a_i}\left(\omega,  0\right)  =
- \frac{N^{2}_c T}{32\pi} \, \frac{i \omega (4+4 \kappa + 3\kappa^2)}{ (2-\kappa) (1+\kappa)^{2}}\,,  \qquad a=1,2,3\,,\\
&\,& G_{J^a_i J^b_i}\left(\omega,  0\right)  = \frac{N^{2}_c T}{32\pi } \frac{i \omega 
\kappa (4+\kappa)}{(2-\kappa) (1+\kappa)^2}\,,  \qquad a\neq b\,,
\end{eqnarray}
where we used the relation $T_0 = 2 T/(2-\kappa) \sqrt{1+\kappa}$
   between the parameter $T_0$ and the physical temperature $T$ in this case.

Applying the Kubo formula \eqref{Kubo-sigma}, we find the matrix of conductivities for the equilibrium state with $(\kappa_1,\kappa_2,\kappa_3) = (\kappa,\kappa,\kappa)$:
\begin{align}
  \sigma_{ab} = \frac{N^2_c T}{16\pi}\, \begin{pmatrix}
 \frac{4+4 \kappa +3\kappa^2}{ 2(2-\kappa) (1+\kappa)^2} & -\frac{\kappa (4+\kappa)}{2(2-\kappa) (1+\kappa)^2}&  -\frac{\kappa (4+\kappa)}{2(2-\kappa) (1+\kappa)^2} \\  -\frac{\kappa (4+\kappa)}{2(2-\kappa) (1+\kappa)^2}\ &   \frac{4+4 \kappa +3\kappa^2}{ 2(2-\kappa) (1+\kappa)^2} & -\frac{\kappa (4+\kappa)}{2(2-\kappa) (1+\kappa)^2}\ \\
  -\frac{\kappa (4+\kappa)}{2(2-\kappa) (1+\kappa)^2} & -\frac{\kappa (4+\kappa)}{2(2-\kappa) (1+\kappa)^2}\ &  \frac{4+4 \kappa +3\kappa^2}{ 2(2-\kappa) (1+\kappa)^2}
  \end{pmatrix} \,.
\label{susco-matrix-3-kappas}
\end{align}

\section{Sound channel fluctuations for $(\kappa_1,\kappa_2,\kappa_3)=(\kappa,0,0)$ }
\label{eoms-1-kappa}
The notations in this Appendix follow the ones used in section \ref{single-kappa-section}. In particular,  for the purposes of this section, $f = (1-u)(1+u+\kappa u)$.

a) Scalar equations:

The equation for $s_1$ reads
\begin{align}
s_1'' &+ \left(\frac{f'}{f} - \frac{1+3\kappa u}{u \mathcal{H}}\right) s_1' + \frac{4\kappa^{1/2}(1+\kappa)u\mathcal{H}}{3f} {a_t^{(1)}}' + \frac{\kappa}{3}\left( H_{ii}' - H_{tt}'\right) + \frac{2\kappa(1+\kappa)u}{3\mathcal{H}f} H_{tt} \nonumber\\ &+ \left[ \frac{\mathcal{H}}{u^2f} + \frac{4\kappa(1+\kappa)u}{3\mathcal{H}^2f} + \frac{\fr{w}^2\mathcal{H} - \fr{q}^2f}{uf^2} + \frac{\kappa}{\mathcal{H}}\left(\frac{1+ 3\kappa u}{u\mathcal{H}} - \frac{f'}{f}\right)\right] s_1 = 0\,.
\end{align}
For $s_j$ with $j\neq 1$, the equations are as follows:
\begin{align}
s_j'' &+ \left(\frac{f'}{f} - \frac{1}{u}\right) s_j' - \frac{2\kappa^{1/2}(1+\kappa)u}{3f} {a_t^{(1)}}' - \frac{\kappa}{6\mathcal{H}}\ \left( H_{ii}' - H_{tt}'\right) \nonumber\\ 
&+ \left(\frac{1}{u^2f} + \frac{\fr{w}^2\mathcal{H} - \fr{q}^2f}{uf^2}\right) s_j- \frac{\kappa(1+\kappa)u}{3\mathcal{H}^2f}  H_{tt} - \frac{2\kappa(1+\kappa)u}{3\mathcal{H}^3f}  s_1 = 0\,.
\end{align}
These can be decoupled from $s_1$, and the gauge and gravity sectors, using the variable $\mathcal{S} = s_2 - s_3$, which obeys the following simple equation:
\begin{equation}
\mathcal{S}'' + \left(\frac{f'}{f} - \frac{1}{u}\right)\mathcal{S}' + \left(\frac{1}{u^2f} + \frac{\fr{w}^2\mathcal{H} - \fr{q}^2f}{uf^2}\right)\mathcal{S} = 0\,.
\end{equation} 
 
b) Gauge field equations:

The equations involving $a_{t,z}^{(1)}$ are
\begin{align}
{a_t^{(1)}}'' + \frac{2\kappa}{\mathcal{H}} {a_t^{(1)}}' &+ \frac{2\kappa^{1/2}}{\mathcal{H}^3} s_1' + \frac{\kappa^{1/2}}{2\mathcal{H}^2}\left( H_{ii}' + H_{tt}' \right) \nonumber\\
&- \frac{\fr{q}}{uf}\left(\fr{q}a_t^{(1)} + \fr{w} a_z^{(1)}\right) - \frac{2\kappa^{3/2}}{\mathcal{H}^4} s_1 = 0\,,
\end{align}
\begin{equation}
\fr{w}\mathcal{H} {a_t^{(1)}}' + \fr{q}f {a_z^{(1)}}' + \frac{2\kappa^{1/2}\fr{w}}{\mathcal{H}^2} s_1 
+ \frac{\kappa^{1/2}\fr{w}}{2\mathcal{H}} \left( H_{ii} + H_{tt} \right) + \frac{\kappa^{1/2}\fr{q}}{\mathcal{H}} H_{tz} = 0\,,
\end{equation}
\begin{equation}
{a_z^{(1)}}'' + \left(\frac{\kappa}{\mathcal{H}} + \frac{f'}{f}\right) {a_z^{(1)}}' + \frac{\kappa^{1/2}}{\mathcal{H}f} H_{tz}' + \frac{\fr{w}\mathcal{H}}{u f^2}\left( \fr{q}\cdot a_t^{(1)} + \fr{w} a_z^{(1)}\right)=0\,.
\end{equation}

For $j\neq 1$, the gauge fields obey the following equations:
\begin{equation}
{a_t^{(j)}}'' - \frac{\fr{q}}{uf}\left(\fr{w} a_z^{(j)} + \fr{q} a_t^{(j)}\right) = 0\,,
\end{equation}
\begin{equation}
{a_t^{(j)}}' + \frac{\fr{q}f}{\fr{w}\mathcal{H}} {a_z^{(j)}}' = 0\,,
\end{equation}
\begin{equation}
{a_z^{(j)}}'' + \left(\frac{f'}{f} - \frac{\kappa}{\mathcal{H}}\right) {a_z^{(j)}}' + \frac{\fr{w}\mathcal{H}}{uf^2}\left(\fr{w}\cdot a_z^{(j)} + \fr{q} a_t^{(j)}\right) = 0\,.
\end{equation}

c) Einstein equations:
\begin{align}
H_{tt}'' &+ \left(-\frac{3}{2u} - \frac{\kappa}{3\mathcal{H}} + \frac{3f'}{2f}\right) H_{tt}' + \left(\frac{1}{2u} + \frac{\kappa}{3\mathcal{H}} - \frac{f'}{2f}\right) H_{ii}' \nonumber \\
&+ \frac{4\kappa^{1/2}(1+\kappa)u}{3f}{a_t^{(1)}}' + \frac{2\kappa}{3u\mathcal{H}f}\left(\frac{2(1+\kappa)u^2}{\mathcal{H}^2}+1\right)s_1 \nonumber\\
&- \frac{\fr{w}\mathcal{H}}{u f^2} \left( \fr{w} H_{ii} + 2\fr{q} H_{tz}\right) + \left(\frac{2\kappa(1+\kappa)u}{3\mathcal{H}^2f} - \frac{\fr{q}^2}{u f}\right)H_{tt} = 0\,,
\end{align}
\begin{align}
\fr{w}H_{ii}' + \fr{q} H_{tz}' + \frac{\kappa \fr{w}}{\mathcal{H}^2} s_1 + \frac{1}{2}\left(\frac{\kappa}{\mathcal{H}} - \frac{f'}{f}\right) \left(\fr{w} H_{ii} + 2\fr{q}H_{tz}\right) = 0\,,
\end{align}
\begin{align}
H_{tz}'' + \left(\frac{\kappa}{\mathcal{H}} - \frac{1}{u}\right) H_{tz}' + \frac{\kappa^{1/2}(1+\kappa)u}{\mathcal{H}}{a_z^{(1)}}' + \frac{\fr{qw}}{uf} H_{aa} =0\,,
\end{align}
\begin{align}
H_{tt}'' &- H_{ii}'' - \left( \frac{\kappa}{6\mathcal{H}} + \frac{f'}{2f}\right) H_{ii}' + \left( -\frac{5\kappa}{6\mathcal{H}} + \frac{3f'}{2f}\right)H_{tt}' + \frac{4\kappa^{1/2}(1+\kappa)u}{3f} {a_t^{(1)}}' \nonumber \\ 
&- \frac{2\kappa}{\mathcal{H}^2}s_1' + \left( \frac{2\kappa^2}{\mathcal{H}^3} + \frac{4\kappa (1+\kappa)u^2}{3u\mathcal{H}^3f} + \frac{2\kappa}{3u\mathcal{H}f}\right) s_1 + \frac{2\kappa(1+\kappa)u}{3\mathcal{H}^2f} H_{tt} = 0\,,
\end{align}
\begin{align}
\fr{q}f\left( H_{tt}' - H_{aa}'\right) &+ \fr{w}\mathcal{H}H_{tz}' + \frac{\fr{q}f}{2}\left(\frac{f'}{f} - \frac{\kappa}{\mathcal{H}}\right) H_{tt} \nonumber\\
&+ \kappa^{1/2}(1+\kappa)u\left(\fr{q} a_t^{(1)} + \fr{w} a_z^{(1)}\right) - \frac{\fr{q}\kappa f}{\mathcal{H}^2} s_1 = 0\,,
\end{align}
\begin{align}
H_{aa}'' &+ \left( - \frac{2}{u} + \frac{\kappa}{3\mathcal{H}} + \frac{f'}{f}\right) H_{aa}' + \left(\frac{\kappa}{3\mathcal{H}} - \frac{1}{u}\right)\left( H_{zz}' - H_{tt}' \right) \nonumber\\
&+ \frac{4\kappa^{1/2}(1+\kappa)u}{3f} {a_t^{(1)}}' + \frac{4\kappa}{3u\mathcal{H}f}\left( \frac{(1+\kappa)u^2}{\mathcal{H}^2} - 1\right)s_1 \nonumber \\
&+ \frac{2\kappa(1+\kappa)u}{3\mathcal{H}^2f} H_{tt} + \frac{\fr{w}^2\mathcal{H} - \fr{q}^2f}{uf^2}H_{aa} = 0\,,
\end{align}
\begin{align}
H_{zz}'' &+ \left(-\frac{3}{2u} + \frac{\kappa}{6\mathcal{H}} + \frac{f'}{f}\right) H_{zz}' + \left(\frac{\kappa}{6\mathcal{H}} - \frac{1}{2u}\right)\left( H_{aa}' - H_{tt}'\right) \nonumber \\ 
&+ \frac{2\kappa^{1/2}(1+\kappa)u}{3f} {a_t^{(1)}}' + \frac{2\kappa}{3u\mathcal{H}f}\left(\frac{(1+\kappa)u^2}{\mathcal{H}^2} - 1)\right) s_1 - \frac{\fr{q}^2}{uf} H_{aa} \nonumber\\
&+ \left( \frac{\kappa(1+\kappa)u}{3\mathcal{H}^2f} + \frac{\fr{q}^2}{u f}\right) H_{tt} + \frac{\fr{w}\mathcal{H}}{u f^2}\left( 2\fr{q} H_{tz} + \fr{w} H_{zz}\right) = 0\,.
\end{align}

%%%%%%%%%%%%%%%%%%%%%%%%%%%%%%%%%%%%%%%%

\section{Sound channel fluctuations for $(\kappa_1,\kappa_2,\kappa_3)=(\kappa,\kappa,0)$ }
\label{fluct-eom-2-kappa}
The notations in this Appendix correspond to the ones used in section \ref{section-two-kappas}.  In particular, for the purposes of this section, $f=(1-u)(1+u + 2\kappa u)$.

a) Scalar field equations
\begin{align}
\mathfrak{s}'' &+ \left(\frac{f'}{f} - \frac{1+3\kappa u}{u \mathcal{H}}\right) \mathfrak{s}' + \frac{2\kappa^{1/2}(1+\kappa)^2u\mathcal{H}}{f} \mathfrak{a}_t' \nonumber \\ 
&+ \left(\frac{\mathfrak{D}(u)}{u f^2} + \frac{1}{u^2\mathcal{H}^2f} + \frac{\kappa(1+3\kappa u)}{u \mathcal{H}^2} - \frac{\kappa f'}{\mathcal{H} f} + \frac{\kappa\left( 3\mathcal{H} + \mathfrak{C}(u) \right)}{u \mathcal{H}^2 f}\right)\mathfrak{s} =0\,,
\end{align}
\begin{align}
s_3'' &+ \left(\frac{f'}{f} - \frac{1}{u}\right) s_3' - \frac{2\kappa^{1/2}(1+\kappa)^2u}{3f} \mathfrak{A}_t' - \frac{H'}{6H}\left(H_{ii}' - H_{tt}'\right) \nonumber \\
&- \frac{2\kappa (1+\kappa)^2 u}{3\mathcal{H}^2f} H_{tt} + \left( \frac{2\kappa(1+\kappa)^2u}{3\mathcal{H}^2f} + \frac{1}{u^2f} + \frac{\mathfrak{D}(u)}{u f^2}\right)s_3  = 0\,.
\end{align}
It should be noted that substituting for $s_3$ in terms of $s_1$ and $s_2$ in the above, and subtracting the result from the sum of the equations for $s_1$ and $s_2$, one obtains the constraint $s_1 + s_2 = 0$.

b) Gauge field equations

For gauge field perturbations $a^{(i)}$, with $i=1,2$, we find the following equations:
\begin{align}
{a_t^{(i)}}'' + \frac{2\kappa}{\mathcal{H}}{a_t^{(i)}}' &+ \frac{2\kappa^{1/2}}{\mathcal{H}^3} s_i' + \frac{\kappa^{1/2}}{2\mathcal{H}^2}\left(H_{ii}' + H_{tt}'\right) \nonumber\\
&- \frac{\mathfrak{q}}{u f}\left( \mathfrak{q} a_t^{(i)} + \mathfrak{w} a_z^{(i)}\right) - \frac{2\kappa^{3/2}}{\mathcal{H}^4} s_i = 0\,,
\end{align}
\begin{equation}
H {a_t^{(i)}}' + \frac{\mathfrak{q}f}{\mathfrak{w}} {a_z^{(i)}}' + \frac{2\kappa^{1/2}}{\mathcal{H}} s_i + \frac{\kappa^{1/2}}{2}\left(H_{ii} + H_{tt} + \frac{2\mathfrak{q}}{\mathfrak{w}} H_{tz}\right) = 0\,,
\end{equation}
\begin{equation}
{a_z^{(i)}}'' + \left(\frac{2\kappa}{\mathcal{H}} - \frac{H'}{H} + \frac{f'}{f}\right){a_z^{(i)}}' + \frac{\kappa^{1/2}}{f} H_{tz}' + \frac{\mathfrak{w} H}{u f^2}\left( \mathfrak{q}\,  a_t^{(i)} + \mathfrak{w} \, a_z^{(i)}\right) = 0\,.
\end{equation}

The gauge field perturbation $a^{(3)}$ obeys:  
\begin{equation}
{a_t^{(3)}}'' -\frac{\mathfrak{q}}{u f}\left( \mathfrak{q} a_t^{(3)} + \mathfrak{w} a_z^{(3)}\right) = 0\,,
\end{equation}
\begin{equation}
\mathfrak{w}H {a_t^{(3)}}' + \mathfrak{q}f {a_z^{(3)}}' = 0\,,
\end{equation}
\begin{equation}
{a_z^{(3)}}'' + \left(\frac{f'}{f} - \frac{H'}{H}\right){a_z^{(3)}}' + \frac{\mathfrak{w} H}{u f^2}\left( \mathfrak{q} a_t^{(3)} + \mathfrak{w} a_z^{(3)}\right) = 0\,.
\end{equation}

Particularly useful are the equations of motion for the difference $\mathfrak{a}_\mu = a_\mu^{(1)} - a_\mu^{(2)}$:
\begin{equation}
\mathfrak{a}_t'' + \frac{2\kappa}{\mathcal{H}} \mathfrak{a}_t' + \frac{2\kappa^{1/2}}{\mathcal{H}^3} \mathfrak{s}'  -\frac{\mathfrak{q}}{u f}\left(\mathfrak{q} \mathfrak{a}_t + \mathfrak{w} \mathfrak{a}_z\right) - \frac{2\kappa^{3/2}}{\mathcal{H}^4} \mathfrak{s} = 0\,,
\end{equation}
\begin{equation}
\mathfrak{a}_t' + \frac{\mathfrak{ q}f}{\mathfrak{w}H} \mathfrak{a}_z' + \frac{2\kappa^{1/2}}{\mathcal{H}^3} \mathfrak{s} =0\,,
\end{equation}
\begin{equation}
\mathfrak{a}_z'' + \left(\frac{2\kappa}{\mathcal{H}} + \frac{f'}{f} - \frac{H'}{H}\right)\mathfrak{a}_z' + \frac{\mathfrak{w} H}{u f^2}\left(\mathfrak{q} \mathfrak{a}_t + \mathfrak{w} \mathfrak{a}_z\right)  = 0\,.
\end{equation}

c) Einstein equations

\begin{align}
H_{tt}'' &+ \left(\frac{3f'}{2f} - \frac{3}{2u} - \frac{H'}{3H}\right)H_{tt}' + \left(\frac{1}{2u} + \frac{H'}{3H} - \frac{f'}{2f}\right)H_{ii}' + \frac{4\kappa^{1/2}(1+\kappa)^2u}{3f} \mathfrak{A}_t' \nonumber \\ 
&- \frac{\mathfrak{w} H}{u f^2}\left( \mathfrak{w} H_{ii} + 2\mathfrak{q} H_{tz}\right)  + \left(\frac{4\kappa(1+\kappa)^2u}{3\mathcal{H}^2f} - \frac{\mathfrak{q}^2}{u f}\right)H_{tt}   = 0\,,
\end{align}
\begin{align}
H_{tz}' + \frac{\mathfrak{w}}{\mathfrak{q}} H_{ii}' + \left(\frac{H'}{H} - \frac{f'}{f}\right) H_{tz} + \frac{\mathfrak{w}}{2\mathfrak{q}}\left(\frac{H'}{H} - \frac{f'}{f}\right) H_{ii} = 0\,,
\end{align}
\begin{align}
H_{tz}'' + \left(\frac{H'}{H} - \frac{1}{u}\right)H_{tz}' + \frac{\kappa^{1/2}(1+\kappa)^2u}{H} \mathfrak{A}_z' + \frac{\mathfrak{w q}}{uf} H_{aa} = 0\,,
\end{align}
\begin{align}
H_{tt}'' - H_{ii}'' &+ \frac{1}{2}\left(-\frac{5H'}{3H} + \frac{3f'}{f}\right)H_{tt}' - \frac{1}{2}\left( \frac{H'}{3H} + \frac{f'}{f}\right)H_{ii}' \nonumber \\
&+ \frac{4\kappa^{1/2}(1+\kappa)^2u}{3f} \mathfrak{A}_t' + \frac{4\kappa(1+\kappa)^2u}{3\mathcal{H}^2f} H_{tt} = 0\,,
\end{align}
\begin{align}
H_{tt}' - H_{aa}' + \frac{\mathfrak{w}H}{\mathfrak{q}f} H_{tz}' &+ \frac{\kappa^{1/2}(1+\kappa)^2u}{f}\left(\mathfrak{A}_t + \frac{\mathfrak{w}}{\mathfrak{q}} \mathfrak{A}_z\right) \nonumber \\
&+ \frac{1}{2}\left(\frac{f'}{f} - \frac{H'}{H}\right)H_{tt} = 0\,,
\end{align}
\begin{align}
H_{aa}'' + \left(-\frac{2}{u} + \frac{H'}{3H} + \frac{f'}{f}\right)H_{aa}' &+ \left(\frac{1}{u} - \frac{H'}{3H}\right)\left( H_{tt}' - H_{zz}'\right) + \frac{4\kappa^{1/2}(1+\kappa)^2u}{3f}\mathfrak{A}_t' \nonumber \\
&+ \frac{4\kappa(1+\kappa)^2u}{3\mathcal{H}^2f} H_{tt} + \frac{\mathfrak{D(u)}}{u f^2} H_{aa} = 0\,,
\end{align}
\begin{align}
H_{zz}'' &+ \left(\frac{H'}{6H} + \frac{f'}{f} - \frac{3}{2u}\right)H_{zz}' + \frac{1}{2}\left(\frac{H'}{3H} - \frac{1}{u}\right)\left(H_{aa}' - H_{tt}'\right) \nonumber\\ 
&+ \frac{2\kappa^{1/2}(1+\kappa)^2u}{3f} \mathfrak{A}_t' - \frac{\mathfrak{q}^2}{uf}H_{aa} + \left(\frac{\mathfrak{q}^2}{uf} + \frac{2\kappa(1+\kappa)^2u}{3\mathcal{H}^2f}\right)H_{tt} \nonumber\\
&+ \frac{2\mathfrak{q w} H}{uf^2} H_{tz}  + \frac{\mathfrak{w}^2\mathcal{H}^2}{u f^2}H_{zz} = 0\,.
\end{align}
Here, we use the following variables:
\begin{align}
\mathfrak{C}(u) &= (2+ \kappa (4+3\kappa))u^2\,,\\
\mathcal{H}(u) &= 1+ \kappa u \,,\\
H(u) &= (1+\kappa u)^2\,, \\
\mathfrak{s}(u) &= s_1 - s_2\,, \\
\mathfrak{a}_\mu &= a_\mu^{(1)} - a_\mu^{(2)}\,, \\
\mathfrak{A}_\mu &= a_\mu^{(1)} + a_\mu^{(2)}\,,  \\
H_{ii} &= H_{aa} + H_{zz}\,.
\end{align}

\section{Sound channel fluctuations for $(\kappa_1,\kappa_2,\kappa_3)=(\kappa,\kappa,\kappa)$ }
\label{fluct-eom-3-kappa}
The notations in this section correspond to the ones used in section \ref{3-equal-kappas-section}. In particular, for 
 the purposes of this section, $f = (1+\kappa u)^3 - (1+\kappa)^3u^2$. In addition, we use the following variables:
\begin{align}
\mathfrak{s}^{(i)} &= 2s_i - \sum_{j\neq i} s_j \,,\\
\mathfrak{a}_\mu^{(i)} &= 2a_\mu^{(i)} - \sum_{j\neq i} a_\mu^{(j)} \,, \label{eq:COMvar}\\
\fr{D} &= \fr{w}^2\mathcal{H}^{3} - \fr{q}^2f \,, \\
\fr{A}_\mu &= a_\mu^{(1)} + a_\mu^{(2)} + a_\mu^{(3)}\,,  \label{yy1}\\
\mathfrak{S} &= s_1 + s_2 + s_3\,.  \label{yy2}
\end{align}

a) Scalar field equations

\begin{align}
\fr{s}_i'' &+ \left(\frac{f'}{f}-\frac{1+3\kappa u}{u\mathcal{H}}\right)\fr{s}_i + \frac{2\kappa^{1/2}(1+\kappa)^3\cdot u \mathcal{H}}{f} \fr{a}_t^{(i)}\nonumber\\ 
&+ \left[\frac{\fr{D}}{uf^2} + \frac{\mathcal{H}}{u^2f} + \frac{2\kappa(1+\kappa)^3u}{\mathcal{H}^2f} - \frac{\kappa}{\mathcal{H}}\left(\frac{f'}{f}-\frac{1+3\kappa u}{u\mathcal{H}}\right)\right] \fr{s}_i = 0\,.
\end{align}

b) Gauge equations  

\begin{align}
{a_z^{(i)}}'' + \left(\frac{f'}{f} - \frac{\kappa}{\mathcal{H}}\right){a_z^{(i)}}' + \frac{\fr{w}\mathcal{H}^3}{u f^2}\left( \fr{q}a_t^{(i)} + \fr{w} a_z^{(i)}\right) + \frac{\kappa^{1/2}\mathcal{H}}{f} H_{tz}' =0\,, \label{xx1}
\end{align}
\begin{align}
{a_t^{(i)}}'' &+ \frac{2\kappa}{\mathcal{H}} {a_t^{(i)}}' + \frac{2\kappa^{1/2}}{\mathcal{H}^3} s_i' + \frac{\kappa^{1/2}}{2\mathcal{H}^2}\left(H_{ii}' + H_{tt}'\right) \nonumber\\
&- \frac{\fr{q}}{uf}\left( \fr{q} a_t^{(i)} + \fr{w} a_z^{(i)}\right) - \frac{2\kappa^{3/2}}{\mathcal{H}^4} s_i  = 0\,,  \label{xx2}
\end{align}
\begin{align}
{a_z^{(i)}}' + \frac{\fr{w}\mathcal{H}^3}{\fr{q}f} {a_t^{(i)}}' + \frac{2\kappa^{1/2}\fr{w}}{\fr{q}f} s_i + \frac{\kappa^{1/2}\mathcal{H}}{f} H_{tz} + \frac{\kappa^{1/2}\fr{w}\mathcal{H}}{2\fr{q}f}\left( H_{ii} + H_{tt}\right) = 0\,. \label{xx3}
\end{align}
Considering linear combinations (\ref{eq:COMvar}), one can show that the fluctuations $\mathfrak{s}_i$ decouple from gravity. The variables
${\mathfrak{a}_t^{(i)}}$, ${\mathfrak{a}_z^{(i)}}$ and $\mathfrak{s}_i$ obey the system of equations:
\begin{align}
{\mathfrak{a}_z^{(i)}}'' + \left(\frac{f'}{f} - \frac{\kappa}{\mathcal{H}}\right) {\mathfrak{a}_z^{(i)}}' + \frac{\mathfrak{w}\mathcal{H}^3}{uf^2}\left( \mathfrak{q}\mathfrak{a}_t^{(i)} + \mathfrak{w}\mathfrak{a}_z^{(i)}\right) = 0\,,
\end{align}
\begin{align}
{\mathfrak{a}_t^{(i)}}'' + \frac{2\kappa}{\mathcal{H}} {\mathfrak{a}_t^{(i)}}' + \frac{2\kappa^{1/2}}{\mathcal{H}^3} \mathfrak{s}_i' - \frac{\mathfrak{q}}{u f}\left( \mathfrak{q} \mathfrak{a}_t^{(i)} + \mathfrak{w} \mathfrak{a}_z^{(i)}\right) - \frac{2\kappa^{3/2}}{\mathcal{H}^4}\mathfrak{s}_i = 0\,,
\end{align}
\begin{align}
{\mathfrak{a}_t^{(i)}}' + \frac{\mathfrak{q}f}{\mathfrak{w}\mathcal{H}^3} {\mathfrak{a}_z^{(i)}}' + \frac{2\kappa^{1/2}}{\mathcal{H}^3} \mathfrak{s}_i = 0\,.
\end{align}
Taking the sum of Eqs.~\eqref{xx1}, \eqref{xx2} and \eqref{xx3},  and noting that $\sum_i s_i = 0$ due to redundancy in the scalar sector (see Appendix \ref{scalar-and-redundancy}), one obtains the following equations for the ``centre of mass'' variables \eqref{yy1} and \eqref{yy2}:
\begin{align}
\mathfrak{A}_z'' + \left(\frac{f'}{f} - \frac{\kappa}{\mathcal{H}}\right) \mathfrak{A}_z' + \frac{3\kappa^{1/2}\mathcal{H}}{f} H_{tz}' + \frac{\mathfrak{w}\mathcal{H}^3}{u f^2}\left( \mathfrak{q} \mathfrak{A}_t + \mathfrak{w} \mathfrak{A}_z\right) = 0\,,
\end{align}
\begin{eqnarray}
\mathfrak{A}_t'' &+& \frac{2\kappa}{\mathcal{H}} \mathfrak{A}_t' + \frac{2\kappa^{1/2}}{\mathcal{H}^3}\mathfrak{S}' + \frac{3\kappa^{1/2}}{2\mathcal{H}^2}\left( H_{ii}' + H_{tt}'\right) - \frac{\mathfrak{q}}{uf}\left( \mathfrak{q} \mathfrak{A}_t + \mathfrak{w} \mathfrak{A}_z\right) \nonumber \\
&-& \frac{2\kappa^{3/2}}{\mathcal{H}^4}\mathfrak{S} = 0\,,
\end{eqnarray}
\begin{align}
\mathfrak{A}_t' + \frac{\mathfrak{q}f}{\mathfrak{w}\mathcal{H}^3} \mathfrak{A}_z' + \frac{2\kappa^{1/2}}{\mathcal{H}^3} \mathfrak{S} + \frac{3\kappa^{1/2}}{2\mathcal{H}^2}\left( H_{ii} + H_{tt} + \frac{2\mathfrak{q}}{\mathfrak{w}} H_{tz}\right) = 0\,.
\end{align}

c) Einstein equations

\begin{align}
H_{tt}'' &+ \left(-\frac{3}{2u} - \frac{\kappa}{\mathcal{H}} + \frac{3f'}{2f}\right) H_{tt}' + \left(\frac{1}{2u} + \frac{\kappa}{\mathcal{H}} - \frac{f'}{2f}\right) H_{ii}'  + \frac{4\kappa^{1/2}(1+\kappa)^3u}{3f} \fr{A}_t' \nonumber \\ &-\frac{\fr{w}\mathcal{H}^3}{uf^2}\left( \fr{w}\, H_{ii} + 2\fr{q}\, H_{tz}\right) + \left( \frac{2\kappa(1+\kappa)^3u}{\mathcal{H}^2f} - \frac{\fr{q}^2}{uf}\right) H_{tt} = 0\,,
\end{align}
\begin{align}
\fr{w} H_{ii}' + \fr{q}H_{tz}' + \frac{1}{2}\left(\frac{3\kappa}{\mathcal{H}} - \frac{f'}{f}\right)\left( \fr{w}\, H_{ii} + 2\fr{q}\, H_{tz}\right) = 0\,,
\end{align}
\begin{align}
H_{tz}'' + \left(\frac{3\kappa}{\mathcal{H}} - \frac{1}{u}\right) H_{tz}' + \frac{\kappa^{1/2}(1+\kappa)^3u}{\mathcal{H}^3} \fr{A}_z' + \frac{\fr{qw}}{uf} H_{aa} = 0\,,
\end{align}
\begin{align}
H_{tt}'' - H_{ii}'' &+ \frac{1}{2}\left(-\frac{5\kappa}{\mathcal{H}} + \frac{3f'}{f}\right) H_{tt}' - \frac{1}{2}\left(\frac{\kappa}{\mathcal{H}} + \frac{f'}{f}\right) H_{ii}' \nonumber\\
&+ \frac{4\kappa^{1/2}(1+\kappa)^3u}{3f}\cdot\fr{A}_t'  + \frac{2\kappa(1+\kappa)^3u}{\mathcal{H}^2f}H_{tt} = 0\,,
\end{align}
\begin{align}
H_{tz}' &+ \frac{\fr{q}f}{\fr{w}\mathcal{H}}\left(H_{tt}' - H_{zz}'\right) + \frac{\fr{q}f}{2\fr{w}\mathcal{H}}\left(\frac{f'}{f} - \frac{3\kappa}{\mathcal{H}}\right) H_{tt} \nonumber \\
&+ \frac{\kappa^{1/2}(1+\kappa)^3u}{\fr{w}\mathcal{H}}\left( \fr{q}\,\fr{A}_t + \fr{w}\,\fr{A}_z\right)  = 0\,,
\end{align}
\begin{align}
H_{aa}'' + \left(-\frac{2}{u}+\frac{\kappa}{\mathcal{H}} + \frac{f'}{f}\right) H_{aa}' &+ \left(\frac{1}{u} - \frac{\kappa}{\mathcal{H}}\right)\left( H_{tt}' - H_{zz}'\right) + \frac{4\kappa^{1/2}(1+\kappa)^3u}{3f}\fr{A}_t' \nonumber \\ 
&+ \frac{\fr{D}}{uf^2} H_{aa} + \frac{2\kappa(1+\kappa)^3u}{\mathcal{H}^2f}H_{tt} = 0\,,
\end{align}
\begin{align}
H_{zz}'' &+ \left(-\frac{3}{2u} + \frac{\kappa}{2\mathcal{H}} + \frac{f'}{f}\right)H_{zz}' + \frac{1}{2}\left( \frac{1}{u} - \frac{\kappa}{\mathcal{H}}\right)\left(H_{tt}' - H_{aa}'\right)\nonumber\\
&+ \frac{2\kappa^{1/2}(1+\kappa)^3u}{3f}\fr{A}_t' + \frac{\fr{w}\mathcal{H}^3}{uf^2} H_{zz}  - \frac{\fr{q}^2}{uf} H_{aa} \nonumber\\
&+ \left(\frac{\kappa(1+\kappa)^3u}{\mathcal{H}^2f} + \frac{\fr{q}^2}{uf}\right) H_{tt} + \frac{2\fr{qw}\mathcal{H}^3}{uf^2} H_{tz} = 0\,.
\end{align}

\bigskip

\bibliographystyle{JHEP}
\bibliography{stu-sound-refs}

\providecommand{\href}[2]{#2}\begingroup\raggedright\begin{thebibliography}{10}

\bibitem{Romatschke:2017ejr}
P.~Romatschke and U.~Romatschke, {\em {Relativistic Fluid Dynamics In and Out
  of Equilibrium}}.
\newblock Cambridge Monographs on Mathematical Physics. Cambridge University
  Press, 5, 2019.

\bibitem{Heinz:2024jwu}
U.~Heinz and B.~Schenke, {\it {Hydrodynamic Description of the Quark-Gluon
  Plasma}},  \href{http://arxiv.org/abs/2412.19393}{{\tt arXiv:2412.19393}}.

\bibitem{Baiotti:2016qnr}
L.~Baiotti and L.~Rezzolla, {\it {Binary neutron star mergers: a review of
  Einstein{\textquoteright}s richest laboratory}},  {\em Rept. Prog. Phys.}
  {\bf 80} (2017), no.~9 096901, [\href{http://arxiv.org/abs/1607.03540}{{\tt
  arXiv:1607.03540}}].

\bibitem{Annala:2019puf}
E.~Annala, T.~Gorda, A.~Kurkela, J.~N\"attil\"a, and A.~Vuorinen, {\it
  {Evidence for quark-matter cores in massive neutron stars}},  {\em Nature
  Phys.} {\bf 16} (2020), no.~9 907--910,
  [\href{http://arxiv.org/abs/1903.09121}{{\tt arXiv:1903.09121}}].

\bibitem{Lovato:2022vgq}
A.~Lovato et~al., {\it {Long Range Plan: Dense matter theory for heavy-ion
  collisions and neutron stars}},  \href{http://arxiv.org/abs/2211.02224}{{\tt
  arXiv:2211.02224}}.

\bibitem{LL6}
L.~D. Landau and E.~M. Lifshitz, {\em Fluid Mechanics}.
\newblock Pergamon, 1987.

\bibitem{rhic}
``{{Relativistic Heavy Ion Collider (RHIC)}}.''
  \url{https://www.bnl.gov/rhic/}.

\bibitem{nica}
``{{Nuclotron-based Ion Collider fAcility (NICA)}}.''
  \url{https://nica.jinr.ru/}.

\bibitem{MPD:2022qhn}
{\bf MPD} Collaboration, V.~Abgaryan et~al., {\it {Status and initial physics
  performance studies of the MPD experiment at NICA}},  {\em Eur. Phys. J. A}
  {\bf 58} (2022), no.~7 140, [\href{http://arxiv.org/abs/2202.08970}{{\tt
  arXiv:2202.08970}}].

\bibitem{fair}
``{{Facility for Antiproton and Ion Research (FAIR)}}.''
  \url{https://www.gsi.de/en/researchaccelerators/fair}.

\bibitem{Goedbloed_Keppens_Poedts_2019}
H.~Goedbloed, R.~Keppens, and S.~Poedts, {\em Magnetohydrodynamics of
  Laboratory and Astrophysical Plasmas}.
\newblock Cambridge University Press, 2019.

\bibitem{Greif:2017byw}
M.~Greif, J.~A. Fotakis, G.~S. Denicol, and C.~Greiner, {\it {Diffusion of
  conserved charges in relativistic heavy ion collisions}},  {\em Phys. Rev.
  Lett.} {\bf 120} (2018), no.~24 242301,
  [\href{http://arxiv.org/abs/1711.08680}{{\tt arXiv:1711.08680}}].

\bibitem{Plumberg:2024leb}
C.~Plumberg et~al., {\it {Conservation of B, S, and Q charges in relativistic
  viscous hydrodynamics solved with smoothed particle hydrodynamics}},  {\em
  Phys. Rev. C} {\bf 111} (2025), no.~4 044905,
  [\href{http://arxiv.org/abs/2405.09648}{{\tt arXiv:2405.09648}}].

\bibitem{Most:2021uck}
E.~R. Most, J.~Noronha, and A.~A. Philippov, {\it {Modelling
  general-relativistic plasmas with collisionless moments and dissipative
  two-fluid magnetohydrodynamics}},  {\em Mon. Not. Roy. Astron. Soc.} {\bf
  514} (2022), no.~4 4989--5003, [\href{http://arxiv.org/abs/2111.05752}{{\tt
  arXiv:2111.05752}}].

\bibitem{Aharony:1999ti}
O.~Aharony, S.~S. Gubser, J.~M. Maldacena, H.~Ooguri, and Y.~Oz, {\it {Large N
  field theories, string theory and gravity}},  {\em Phys. Rept.} {\bf 323}
  (2000) 183--386, [\href{http://arxiv.org/abs/hep-th/9905111}{{\tt
  hep-th/9905111}}].

\bibitem{callen-greene}
R.~F. Greene and H.~B. Callen, {\it On the formalism of thermodynamic
  fluctuation theory},  {\em Phys. Rev.} {\bf 83} (Sep, 1951) 1231--1235.

\bibitem{Kubo:1991}
R.~Kubo, M.~Toda, and N.~Hashitsume, {\em Statistical Physics II:
  Nonequilibrium Statistical Mechanics}, vol.~31 of {\em Springer Series in
  Solid-State Sciences}.
\newblock Springer-Verlag, Berlin, Heidelberg, 2nd~ed., 1991.

\bibitem{zubarev-book}
D.~N. Zubarev, {\em {Non-equilibrium Statistical Thermodynamics}}.
\newblock Plenum Press, New York, 1974.

\bibitem{kvasnikov-book-3}
I.~A. Kvasnikov, {\em {Thermodynamics and Statistical Physics: A Theory of
  Non-equilibrium Systems (in Russian)}}.
\newblock Editorial URSS, Moscow, 2003.

\bibitem{McQuarrie:2000}
D.~A. McQuarrie, {\em Statistical Mechanics}.
\newblock University Science Books, Sausalito, CA, 2000.

\bibitem{pathria-book}
R.~Pathria and P.~D. Beale, {\em {Statistical Mechanics}}.
\newblock Academic Press, New York, 2021.

\bibitem{Einstein-1910}
A.~Einstein, {\it Theorie der opaleszenz von homogenen fl\"{u}ssigkeiten und
  fl\"{u}ssigkeitsgemischen in der n\"{a}he des kritischen zustandes},  {\em
  Annalen der Physik} {\bf 338} (1910), no.~16 1275--1298,
  [\href{http://arxiv.org/abs/https://onlinelibrary.wiley.com/doi/pdf/10.1002/andp.19103381612}{{\tt
  https://onlinelibrary.wiley.com/doi/pdf/10.1002/andp.19103381612}}].

\bibitem{Kovtun:2012rj}
P.~Kovtun, {\it {Lectures on hydrodynamic fluctuations in relativistic
  theories}},  {\em J. Phys. A} {\bf 45} (2012) 473001,
  [\href{http://arxiv.org/abs/1205.5040}{{\tt arXiv:1205.5040}}].

\bibitem{Maldacena:1997re}
J.~M. Maldacena, {\it {The Large $N$ limit of superconformal field theories and
  supergravity}},  {\em Adv. Theor. Math. Phys.} {\bf 2} (1998) 231--252,
  [\href{http://arxiv.org/abs/hep-th/9711200}{{\tt hep-th/9711200}}].

\bibitem{Gubser:1998bc}
S.~S. Gubser, I.~R. Klebanov, and A.~M. Polyakov, {\it {Gauge theory
  correlators from noncritical string theory}},  {\em Phys. Lett. B} {\bf 428}
  (1998) 105--114, [\href{http://arxiv.org/abs/hep-th/9802109}{{\tt
  hep-th/9802109}}].

\bibitem{Witten:1998qj}
E.~Witten, {\it {Anti-de Sitter space and holography}},  {\em Adv. Theor. Math.
  Phys.} {\bf 2} (1998) 253--291,
  [\href{http://arxiv.org/abs/hep-th/9802150}{{\tt hep-th/9802150}}].

\bibitem{Yamada:2006rx}
D.~Yamada and L.~G. Yaffe, {\it {Phase diagram of N=4 super-Yang-Mills theory
  with R-symmetry chemical potentials}},  {\em JHEP} {\bf 09} (2006) 027,
  [\href{http://arxiv.org/abs/hep-th/0602074}{{\tt hep-th/0602074}}].

\bibitem{Cai:1998ji}
R.-G. Cai and K.-S. Soh, {\it {Critical behavior in the rotating D-branes}},
  {\em Mod. Phys. Lett. A} {\bf 14} (1999) 1895--1908,
  [\href{http://arxiv.org/abs/hep-th/9812121}{{\tt hep-th/9812121}}].

\bibitem{Cvetic:1999ne}
M.~Cvetic and S.~S. Gubser, {\it {Phases of R charged black holes, spinning
  branes and strongly coupled gauge theories}},  {\em JHEP} {\bf 04} (1999)
  024, [\href{http://arxiv.org/abs/hep-th/9902195}{{\tt hep-th/9902195}}].

\bibitem{Cvetic:1999rb}
M.~Cvetic and S.~S. Gubser, {\it {Thermodynamic stability and phases of general
  spinning branes}},  {\em JHEP} {\bf 07} (1999) 010,
  [\href{http://arxiv.org/abs/hep-th/9903132}{{\tt hep-th/9903132}}].

\bibitem{Harmark:1999xt}
T.~Harmark and N.~A. Obers, {\it {Thermodynamics of spinning branes and their
  dual field theories}},  {\em JHEP} {\bf 01} (2000) 008,
  [\href{http://arxiv.org/abs/hep-th/9910036}{{\tt hep-th/9910036}}].

\bibitem{Cvetic:1999xp}
M.~Cvetic, M.~J. Duff, P.~Hoxha, J.~T. Liu, H.~Lu, J.~X. Lu,
  R.~Martinez-Acosta, C.~N. Pope, H.~Sati, and T.~A. Tran, {\it {Embedding AdS
  black holes in ten-dimensions and eleven-dimensions}},  {\em Nucl. Phys. B}
  {\bf 558} (1999) 96--126, [\href{http://arxiv.org/abs/hep-th/9903214}{{\tt
  hep-th/9903214}}].

\bibitem{Cvetic:2000nc}
M.~Cvetic, H.~Lu, C.~N. Pope, A.~Sadrzadeh, and T.~A. Tran, {\it {Consistent
  SO(6) reduction of type IIB supergravity on $S^5$}},  {\em Nucl. Phys. B}
  {\bf 586} (2000) 275--286, [\href{http://arxiv.org/abs/hep-th/0003103}{{\tt
  hep-th/0003103}}].

\bibitem{Behrndt:1998jd}
K.~Behrndt, M.~Cvetic, and W.~Sabra, {\it {Nonextreme black holes of
  five-dimensional N=2 AdS supergravity}},  {\em Nucl. Phys. B} {\bf 553}
  (1999) 317--332, [\href{http://arxiv.org/abs/hep-th/9810227}{{\tt
  hep-th/9810227}}].

\bibitem{Gubser:2000ec}
S.~S. Gubser and I.~Mitra, {\it {Instability of charged black holes in Anti-de
  Sitter space}},  {\em Clay Math. Proc.} {\bf 1} (2002) 221,
  [\href{http://arxiv.org/abs/hep-th/0009126}{{\tt hep-th/0009126}}].

\bibitem{Gubser:2000mm}
S.~S. Gubser and I.~Mitra, {\it {The Evolution of unstable black holes in
  anti-de Sitter space}},  {\em JHEP} {\bf 08} (2001) 018,
  [\href{http://arxiv.org/abs/hep-th/0011127}{{\tt hep-th/0011127}}].

\bibitem{Son:2006em}
D.~T. Son and A.~O. Starinets, {\it {Hydrodynamics of r-charged black holes}},
  {\em JHEP} {\bf 03} (2006) 052,
  [\href{http://arxiv.org/abs/hep-th/0601157}{{\tt hep-th/0601157}}].

\bibitem{Buchel:2010gd}
A.~Buchel, {\it {Critical phenomena in N=4 SYM plasma}},  {\em Nucl. Phys. B}
  {\bf 841} (2010) 59--99, [\href{http://arxiv.org/abs/1005.0819}{{\tt
  arXiv:1005.0819}}].

\bibitem{Gentle:2012rg}
S.~A. Gentle and B.~Withers, {\it {Superconducting instabilities of R-charged
  black branes}},  {\em JHEP} {\bf 10} (2012) 006,
  [\href{http://arxiv.org/abs/1207.3086}{{\tt arXiv:1207.3086}}].

\bibitem{Henriksson:2019zph}
O.~Henriksson, C.~Hoyos, and N.~Jokela, {\it {Novel color superconducting
  phases of $\cal{N}$ = 4 super Yang-Mills at strong coupling}},  {\em JHEP}
  {\bf 09} (2019) 088, [\href{http://arxiv.org/abs/1907.01562}{{\tt
  arXiv:1907.01562}}].

\bibitem{Anabalon:2024lgp}
A.~Anabalon and J.~Oliva, {\it {Plasma-Plasma Third Order Phase Transition from
  Type IIB Supergravity}},  {\em Phys. Rev. Lett.} {\bf 133} (2024), no.~12
  121601, [\href{http://arxiv.org/abs/2405.04611}{{\tt arXiv:2405.04611}}].

\bibitem{Gladden:2024ssb}
L.~Gladden, V.~Ivo, P.~Kovtun, and A.~O. Starinets, {\it {Instability in N=4
  supersymmetric Yang-Mills theory at finite density}},  {\em Phys. Rev. D}
  {\bf 111} (2025), no.~8 086030, [\href{http://arxiv.org/abs/2412.12353}{{\tt
  arXiv:2412.12353}}].

\bibitem{Reall:2001ag}
H.~S. Reall, {\it {Classical and thermodynamic stability of black branes}},
  {\em Phys. Rev. D} {\bf 64} (2001) 044005,
  [\href{http://arxiv.org/abs/hep-th/0104071}{{\tt hep-th/0104071}}].

\bibitem{Buchel:2005nt}
A.~Buchel, {\it {A Holographic perspective on Gubser-Mitra conjecture}},  {\em
  Nucl. Phys. B} {\bf 731} (2005) 109--124,
  [\href{http://arxiv.org/abs/hep-th/0507275}{{\tt hep-th/0507275}}].

\bibitem{Emparan:2008eg}
R.~Emparan and H.~S. Reall, {\it {Black Holes in Higher Dimensions}},  {\em
  Living Rev. Rel.} {\bf 11} (2008) 6,
  [\href{http://arxiv.org/abs/0801.3471}{{\tt arXiv:0801.3471}}].

\bibitem{Son:2002sd}
D.~T. Son and A.~O. Starinets, {\it {Minkowski space correlators in AdS / CFT
  correspondence: Recipe and applications}},  {\em JHEP} {\bf 09} (2002) 042,
  [\href{http://arxiv.org/abs/hep-th/0205051}{{\tt hep-th/0205051}}].

\bibitem{Kovtun:2005ev}
P.~K. Kovtun and A.~O. Starinets, {\it {Quasinormal modes and holography}},
  {\em Phys. Rev.} {\bf D72} (2005) 086009,
  [\href{http://arxiv.org/abs/hep-th/0506184}{{\tt hep-th/0506184}}].

\bibitem{Mas:2006dy}
J.~Mas, {\it {Shear viscosity from R-charged AdS black holes}},  {\em JHEP}
  {\bf 03} (2006) 016, [\href{http://arxiv.org/abs/hep-th/0601144}{{\tt
  hep-th/0601144}}].

\bibitem{Kovtun:2003wp}
P.~Kovtun, D.~T. Son, and A.~O. Starinets, {\it {Holography and hydrodynamics:
  Diffusion on stretched horizons}},  {\em JHEP} {\bf 10} (2003) 064,
  [\href{http://arxiv.org/abs/hep-th/0309213}{{\tt hep-th/0309213}}].

\bibitem{Kovtun:2004de}
P.~Kovtun, D.~T. Son, and A.~O. Starinets, {\it {Viscosity in strongly
  interacting quantum field theories from black hole physics}},  {\em Phys.
  Rev. Lett.} {\bf 94} (2005) 111601,
  [\href{http://arxiv.org/abs/hep-th/0405231}{{\tt hep-th/0405231}}].

\bibitem{Buchel:2003tz}
A.~Buchel and J.~T. Liu, {\it {Universality of the shear viscosity in
  supergravity}},  {\em Phys. Rev. Lett.} {\bf 93} (2004) 090602,
  [\href{http://arxiv.org/abs/hep-th/0311175}{{\tt hep-th/0311175}}].

\bibitem{Starinets:2008fb}
A.~O. Starinets, {\it {Quasinormal spectrum and the black hole membrane
  paradigm}},  {\em Phys. Lett. B} {\bf 670} (2009) 442--445,
  [\href{http://arxiv.org/abs/0806.3797}{{\tt arXiv:0806.3797}}].

\bibitem{Callen}
H.~B. Callen, {\em Thermodynamics and introduction to thermostatistics}.
\newblock Wiley, 1985.

\bibitem{Gubser:1998jb}
S.~S. Gubser, {\it {Thermodynamics of spinning D3-branes}},  {\em Nucl. Phys.
  B} {\bf 551} (1999) 667--684,
  [\href{http://arxiv.org/abs/hep-th/9810225}{{\tt hep-th/9810225}}].

\bibitem{Chamblin:1999tk}
A.~Chamblin, R.~Emparan, C.~V. Johnson, and R.~C. Myers, {\it {Charged AdS
  black holes and catastrophic holography}},  {\em Phys. Rev.} {\bf D60} (1999)
  064018, [\href{http://arxiv.org/abs/hep-th/9902170}{{\tt hep-th/9902170}}].

\bibitem{Bergshoeff:2004kh}
E.~Bergshoeff, S.~Cucu, T.~de~Wit, J.~Gheerardyn, S.~Vandoren, and
  A.~Van~Proeyen, {\it {N = 2 supergravity in five-dimensions revisited}},
  {\em Class. Quant. Grav.} {\bf 21} (2004) 3015--3042,
  [\href{http://arxiv.org/abs/hep-th/0403045}{{\tt hep-th/0403045}}].

\bibitem{Gunaydin:1999zx}
M.~Gunaydin and M.~Zagermann, {\it {The Gauging of five-dimensional, N=2
  Maxwell-Einstein supergravity theories coupled to tensor multiplets}},  {\em
  Nucl. Phys. B} {\bf 572} (2000) 131--150,
  [\href{http://arxiv.org/abs/hep-th/9912027}{{\tt hep-th/9912027}}].

\bibitem{Ceresole:2000jd}
A.~Ceresole and G.~Dall'Agata, {\it {General matter coupled N=2, D = 5 gauged
  supergravity}},  {\em Nucl. Phys. B} {\bf 585} (2000) 143--170,
  [\href{http://arxiv.org/abs/hep-th/0004111}{{\tt hep-th/0004111}}].

\bibitem{Gunaydin:1983bi}
M.~Gunaydin, G.~Sierra, and P.~K. Townsend, {\it {The Geometry of N=2
  Maxwell-Einstein Supergravity and Jordan Algebras}},  {\em Nucl. Phys. B}
  {\bf 242} (1984) 244--268.

\bibitem{Horowitz:1998ha}
G.~T. Horowitz and R.~C. Myers, {\it {The AdS / CFT correspondence and a new
  positive energy conjecture for general relativity}},  {\em Phys. Rev. D} {\bf
  59} (1998) 026005, [\href{http://arxiv.org/abs/hep-th/9808079}{{\tt
  hep-th/9808079}}].

\bibitem{Lagrange1762}
J.-L. Lagrange, {\it Nouvelles recherches sur la nature et la propagation du
  son},  {\em M{\'e}langes de Turin} {\bf 2} (1762) 263--265. Reprinted in
  \emph{Oeuvres de Lagrange}, vol.~1, Gauthier-Villars, Paris, 1867.

\bibitem{Lagrange2007}
J.-L. Lagrange, {\em Analytical Mechanics}.
\newblock Dover Publications, Mineola, NY, 2007.
\newblock Unabridged republication of the 1898 English translation by A.~E.~H.
  Love, based on the 1811 French edition of \emph{M{\'e}canique Analytique}.

\bibitem{Policastro:2002tn}
G.~Policastro, D.~T. Son, and A.~O. Starinets, {\it {From AdS / CFT
  correspondence to hydrodynamics. 2. Sound waves}},  {\em JHEP} {\bf 12}
  (2002) 054, [\href{http://arxiv.org/abs/hep-th/0210220}{{\tt
  hep-th/0210220}}].

\bibitem{Baier:2007ix}
R.~Baier, P.~Romatschke, D.~T. Son, A.~O. Starinets, and M.~A. Stephanov, {\it
  {Relativistic viscous hydrodynamics, conformal invariance, and holography}},
  {\em JHEP} {\bf 04} (2008) 100, [\href{http://arxiv.org/abs/0712.2451}{{\tt
  arXiv:0712.2451}}].

\bibitem{chiral-life}
V.~Almeida~Ivo, P.~K. Kovtun, and A.~O. Starinets, {\it {The complex chiral
  life of hydrodynamic modes in strongly coupled ${\cal N}=4$ SYM theory, {\it
  to appear}}},  \href{http://arxiv.org/abs/2025}{{\tt 2025}}.

\bibitem{Tester-Modell}
J.~W. Tester and M.~Modell, {\em Thermodynamics and its applications}.
\newblock Prentice Hall, 1997.

\bibitem{Policastro:2002se}
G.~Policastro, D.~T. Son, and A.~O. Starinets, {\it {From AdS / CFT
  correspondence to hydrodynamics}},  {\em JHEP} {\bf 09} (2002) 043,
  [\href{http://arxiv.org/abs/hep-th/0205052}{{\tt hep-th/0205052}}].

\bibitem{Maeda:2008hn}
K.~Maeda, M.~Natsuume, and T.~Okamura, {\it {Dynamic critical phenomena in the
  AdS/CFT duality}},  {\em Phys. Rev. D} {\bf 78} (2008) 106007,
  [\href{http://arxiv.org/abs/0809.4074}{{\tt arXiv:0809.4074}}].

\bibitem{Buchel:2009mf}
A.~Buchel and C.~Pagnutti, {\it {Transport at criticality}},  {\em Nucl. Phys.
  B} {\bf 834} (2010) 222--236, [\href{http://arxiv.org/abs/0912.3212}{{\tt
  arXiv:0912.3212}}].

\bibitem{Natsuume:2010bs}
M.~Natsuume and T.~Okamura, {\it {Dynamic universality class of large-N gauge
  theories}},  {\em Phys. Rev. D} {\bf 83} (2011) 046008,
  [\href{http://arxiv.org/abs/1012.0575}{{\tt arXiv:1012.0575}}].

\bibitem{DeWolfe:2011ts}
O.~DeWolfe, S.~S. Gubser, and C.~Rosen, {\it {Dynamic critical phenomena at a
  holographic critical point}},  {\em Phys. Rev. D} {\bf 84} (2011) 126014,
  [\href{http://arxiv.org/abs/1108.2029}{{\tt arXiv:1108.2029}}].

\bibitem{Buchel:2010wk}
A.~Buchel and C.~Pagnutti, {\it {Correlated stability conjecture revisited}},
  {\em Phys. Lett. B} {\bf 697} (2011) 168--172,
  [\href{http://arxiv.org/abs/1010.5748}{{\tt arXiv:1010.5748}}].

\bibitem{Buchel:2025tjq}
A.~Buchel, {\it {The ordered phase of charged N=4 SYM plasma from STU}},
  \href{http://arxiv.org/abs/2501.12403}{{\tt arXiv:2501.12403}}.

\bibitem{Gouteraux:2024adm}
B.~Gout\'eraux and E.~Mefford, {\it {Linear dynamical stability and the laws of
  thermodynamics}},  \href{http://arxiv.org/abs/2407.07939}{{\tt
  arXiv:2407.07939}}.

\bibitem{Yaffe:2017axl}
L.~G. Yaffe, {\it {Large $N$ phase transitions and the fate of small
  Schwarzschild-AdS black holes}},  {\em Phys. Rev. D} {\bf 97} (2018), no.~2
  026010, [\href{http://arxiv.org/abs/1710.06455}{{\tt arXiv:1710.06455}}].

\bibitem{Buchel:2025cve}
A.~Buchel, {\it {The ordered phase of charged N=4 SYM plasma}},
  \href{http://arxiv.org/abs/2501.01856}{{\tt arXiv:2501.01856}}.

\bibitem{Buchel:2025ves}
A.~Buchel, {\it {On the relevance of GIKS instability of charged N=4 SYM
  plasma}},  \href{http://arxiv.org/abs/2502.11354}{{\tt arXiv:2502.11354}}.

\bibitem{Armas:2025fvo}
J.~Armas and G.~Batzios, {\it {Chiral anomaly from (anomalous) spin
  hydrodynamics}},  \href{http://arxiv.org/abs/2505.01843}{{\tt
  arXiv:2505.01843}}.

\bibitem{Choi:2024xnv}
S.~Choi, D.~Jain, S.~Kim, V.~Krishna, E.~Lee, S.~Minwalla, and C.~Patel, {\it
  {Dual Dressed Black Holes as the end point of the Charged Superradiant
  instability in ${\cal N} = 4$ Yang Mills}},
  \href{http://arxiv.org/abs/2409.18178}{{\tt arXiv:2409.18178}}.

\bibitem{Gubser:2008px}
S.~S. Gubser, {\it {Breaking an Abelian gauge symmetry near a black hole
  horizon}},  {\em Phys. Rev. D} {\bf 78} (2008) 065034,
  [\href{http://arxiv.org/abs/0801.2977}{{\tt arXiv:0801.2977}}].

\bibitem{Frobenius1877}
G.~Frobenius, {\it Ueber lineare substitutionen und bilineare formen.},  {\em
  Journal f\"{u}r die reine und angewandte Mathematik} {\bf 84} (1877) 1--63.

\bibitem{BiedenharnL.C.1963OtRo}
L.~C. Biedenharn, {\it On the representations of the semisimple lie groups. i.
  the explicit construction of invariants for the unimodular unitary group in n
  dimensions},  {\em Journal of mathematical physics} {\bf 4} (1963), no.~3
  436--445.

\bibitem{KleinAbraham1963IOot}
A.~Klein, {\it Invariant operators of the unitary unimodular group in n
  dimensions},  {\em Journal of mathematical physics} {\bf 4} (1963), no.~10
  1283--1284.

\bibitem{deAzcarraga:1997ya}
J.~A. de~Azcarraga, A.~J. Macfarlane, A.~J. Mountain, and J.~C. Perez~Bueno,
  {\it {Invariant tensors for simple groups}},  {\em Nucl. Phys. B} {\bf 510}
  (1998) 657--687, [\href{http://arxiv.org/abs/physics/9706006}{{\tt
  physics/9706006}}].

\bibitem{Grozdanov:2016fkt}
S.~Grozdanov and A.~O. Starinets, {\it {Second-order transport, quasinormal
  modes and zero-viscosity limit in the Gauss-Bonnet holographic fluid}},  {\em
  JHEP} {\bf 03} (2017) 166, [\href{http://arxiv.org/abs/1611.07053}{{\tt
  arXiv:1611.07053}}].

\end{thebibliography}\endgroup

\end{document}